\documentclass[11pt,a4paper]{article}
\pdfoutput=1
\usepackage{jheppub}

\usepackage{placeins}
\usepackage{lmodern}
\usepackage[T1]{fontenc}
\usepackage[utf8]{inputenc}
\usepackage{epsf}
\usepackage{float}
\usepackage{rotating}
\usepackage{diagbox}
\usepackage{slashed}

\usepackage[usenames,dvipsnames]{xcolor}

\let\originalleft\left
\let\originalright\right
\renewcommand{\left}{\mathopen{}\mathclose\bgroup\originalleft}
\renewcommand{\right}{\aftergroup\egroup\originalright}

\newcommand{\eq}{\begin{equation}}
\newcommand{\eeq}{\end{equation}}

\newcommand{\nc}{\newcommand}
\nc{\rnc}{\renewcommand}
\rnc{\d}{\mathrm{d}}
\nc{\D}{\partial}
\nc{\K}{\kappa}
\nc{\bK}{\bar{\K}}
\nc{\bk}{\bar{k}}
\nc{\vbq}{\vec{\bar{q}}}
\nc{\g}{\gamma}
\nc{\lrarrow}{\leftrightarrow}
\nc{\rg}{\sqrt{g}}
\rnc{\[}{\begin{equation}}
\rnc{\]}{\end{equation}}
\nc{\bea}{\begin{eqnarray}}
\nc{\eea}{\end{eqnarray}}
\nc{\nn}{\nonumber}
\rnc{\(}{\left(}
\rnc{\)}{\right)}
\nc{\q}{\vec{q}}
\nc{\x}{\vec{x}}
\rnc{\a}{\hat{a}}
\nc{\ep}{\epsilon}
\nc{\tto}{\rightarrow}
\rnc{\inf}{\infty}
\rnc{\Re}{\mathrm{Re}}
\rnc{\Im}{\mathrm{Im}}
\nc{\z}{\zeta}
\nc{\mA}{\mathcal{A}}
\nc{\mB}{\mathcal{B}}
\nc{\mC}{\mathcal{C}}
\nc{\mD}{\mathcal{D}}
\nc{\mN}{\mathcal{N}}
\rnc{\H}{\mathcal{H}}
\rnc{\L}{\mathcal{L}}
\nc{\<}{\langle}
\rnc{\>}{\rangle}
\nc{\fnl}{f_{NL}}
\nc{\gnl}{g_{NL}}
\nc{\fnleq}{f_{NL}^{equil.}}
\nc{\fnlloc}{f_{NL}^{local}}
\nc{\vphi}{\varphi}
\nc{\Lie}{\pounds}
\nc{\half}{\frac{1}{2}}
\nc{\bOmega}{\bar{\Omega}}
\nc{\bLambda}{\bar{\Lambda}}
\nc{\dN}{\delta N}
\nc{\gYM}{g_{\mathrm{YM}}}
\nc{\geff}{g_{\mathrm{eff}}}
\nc{\tr}{\mathrm{tr}}
\nc{\oa}{\stackrel{\leftrightarrow}}
\nc{\IR}{{\rm IR}}
\nc{\UV}{{\rm UV}}
\nc{\gth}{\lambda}

\begin{document}
\title{Two-point function of the energy-momentum tensor and  generalised conformal structure}

\author[a]{Claudio Corian\`{o}}
\author[b]{Luigi Delle Rose}
\author[c]{Kostas Skenderis}

\affiliation[a]{INFN-Lecce and Dipartimento di Matematica e Fisica, Universit\`{a} del Salento, Via Arnesano, 73100 Lecce, Italy} 
\affiliation[b]{INFN, Sezione di Firenze and Department of Physics and Astronomy, University of Florence,
Via G. Sansone 1, 50019 Sesto Fiorentino, Italy}
\affiliation[e]{STAG Research Centre and Mathematical Sciences, Highfield, University of Southampton, SO17 1BJ Southampton, UK}

\emailAdd{claudio.coriano@le.infn.it}
\emailAdd{luigi.dellerose@fi.infn.it}
\emailAdd{K.Skenderis@soton.ac.uk}

\abstract{Theories with generalised conformal structure contain a dimensionful parameter, which appears 
as an overall multiplicative factor in the action. Examples of such theories are gauge theories coupled to massless scalars and fermions with Yukawa interactions and quartic couplings for the scalars in spacetime dimensions other than 4. Many properties of such theories are similar to that of conformal field theories (CFT), and in particular their 2-point functions take the same form as in CFT but with the normalisation constant now replaced by a function of the effective dimensionless coupling  $g$ constructed from the dimensionful parameter and the distance separating the two operators.
Such theories appear in holographic dualities involving non-conformal branes and this behaviour of 
the correlators has already been observed at strong coupling. Here we present a perturbative computation of the two-point function of the energy-momentum tensor to two loops in dimensions $d= 3, 5$, confirming the expected structure and  determining the corresponding functions of $g$ to this order, including the effects of renormalisation. We also discuss the d=4 case for comparison. The results for $d=3$ are relevant for holographic cosmology, and in this case we also study the effect of a $\Phi^6$ coupling, which while marginal in the usual sense it is irrelevant from the perspective of the generalised conformal structure. Indeed, the effect of such coupling in the 2-point function is washed out in the IR but it modifies the UV. }

\maketitle
\newpage

\section{Introduction}

The space of quantum field theories contains distinguished points describing end-points of renormalisation group (RG) flow, where the theory becomes\footnote{Strictly speaking, vanishing of beta functions only implies scale invariance, 
but it turns out that often the theory at the fixed point is a CFT, see 
\cite{Zamolodchikov:1986gt, Polchinski:1987dy, Luty:2012ww, Dymarsky:2013pqa, Bzowski:2014qja, Nakayama:2013is}  for a sample of works regarding the issue of scale versus conformal invariance in dimension $d=2, 4$ and \cite{Jackiw:2011vz, ElShowk:2011gz} for a counterexample in $d \neq 4$: Maxwell theory. We note that this counterexample is a theory with generalised conformal structure.} 
 a conformal field theory (CFT). At the fixed point the structure of correlators is highly constrained and in particular the 2- and 3-point functions are uniquely determined up to constants \cite{DiFrancesco:1997nk}.  Away from the fixed point, the structure of correlators  is far less constrained and in general it is determined 
by case-by-case computations.  In this paper we will discuss a class of quantum field theories that sit in between 
the case of general QFTs and CFTs: this is the case of QFTs with generalised conformal structure. 

Theories with generalised conformal structure
have a dimensionful parameter, which appears in the action only as an overall parameter. This implies that the elementary fields can be assigned a scaling dimension such that all terms in the action scale the same way and all 
other parameters that enter in the action are dimensionless. Examples of such theories are gauge theories coupled to massless scalars and fermions, with Yukawa coupling and quartic couplings for the scalars. After appropriate rescaling of all fields one may arrange such that the Yang-Mills (YM) coupling constant (which is dimensionful in dimensions other than 4) appears only as an overall constant in the action. Assigning ``four-dimensional'' scaling to all fields, {\it i.e.} dimension 1 for gauge fields and scalars and dimension 3/2 for  fermions, all terms in the action have dimension 4.
Examples of such theories are maximally supersymmetric YM theories and it is in this context where generalised conformal symmetry was first introduced \cite{Jevicki:1998yr, Jevicki:1998ub}. It was observed that if one promotes the YM coupling constant to a field that transforms appropriately under conformal transformations then these theories are conformally invariant.  Relatedly, these theories can be coupled to background gravity in a Weyl invariant way, provided the coupling constant also transforms appropriately under Weyl transformations \cite{Kanitscheider:2008kd}.

While this is not a {\it bona fide} symmetry, it still constraints the structure of the correlators of the theory \cite{Kanitscheider:2008kd}. In particular, 2-point functions take the same form as in CFTs, except that now 
the constants become functions of the effective dimensionless coupling,
\bea 
\< O_\Delta(x) O_\Delta(0)\> = \frac{\tilde{c}_\Delta(\tilde{g})}{x^{2 \Delta}}
\eea
where $\tilde{g} =  g_{YM}^2 x^{4-d}$, or in momentum space, which we will use throughout this paper,
\bea \label{corr_GCS}
\< O_\Delta(q) O_\Delta(-q)\> = q^{2 \Delta-d} c_\Delta(g)
\eea
with $g = g_{YM}^2 / q^{4-d}$, and we suppress a momentum conserving delta function.
$\Delta$ is the dimension associated with the generalised conformal structure and $c_\Delta(g)$ is a general 
function of $g$ (and similar for $\tilde{c}_\Delta(\tilde{g})$). In CFTs $c_\Delta$ is a constant (in general may depend on exactly marginal couplings).
In perturbation theory, $g \ll 1$ and 
\bea \label{c_weak}
c_\Delta(g) = c_1  + c_2 g + \cdots 
\eea
with $c_i$ constants that may be obtained by an $i$-loop computation. So the dependence of the correlator on the momentum $q$ is predetermined (similar to CFTs) and it is only the constants $c_i$ that depend on which theory one is considering. It is the purpose of this paper to confirm this picture and compute the constants $c_i$, which we will call generalised conformal structure constants (GCSC), for the class of theories we consider. We emphasise that all computations that we present here are compatible with standard QFT expectations and do not requite any mentioning of general generalised conformal structure. Generalised conformal structure however provides a new view on these results. For example, the implications of dimensional analysis are   reinterpreted as that of generalised scale invariance.

Note that since $g$ depends on $q$ the question of whether perturbation theory is valid depends on the energy scales that we probe. For $d<4$ the theory is asymptotically free, {\it i.e.} $g \to 0$ for $q \to \infty$, so  the expansion in (\ref{c_weak}) is justified in the UV region, and for $d>4$ the theory is free in the IR and we only expect (\ref{c_weak}) 
to be valid in the IR.  

Quantum corrections could still modify (\ref{c_weak}), even in the perturbative regime. We will use dimensional regularisation to address this issue to 2-loops. Since the theory is massless, there are no infinities at 1-loop in odd dimensions, so when $d=3, 5$ the first correction appears at 2-loops and gives rise to a logarithmic correction,
\bea \label{c_weak_odd}
c_\Delta(g) = c_1  + c_2 g + \tilde{c}_2 g \log g + \cdots 
\eea
where the log is due to UV and IR divergences in $d=3$ and due to UV divergences in $d=5$.
We will not discuss the $d<3$ cases, which have severe IR singularities. We note however these dimensions include important models such as the D0 and D1 branes and  the SYK model \footnote{see in particular
\cite{Maldacena:2016hyu} for a relevant discussion of this model and \cite{Taylor:2017dly} for the connection to generalised conformal symmetry.}. In even dimensions, there are singularities already at 1-loop. Since we exclude $d<3$, the first case to discuss is $d=4$. In this case however  there is no generalised conformal structure:  $g_{YM}^2$ is dimensionless and the perturbative expansion does not determine the form of the momentum dependence. Nevertheless, as QFT in $d=4$ is textbook material this case serves as benchmark for the $d=3$ and $d=5$ cases. Moreover, to our knowledge the renormalised 2-point of the energy-momentum tensor  for the general class of theories we discuss here has not appeared before. The next case is $d=6$. An example would be  D5-branes 
but it is known that at least at strong coupling this case is special (see for example\cite{Itzhaki:1998dd, Boonstra:1998mp}), and we will not discuss it here.

In the opposite regime (IR for $d<4$ and UV for $d>4$) the effective coupling $g$ becomes strong, and one 
may question whether the generalised conformal structure would survive in this regime. Remarkably, in the cases 
where there is a working gauge/gravity duality  \cite{Itzhaki:1998dd, Boonstra:1998mp} the dual supergravity solution exhibits generalised conformal structure \cite{Jevicki:1998yr, Jevicki:1998ub, Kanitscheider:2008kd}. In such cases
the correlators still take the form (\ref{corr_GCS}) but now $c_\Delta(g)$ has a strong-coupling expansion. In these strong-coupling examples the generalised conformal structure is further linked with a strong-coupling fixed point but in ``fractional  number of dimensions'': the bulk action and the solutions can be obtained from a higher dimensional AdS via a generalised dimensional reduction (compactification over a torus and then continuation in the dimension of 
the torus) \cite{Kanitscheider:2009as}.

We emphasise that in all cases (\ref{corr_GCS}) is valid only for a limited range of momenta. 
For example, the $\Phi^4$ $O(N)$ vector model in $d=3$ is governed by generalised conformal structure 
for a range of momenta near the UV fixed point, but it flows to a non-trivial fixed point in the IR.
In the gauge/gravity examples discussed in \cite{Itzhaki:1998dd, Boonstra:1998mp,Kanitscheider:2008kd}
one takes the large $N$ limit while keeping fixed and large the effective 
't Hooft coupling $\lambda = g N$ but still small relative to $N$ such that the dilaton is small. However, there is always a regime (a range of momenta) in which
the dilaton becomes large and the theory exits the phase governed by generalised conformal structure.
For example, in the case of D2 and D4 branes the strong dilaton regime takes us to M-theory with the D2 and D4 branes lifted to M2 and M5 branes and correspondingly the D2 theory flows in the IR to the ABJM theory and the D4 theory becomes in the UV the (2,0) theory. 

In this paper we will discuss the perturbative computation of the 2-point function of the energy-momentum tensor to 2-loops. The original motivation for this computation was its application to holographic cosmology. Three dimensional 
QFTs with generalised conformal structure were proposed in \cite{McFadden:2009fg} as holographic models describing a non-geometric very early Universe. The 1-loop computation was discussed in \cite{McFadden:2010na} and the structure of the 2-point function to 2-loops in \cite{Easther:2011wh}. The same paper contained a custom-fit of these models to WMAP and found that these models are compatible to CMB data and competitive to $\Lambda$CDM. 
With the view to comparison to PLANCK data, a precise 2-loop computation was needed. The result of the 2-loop computation was reported (without derivation) in \cite{Afshordi:2016dvb}, which discusses the custom-fit of these models  to  PLANCK data (see also \cite{Afshordi:2017ihr}), again finding that these models are competitive to $\Lambda$CDM. Another purpose of this paper is to provide the technical details that led to the results used in \cite{Afshordi:2016dvb}.

Working with dimensional regularisation, the regularised computation may be used in different dimensions.
To renormalise the 2-point function of the energy-momentum tensor, one first needs to renormalise the 2-point functions of elementary fields. This computation also serves to illustrate   (\ref{corr_GCS}) but now with $O_\Delta$ being an elementary  field (scalar, fermion or gauge field). This computation may also be used to justify the assignments of dimensions to the 
elementary fields under generalised conformal structure. The three cases we discuss ($d=3, 4, 5$) cover super-renormalisable, renormalisable and non-renormalisable theories. Yet up to 2-loops the computations can be done in parallel. 

The perturbative computation requires the evaluation of 2-loop tensor  integrals. We developed a tensor reduction to scalar integrals implementing in the TARCER package \cite{Mertig:1998vk} an algorithm proposed by Tarasov \cite{Tarasov:1997kx,Tarasov:1996br}. The results for the integrals may be of general use and are listed in appendix \ref{tensrid}.

Perturbative analysis of correlators of the energy-momentum tensor has been done before but mostly in the context of conformal field theories. 
Previous perturbative results (almost all 1-loop) were reported for $d=3$ in \cite{McFadden:2009fg, McFadden:2010na, McFadden:2010vh, Maldacena:2011nz, Bzowski:2011ab, Coriano:2012hd} and for $d=4$ in \cite{Armillis:2009pq,Armillis:2010qk, Coriano:2018bbe, Giannotti:2008cv, Coriano:2011zk, Bzowski:2013sza, Bzowski:2015pba, Coriano:2018bsy}. The analysis of correlation functions of elementary fields have been performed in the past in various gauges, up to 2-loop level, most notably the background field gauge  \cite{Jack:1982hf,Jack:1982sn,Jack:1983sk}, focused around the case $d=4$. We are not aware of any similar perturbative computations of energy-momentum correlators in $d>4$.

Returning to holographic cosmology, one outstanding question is how to exit from the non-geometric phase to Einstein gravity. This would be the analogue of the reheating phase of conventional inflationary models. 
Recall that  time evolution is mapped to inverse RG flow in holographic cosmology, and as discussed in \cite{Easther:2011wh} 
in order to exit from the non-geometric phase  we would need to change the UV of the holographic theory. Here we take a  first step towards  building such model: we add a $\Phi^6$ terms in the Lagrangian and compute its contribution at low energies. While such term is marginal in the usual sense, it is irrelevant relative to the generalised conformal structure. Indeed, we will see that it induces a beta function for the quartic coupling and we will discuss its contribution to the 2-point function of the energy-momentum tensor.

This paper is organised as follows. 
In section \ref{defdef}  we introduce the QFT we will analyse and discuss our conventions. Then in section \ref{sec:prelim}
we discuss the UV structure of the correlators and outline the tensor reduction method we used to calculate the relevant 2-loop diagrams.
Section \ref{gcs} is devoted to the analysis of the 2-point of elementary fields;
we present their expressions first for general dimensions and then in $d=3,4$ and $5$ dimensions, discussing in each case specific aspects of their renormalisation. We then turn to the study of the TT correlator in section \ref{TTsection}
and discuss how to renormalise it in section \ref{renTT}. In section \ref{hol} we address the application of our results to holographic cosmology, followed by an analysis of the implications of the addition of a  $\Phi^6$ term to the action in section \ref{rgflows}.  We conclude in section \ref{conclusions} with a discussion of our results. Appendix \ref{appendixA} contains details of the 2-loop computations and appendix \ref{tensrid} the technical details of the tensor reduction and the list of all 2-loop integrals computed using it.

\section{The model}
\label{defdef}

We consider an $SU(N)$ Yang-Mills theory with coupling constant $\gYM$, coupled to massless scalars and fermions, all transforming in the adjoint of $SU(N)$, with generators $(T^a)^{b c}=-i f^{a b c}$, in terms of the $SU(N)$ structure constants $f^{a b c}$. 
The model contains a single gauge field $A$, $\mN_\Phi$ scalars $\Phi^M$ $(M = 1, \ldots, \mN_\Phi)$ and $\mN_\psi$ fermions $\psi^L$ $(L = 1, \ldots, \mN_\psi)$. The numbers of scalars and fermions will be kept arbitrary, as well as the Yukawa interactions of the fermions with the scalars. For the scalars we will introduce generic quartic couplings that will be specialised below.
All the fields are given by $\varphi = \varphi^a T^a$ with group generators normalised as $\tr T^a T^b = 1/2 \, \delta^{ab}$.
The (Euclidean) action is defined as 
\bea
\label{eq.action}
S &=& \frac{2}{\gYM^2} \int d^d x \, \tr \left[ \frac{1}{4} F_{ij} F_{ij} + \frac{1}{2 \xi} (\partial_i A_i)^2 + \partial_i \bar{c} \mD_i c +\frac{1}{2} (\mD \Phi^J)^2 
+ \bar{\psi}^L \slashed{\mD} \psi^L \right. \nn \\
&& \left. \qquad \qquad \qquad + \sqrt{2} \, \mu_{M L_1 L_2} \Phi^M \bar{\psi}^{L_1} \psi^{L_2} \right]   + \frac{1}{4!} \lambda^{(1)}_{M_1 M_2 M_3 M_4} \tr \, \Phi^{M_1} \Phi^{M_2} \Phi^{M_3} \Phi^{M_4}  \nn \\
&& \qquad \qquad \qquad  + \frac{1}{4! \, N} \lambda^{(2)}_{M_1 M_2 M_3 M_4} \tr \, \Phi^{M_1} \Phi^{M_2} \, \,  \tr \, \Phi^{M_3} \Phi^{M_4},
\eea
where 
\begin{align}
\mD_i\phi\equiv \mD_i^{a b}\Phi^b=(\delta^{a b}\partial_i- i  (T^c)^{a b} A_i^c)\Phi^b,
\end{align}
\begin{align}
F_{ij}=F_{i j}^a T^a,\qquad  F_{i j}^a=\partial_i A_j^a - \partial_j A_i^a +  f^{a b c} A_i^b A_j^c.
\end{align}
The fields $c_i$ and $\bar{c}_i$ are the ghost and antighost fields, appearing in the Faddeev-Popov terms in the Lagrangian \eqref{eq.action}.
 Notice that we have included a covariant gauge-fixing and we have adopted the Feynman-'t Hooft gauge $\xi = 1$. 
 The Yang-Mills coupling has mass dimension $(4-d)$, while the Yukawa and the quartic-scalar couplings are dimensionless in any spacetime dimension. We assume a completely symmetric quartic-scalar coupling. This automatically selects a completely symmetric gauge structure in the interaction vertex, namely
\begin{align}
\textrm{Str} \, T^{a_1} T^{a_2} T^{a_3} T^{a_4} &= \frac{1}{4!} \sum_{\pi} \tr \, T^{a_{\pi(1)}} T^{a_{\pi(2)}} T^{a_{\pi(3)}} T^{a_{\pi(4)}} \,, \nonumber \\
\textrm{Str}[ T^{a_1} T^{a_2}][ T^{a_3} T^{a_4}] &=  \frac{1}{4!} \sum_{\pi} \tr \, T^{a_{\pi(1)}} T^{a_{\pi(2)}} \, \tr \, T^{a_{\pi(3)}} T^{a_{\pi(4)}} \,,
\end{align}
where the sum is over all permutations of the indices and
\bea
\tr \, T^a T^b T^c T^d = \frac{1}{4 N} \delta^{ab} \delta^{cd} + \frac{1}{8} \left[ d^{abs} d^{cds} - f^{abs} f^{cds} + i \left( d^{abs} f^{cds} + f^{abs} d^{cds}\right)  \right] \,.
\eea
On the other hand, in the Yukawa interaction only the antisymmetric component of the gauge structure is to be taken into account
\bea
\textrm{Atr} \, T^a T^b T^c = \frac{i}{4} f^{abc} \,.
\eea 
We work with the Wick rotated QFT (with a metric of positive definite signature) and we normalise the $\gamma$ matrices as $\tr \gamma_i \gamma_j = - 2^{[d/2]} \delta_{ij}$ $( \{\gamma_i,\gamma_j\}=-2 \delta_{ij}{\bf 1}_d)$,
where $[a]$ is the integer part of $a$ and the negative sign is a consequence of the Euclidean signature.

We first present all the results in an arbitrary dimension $d$, specialising to definite $d$ only at the end. In particular we consider the $d=3,4,5$ cases as an example of a super-renormalisable, renormalisable and non-renormalisable theory respectively, discussing in detail the structure of the singularities, both infrared and ultraviolet, in each case. The $d = 3$ case has also an important application in the computation of the power-spectrum of the cosmological perturbations in the holographic cosmological models. We will use dimensional regularisation in the $\overline{\textrm{MS}}$ scheme with modified minimal subtraction.

\section{UV structure and Feynman integrals}
\label{sec:prelim}

Before presenting the explicit results for the 2-point functions, we will first discuss in this section 
what we expect based on power counting. We will also outline the computation of the Feynman integrals 
and the present the basis of integrals relevant for our computation.

\subsection{Power-counting}

We are interested in computing the 2-point function of the energy-momentum tensor to 2-loops. 
This computation leads to infinities that need to be renormalised. The first step in this process is to take into account the 
renormalisation of elementary fields. After this step  there are generally still infinities because the energy-momentum tensor is a composite operator; these should be subtracted using new counterterms that involve the source that couples to the composite operator, {\it i.e.} the background metric in our case. 

Renormalisation of elementary field at some loop order would affect the renormalisation of the energy-momentum tensor at higher loops. Thus, for the computation of the 2-point function of the energy-momentum tensor at 2-loops we only need the renormalisation of elementary fields at 1-loop. Moreover, since interactions start contributing to this computation from 2-loops on, we do not need to discuss the renormalisation of 3- and higher-point functions of elementary fields.
It follows that for the computation of the 2-point function of the energy-momentum tensor at 2-loops it would suffice to renormalise the 2-point functions at 1-loop order. Nevertheless, we will discuss this computation to 2-loops, as this computation also serves to illustrate the generalised conformal structure.

There is one additional issue to check:  renormalisation may induce additional terms beyond the ones listed in (\ref{eq.action}) that would affect the computation of interest.
On general grounds, the UV behaviour of the different diagrams can be obtained using the superficial degree of divergence $D$ 
\begin{align}
\label{eq.superficialD}
D &=  d - \sum_f E_f (s_f + \frac{d}{2} -1 ) - \sum_i N_i \Delta_i, \\
& = \sum_f I_f (2 s_f + d - 2) + \sum_i N_i (d_i - d) + d \nonumber 
\end{align}
where the two expressions are linked via the standard identify 
\begin{equation} \label{top_id}
2 I_f + E_f = \sum_i N_i n_{if}
\end{equation}
and  we follow the conventions in \cite{Weinberg:1995mt}. 
In particular,
$f$ sums over fields and $i$ over the interactions, while $s_f$ takes into account the contribution of the field propagators ($s_f = 0$ for scalars and gauge bosons and $s_f = 1/2$ for fermions). $N_i$ represents the number of interactions of type $i$. $\Delta_i = d - d_i - \sum_f n_{if} (s_f + \frac{d}{2} -1)$ is the dimension of the interaction of type $i$,  $d_i$ denotes the number of derivatives and
$n_{if}$ the number of fields of type $f$ in the interaction of this type. $E_f$ are the number of external lines of field type $f$.  For the classification of the diagrams it is also useful to determine the number of loops which is given by
\begin{align}
\label{eq.loop}
L &=   \sum_i N_i \left( \frac{1}{2} \sum_f n_{if} - 1 \right)- \frac{1}{2} \sum_f E_f + 1 \,. 
\end{align}

One may check, using these formulas, that power counting implies that there are superficially divergent diagrams associated with 3-point functions of scalars (in all dimensions of interest). Were such diagrams non-zero,  renormalisation would induce $\tr\, \Phi^3$ terms in the action, thus invalidating the generalised conformal structure, already at leading order. So our first task is to examine whether such terms are generated. 

A $\tr\, \Phi^3$ coupling is odd under $\Phi \to -\Phi$, 
and all terms in the action (\ref{eq.action}) are even under this transformation except for the Yukawa couplings, $\Phi \bar \Psi \Psi$. It follows that such term could only be generated by diagrams that involve an odd number of Yukawa couplings, call this number $N_Y$. Applying (\ref{top_id}) to the fermions of the diagrams with 3 extrernal scalar lines we find
\begin{equation}
I_{\Psi} = N_Y + N_{A \Psi^2}
\end{equation}
where $N_{A \Psi^2}$ is the  number of gauge-fermion vertices. Since $N_Y$ is odd, so is the sum of $I_{\Psi}+N_{A \Psi^2}$ and since each massless fermion propagator and each gauge-fermion vertex contributes one gamma matrix, 
each fermionic loop will involve a trace of an odd number of gamma matrices and therefore will be zero, and thus 
no $\tr\, \Phi^3$ coupling is generated.  The same argument implies that no higher odd power of $\Phi$ is generated either.  

When $d=3$, $\Delta_i > 0$ for all the interactions in (\ref{eq.action}) and the theory is super-renormalisable, while when $d=4$ $\Delta_i=0$ and the theory is renormalisable by power-counting, so no new interactions beyond those listed in 
(\ref{eq.action}) will be generated. When $d=5$ however $\Delta_i<0$ and the theory is non-renormalisable, and additional higher dimension terms will be generated. One should view the results we derive here as valid at energies low compared to the scale set by the lowest such higher dimension operator.

\subsection{Tensor reduction}

In this subsection we describe the method we used to calculate the relevant Feynman integrals. To our knowledge the 2-loop reduction formulas  are new and are tabulated in Appendix \ref{tensrid}.

The 1- and 2- loop diagrams have been computed exploiting the technique of tensor reduction to 1- and 2-loop scalar integrals. We briefly go through details of the computation highlighting the most critical steps. 
The 1-loop tensor reduction of the 2-point functions is straightforward. By direct inspection of the diagrams contributing to the 2-point functions of the fields and the $TT$ correlator, it is easy to realise that the highest rank needed in the computation is 4. In this case, all the scalar coefficients arising from the Lorentz-covariant decomposition of a tensor integral can be reduced by algebraic manipulations to the main scalar integral $B_0$
\bea
B_0 =  \int \frac{d^d k_1 }{(2 \pi)^d} \frac{1}{k_1^2 \, (k_1+p)^2  } = \frac{p^{d-4}}{(4 \pi)^{d/2}} G_1 \,, 
\eea 
where $G_1$ is given by
\bea
\label{Eq.G1}
G_1 &=& \frac{\Gamma(2-d/2) \Gamma(d/2-1)^2}{\Gamma(d-2)}.
\eea 

The tensor reduction of the 2-loop diagrams is more involved for several reasons. Firstly, the highest rank of the tensor integrals appearing in the computation of the energy-momentum tensor 2-point function is 6. 
Secondly, the presence of two integration momenta provides different tensor expansions for a given rank. For the same reason and differently from the 1-loop case, one cannot rely on a fully-symmetrised tensor basis.
The 2-loop tensor decomposition described here can be (tediously) extended to tensor integrals of arbitrary rank.  
The scalar coefficients can be written as $f(d) \, p^n$ where $n$ is fixed by the mass dimensions of the original integral and of the corresponding element of the tensor basis, while $f(d)$ is a complicated function of the spacetime dimensions. 

The scalar coefficients $f(d)$ can be further simplified by expanding them onto a minimal basis of scalar integrals. For such purpose we employed the algorithm proposed by Tarasov \cite{Tarasov:1997kx,Tarasov:1996br} and implemented in the TARCER package \cite{Mertig:1998vk}. On general grounds, the algorithm allows to reduce the 2-loop 2-point integral
\bea
\int  \frac{d^d k_1}{(2 \pi)^d} \frac{d^d k_2}{(2 \pi)^d} \frac{(k_1^2)^{n_1} (k_2^2)^{n_2} (p \cdot k_1)^{n_3} (p \cdot k_2)^{n_4} (k_1 \cdot k_2)^{n_4}}{(k_1^2 - m_1^2)^{\nu_1} (k_2^2 - m_2^2)^{\nu_2} (k_3^2 - m_3^2)^{\nu_3} (k_4^2 - m_4^2)^{\nu_4} (k_5^2 - m_5^2)^{\nu_5}} ,
\eea
with $k_3 = k_1 + p$, $k_4 = k_2 + p$ and $k_5 = k_1 - k_2$, into a linear combination of simpler scalar integrals in which the integration momenta have been removed from the numerator. 
This is achieved by exploiting standard algebraic manipulations first, in which irreducible numerators (where only powers of $p \cdot k_1$ and $p \cdot k_2$ appear) are obtained, and then enforcing the algorithm described in \cite{Tarasov:1997kx,Tarasov:1996br}.

In the massless case and with $\nu_i = 0,1$ (realised in our calculations) it is possible to show by direct computation that the basis is populated by only two elements, namely, $(J_0, B_0^2)$ where
\bea
J_0 = \int \frac{d^d k_1}{(2 \pi)^d} \frac{ d^d k_2}{(2 \pi)^d} \frac{1}{k_1^2 \, k_4^2 \, k_5^2 } =  \frac{p^{2d-6}}{(4 \pi)^{d}} G_2 \,,
\eea
is a genuine 2-loop topology while $B_0^2$ is just the square of 1-loop scalar 2-point function. The loop function $G_2$ is defined as 
\bea
\label{Eq.G2}
G_2 &=& \frac{\Gamma(3-d) \Gamma(d/2-1)^3}{\Gamma(3d/2-3)} .
\eea
From the previous expressions it is clear that $G_1$ in \eqref{Eq.G1} develops a singularity only in even dimensions, while, in the odd-dimensional case, $G_1$ is finite but $G_2$ in \eqref{Eq.G2} diverges. Therefore for even $d$, 1- and 2-loop corrections to the 2-point functions are both divergent with, respectively, a single and a double pole in $1/(d-2k)$. For odd $d$, only 2-loop corrections are singular, with a single pole in $1/(d-(2k+1))$.

\section{2-point functions of elementary fields at 1- and 2-loop level}

\begin{figure}
\centering
\includegraphics[scale=0.45]{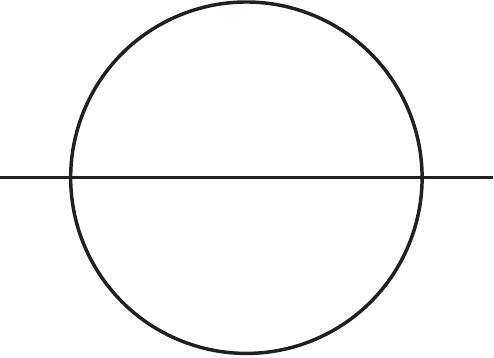} \hspace{0.5cm}
\includegraphics[scale=0.45]{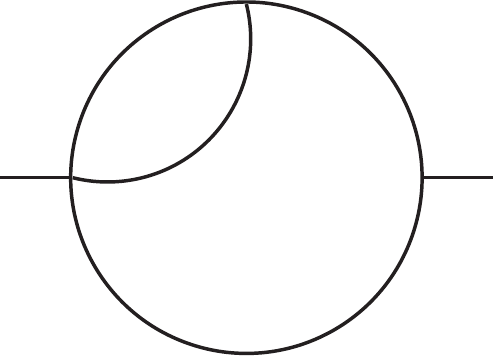} \hspace{0.5cm}
\includegraphics[scale=0.45]{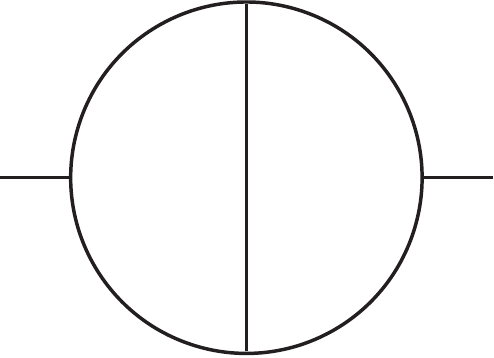} \hspace{0.5cm}
\includegraphics[scale=0.45]{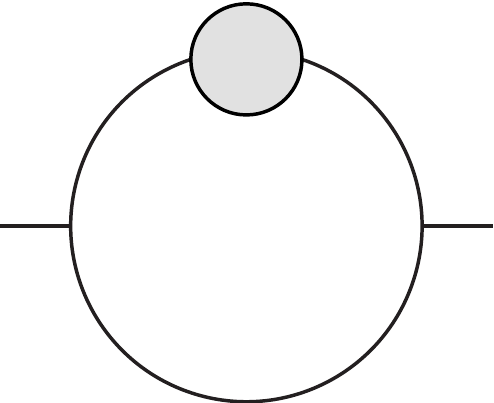} \hspace{0.5cm}
\includegraphics[scale=0.45]{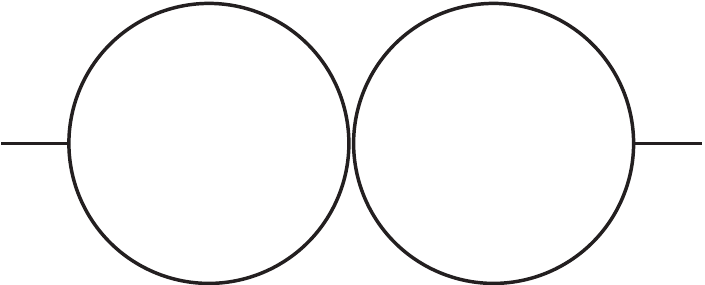}
\caption{Topologies of 2-loop contributions to the 2-point functions of the elementary fields. The blob in the fourth diagram represents an insertion of a 1-loop self-energy. \label{Fig.PHI2L}}
\end{figure}

\label{gcs}
Before discussing the $\< TT \>$ correlator, we present the momentum space results for the 2-point functions, up to 2-loop order, of the fundamental fields, namely the gauge field ($A$), the fermions ($\Psi$) and the scalars ($\Phi$). 
The topologies of the corresponding diagrams are shown in Fig.\,\ref{Fig.PHI2L}.
The 2-point functions can be fixed by generalised conformal invariance as
\bea
\langle \varphi(q) \varphi(-q) \rangle = q^{2 \Delta - d} c_\Delta(g)
\eea
where $\varphi = \{ A, \Phi, \psi \}$, with $\Delta = \{ 1, 1, 3/2\}$ their 4-dimensional mass dimensions and $g = g_{YM}^2 / q^{4-d}$ the dimensionless coupling constant. In the perturbative regime, the function $c_\Delta(g)$ can be expanded as
\bea
c_\Delta(g) = c_0 g + c_1 g^2 + c_2 g^3 + \ldots
\eea
where $c_i$ is the $i$-loop contribution. 

The $c_i$ are expressed in terms of the self-energies (the amputated 2-point functions) which, 
on general ground, can be decomposed as
\bea
\Pi^{ab}_{A \, ij}(q) &=& \delta^{a b}  \, \Pi_{A \, ij}(q) = g^{-1} \delta^{a b} \, q^{d-2}  \left( \gth \, \Pi_{A \, ij}^{(1)} + \gth^2 \, \Pi_{A \, ij}^{(2)} + \ldots \right)  , \\
{\Pi_{\psi}}^{ab}_{L_1 L_2}(q) &=& - i \delta^{a b} q\!\!\!/ \, \Pi_{\psi \, L_1 L_2}(q) = - i g^{-1} \, \delta^{a b} q\!\!\!/  \, q^{d-4} \left( \gth \, \Pi_{\psi \, L_1 L_2}^{(1)} +  \gth^2 \, \Pi_{\psi \, L_1 L_2}^{(2)} + \ldots \right) , \\
\label{eq:selfenergyscalar}
{\Pi_{\Phi}}^{ab}_{M_1 M_2}(q) &=& \delta^{ab} \, \Pi_{\Phi \, M_1 M_2}(q) = g^{-1} \, \delta^{ab} \, q^{d-2} \left( \gth \, \Pi_{\Phi \, M_1 M_2}^{(1)} + \gth^2 \, \Pi_{\Phi \, M_1 M_2}^{(2)} + \ldots \right) ,
\eea
where we expressed the answer in terms of the effective 't Hooft coupling $\lambda = g N$ so that the structure of the large $N$ limit is clear.  Here the ellipses stand for higher loop corrections. Lower-case latin letters $a,b$ denote the gauge indices in the adjoint representation which appear in the factorised Kronecker delta. On the other hand, upper-case latin letters are used for describing the flavour structure. From each term in the equations above we have extracted the momentum dependence and the dimensionless coupling constant $g$, so that the $\Pi$ coefficients are functions of the spacetime dimension $d$ and of the dimensionless couplings $\lambda^{(1)}, \lambda^{(2)}$ of the single and double trace quartic terms and the Yukawa couplings $\mu$.  Gauge invariance fixes the structure of the gauge self-energy as $\Pi_{A \, ij} = \pi_{ij} \, \Pi_{A} $ where
the tensor $\pi_{ij}$ is the usual transverse projection tensor defined as
\bea
\pi_{ij} = \delta_{ij} - \frac{q_i q_j}{q^2}.
\eea
Having introduced the decomposition of the self-energies, we can detail the perturbative expansion of the 2-point functions, which in the three cases are given by
\bea
\langle A^a_i(q) A^b_j(-q) \rangle &=&   g \, q^{2-d}  \, \delta^{ab} \sum_{k=0} \gth^{k} c^A_{k \, ij}   \nn \\
&=& g \, q^{2-d}  \, \delta^{ab} \left\{  \delta_{ij} + \gth \, \Pi^{(1)}_A \, \pi_{ij} + \gth^2 \left[ \left( \Pi^{(1)}_A \right)^2  +  \Pi^{(2)}_A   \right] \pi_{ij}   + \mathcal O(\gth^3)  \right\} \,, \nn \\
\langle \bar \psi^a_{L_1}(q) \psi^b_{L_2}(-q) \rangle &=&  -i g \, q\!\!\!/ \, q^{2-d} \, \delta^{ab} \sum_{k=0} \gth^{k} c^\psi_{k \, L_1 L_2} \nn \\
&=& -i g \, q\!\!\!/ \, q^{2-d} \, \delta^{ab}  \left\{  \delta_{L_1 L_2}   + \gth \, \Pi^{(1)}_{\psi \, L_1 L_2}  + \gth^2 \left[ \Pi^{(1)}_{\psi \, L_1 L_3} \Pi^{(1)}_{\psi \, L_3 L_2}  +  \Pi^{(2)}_{\psi \, L_1 L_2}   \right]    + \mathcal O(\gth^3) \right\}   \,, \nn \\
\langle  \Phi^a_{M_1}(q) \Phi^b_{M_2}(-q) \rangle &=&  g \, q^{2-d}  \, \delta^{ab} \sum_{k=0} \gth^{k} c^\Phi_{k \, M_1 M_2}   \nn \\
&=& g \, q^{2-d}  \, \delta^{ab} \left\{   \delta_{M_1 M_2} + \gth \, \Pi^{(1)}_{\Phi \, M_1 M_2}  + \gth^2 \left[ \Pi^{(1)}_{\Phi \, M_1 M_3} \Pi^{(1)}_{\Phi \, M_3 M_2}  +  \Pi^{(2)}_{\Phi \, M_1 M_2}   \right]   + \mathcal O(\gth^3)\right\}    \nn \\
\eea
where the summation runs over the perturbative orders covered by the expansion, and where the coefficients are
\bea
c_{0 \, I_1 I_2}^{\varphi} = \delta_{I_1 I_2} \,, \qquad c_{1 \, I_1 I_2}^{\varphi} = \Pi^{(1)}_{\varphi \, I_1 I_2}  \,, \qquad c_{2 \, I_1 I_2}^{\varphi} = \Pi^{(1)}_{\varphi \, I_1 I_3} \Pi^{(1)}_{\varphi \, I_3 I_2}  +  \Pi^{(2)}_{\varphi \, I_1 I_2} \,,
\eea
with the index $I=\{i, L, M\}$ for $\varphi = \{A, \psi, \Phi\}$. \\

We proceed by presenting the expressions of the 
scalar form factors at 1- and 2-loop level $\Pi^{(1,2)}$ in the three cases.
\begin{itemize}
\item{\bf One-loop}\\
At 1-loop order and for arbitrary $d$ dimensions, the 2-point self-energies take the form
\bea
\Pi_A^{(1)} &=&   \left[ 3d-2 + 2(2-d) \mN_\psi  -  \mN_\Phi   \right] \frac{1}{2(d-1)}  \frac{G_1}{(4 \pi)^{d/2}} , \label{PiGauge1Loop} \\
\Pi_{\psi \, L_1 L_2}^{(1)} &=&  \left[ -   \frac{d-2}{2} \, \delta_{L_1 L_2}    +    \frac{1}{4}  \, \mu^{(0)}_{L_1 L_2}    \right]   \frac{G_1}{(4 \pi)^{d/2}}  ,  \label{PiFermion1Loop}  \\
\Pi_{\Phi \, M_1 M_2}^{(1)} &=&  \left[ 2 \,   \delta_{M_1 M_2}    +    \frac{1}{2} \, \mu^{(0)}_{M_1 M_2}  \right]  \frac{G_1}{(4 \pi)^{d/2}} ,   \label{PiScalar1Loop}
\eea
where $\mN_\Phi$ counts all the scalar fields and we have defined 
\bea
\mu^{(0)}_{L_1 L_2} &=& \mu_{M L_1 L_3} \, \mu_{M L_3 L_2} \,, \qquad  L\to \textrm{fermion flavours}\,, \nn \\
\mu^{(0)}_{M_1 M_2} &=& \mu_{M_1 L_1 L_2} \, \mu_{M_2 L_2 L_1}\,, \qquad M\to \textrm{scalar flavours} \,.
\eea
At two loop level we will be needing additional definitions of such products, which can be found in 
\eqref{refquartic}. Notice that we have introduced the same notation $\mu^{(0)}$ to denote two different contractions of two Yukawa couplings. There is no risk of ambiguity as we always use the latin letters $L$ and $M$ to represent fermionic and scalar flavour indices, respectively. 

\item{\bf Two-loop}
\end{itemize}
Moving to 2-loop level, the corrections to the scalar form factors in the expansion of the self-energies of all the fields are given by
\bea
\Pi_A^{(2)} &=&  \frac{1}{(4 \pi)^d}\left\{ \alpha_{A0}\left[ \alpha_{A1} G_1^2 + \alpha_{A2} G_2\right] +\mu_Y^2  \left[ \frac{1}{4} G_1^2 + \frac{8-3d}{2 (d-4)} G_2\right] \right\}
\label{pi0}
\eea

\bea
\Pi_{\psi \, L_1 L_2}^{(2)} &=&  \frac{1}{(4 \pi)^d} \bigg\{\alpha_{\psi 0} \left( \alpha_{\psi 1}G_1^2 + \alpha_{\psi 2} G_2 \right) \delta_{L_1 L_2}  +    \frac{\mu^{(0)}_{L_1 L_2}}{4 (d-4)^2} \left[ \alpha_{\psi 3} G_1^2  +\alpha_{\psi 4} G_2 \right] \nn \\
&& +  \frac{1}{16 (d-4)} \left[ \mu^{(1)}_{L_1 L_2} (d -4 ) G_1^2  + 4  ((d -2 ) \mu^{(2)}_{L_1 L_2}  + (d -3) \mu^{(3)}_{L_1 L_2} ) G_2 \right]
\bigg\} ,   
\label{pi1}
\eea

\bea
\Pi_{\Phi \, M_1 M_2}^{(2)} &=& \frac{1}{(4 \pi)^d} \bigg\{ \alpha_{\Phi 0} \left[  
 \alpha_{\Phi 1}  G_1^2      
+ \alpha_{\Phi 2}G_2
\right] \delta_{M_1 M_2} +  \frac{\mu^{(0)}_{M_1 M_2} }{4 (d-4)^2} \left[ 
\alpha_{\Phi 3} G_1^2  
 + \alpha_{\Phi 4} G_2 \right] \nn \\
&& +   \frac{1}{8 (d-4)} \left[ \mu^{(5)}_{M_1 M_2} (d -4 ) (G_1^2 - 2 G_2)  + 4 \, \mu^{(6)}_{M_1 M_2}  (d -2 ) G_2  \right] \nn \\
&& + \frac{1}{144} \lambda^{(0)}_{M_1 M_2}     G_2
\bigg\} ,
\label{pi2}
\eea
where $\alpha_{A i}, \alpha_{\psi j},\alpha_{\Phi k}$ which are functions both of 
the dimension $d$ and the field multiplicities $\mathcal{N}_\psi$, $\mathcal{N}_\Phi$
and are given in appendix \ref{selfies}. $\lambda^{(0)}$, $\mu_Y^2$ and $\mu^{(i)}$ are quadratic and quartic products 
of the couplings $\lambda$ and $\mu$ defined in \eqref{eq.action}, of the form 
\bea
\label{thisone}
\lambda^{(0)}_{M_1 M_2} &=& \lambda^{(1)}_{M_1 M_3 M_4 M_5} \lambda^{(1)}_{M_2 M_3 M_4 M_5}   \left( 1 - \frac{6}{N^2} +\frac{18}{N^4} \right)   +  \lambda^{(2)}_{M_1 M_3 M_4 M_5} \lambda^{(2)}_{M_2 M_3 M_4 M_5}  
\frac{2}{N^2} \left( 1 + \frac{1}{N^2}  \right)  \nn \\
&+& \left( \lambda^{(1)}_{M_1 M_3 M_4 M_5} \lambda^{(2)}_{M_2 M_3 M_4 M_5} +  \lambda^{(1)}_{M_2 M_3 M_4 M_5} \lambda^{(2)}_{M_1 M_3 M_4 M_5}   \right)  \frac{2}{N^2} \left( 2 - \frac{3}{N^2}  \right)     , \nn \\
\mu_Y^2 &=& \mu_{M L_1 L_2} \, \mu_{M L_2 L_1} , \nn \\
\mu^{(1)}_{L_1 L_2} &=&\mu_{M_1 L_1 L_3} \, \mu_{M_2 L_3 L_4} \, \mu_{M_1 L_4 L_5} \, \mu_{M_2 L_5 L_2} ,  
\eea
with the remaining $\mu^{(i)}$ differing from the way the indices of the $\mu$'s are contracted and can be found in appendix \ref{selfies}.
\subsection{Generalised conformal structure constants: Results in $d = 3$}

 This case represents an example of a super-renormalisable $SU(N)$ theory in which the gauge coupling constant $\gYM$ is dimensionful with mass dimension $1/2$. 

\begin{figure}[t]
\centering
\includegraphics[scale=1]{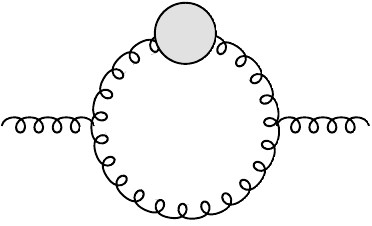} \hspace{2cm}
\includegraphics[scale=1]{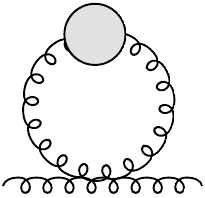} 
\caption{IR divergent contributions to the 2-loop self-energy of the gauge field. The blob represents the 1-loop correction. \label{Fig.IRdiv}}
\end{figure}

In $d=3$, $G_1$ is finite but $G_2$ develops a singularity parametrised, in dimensional regularisation, by a single pole in $d-3$. Concerning the self-energy of the gauge field,  one can show using power-counting arguments that all diagrams but one are UV divergent. The exception is the last diagram  in Fig.\,\ref{Fig.PHI2L}, which is the product of two UV-finite 1-loop bubbles. Actually, the UV singularity cancels in the full 2-loop result and only an IR divergence survives. This can be easily proven introducing a small mass regulator to control the small momentum behaviour of the correlator. The use of the regulator is particularly useful to disentangle the poles of dimensional regularisation which, otherwise, would hide their UV or IR nature in the $\epsilon$ expansion. 
The two IR divergent contributions in the perturbative expansion of the 2-loop 2-point function of the gauge fields are depicted in Fig.\,\ref{Fig.IRdiv}. Similarly, the 2-point function of the fermion fields develops an IR singularity. \\
The structure of the self-energy of the scalar fields is instead different, because the UV divergence does not cancel and must be removed by a suitable mass counterterm as shown below.

In $d=3$ dimensions, the 2-point functions of the fundamental fields, up to 2-loop in perturbation theory, are given by
\bea
\label{eq.3Dconform}
\langle A^a_i(q) A^b_j(-q) \rangle &=&  g \, q^{-1}  \, \delta^{ab}  \left\{  \delta_{ij} +  \pi_{ij} \left[  \gth \, c_1^A \,  + \gth^2  \, c_2^A   +   \gth^2  \, \log \gth \,\, \tilde c_{2 \, \IR }^A \right] + \mathcal O(\gth^3) \right\} , \nn \\
\langle \bar \psi^a_{L_1}(q) \psi^b_{L_2}(-q) \rangle &=&  -i g \, q\!\!\!/ \, q^{-1} \, \delta^{ab} \left\{  \delta_{L_1 L_2} + \gth \, c_{1 \, L_1 L_2}^\psi  + \gth^2  \, c_{2 \, L_1 L_2}^\psi   +   \gth^2  \, \log \gth \,\, \tilde c_{2 \, \IR \, L_1 L_2}^\psi  + \mathcal O(\gth^3)  \right\} , \nn \\
\langle \Phi^a_{M_1}(q) \Phi^b_{M_2}(-q) \rangle &=&  g \, q^{-1}  \, \delta^{ab} \left\{  \delta_{M_1 M_2} + \gth \, c_{1 \, M_1 M_2}^\Phi  + \gth^2  \, c_{2 \, M_1 M_2}^\Phi   +   \gth^2  \, \log \gth \,\, \tilde c_{2 \, \IR \, M_1 M_2}^\Phi  \right. \nn \\  
&+& \left.   \gth^2  \, \log \gth \,\, \tilde c_{2 \, \UV \, M_1 M_2}^\Phi  + \mathcal O(\gth^3)\right\} , 
\eea
where $\gth \equiv g N$ and the explicit results for the 1- and 2-loop  generalised conformal structure constants of the gauge field are
\bea
c_{1}^{A} &=& \frac{1}{32}   \left[ 7 - 2 \mN_\psi  -  \mN_\Phi   \right]  , \nn \\
\tilde c_{2 \, \IR}^A &=& \frac{1}{32 \pi^2} \left[ (-7 + 2 \mN_\psi + 5 \mN_\Phi + \mu_Y^2  \right] , \nn \\
c_2^A &=& - \frac{\tilde c_{2 \, \IR}^A}{2} \left( \frac{1}{ \omega} + 2 \log \frac{\gYM^2 N}{\mu_{\rm{IR}}} \right) + 
\frac{1}{1024 \pi^2} \left[
-16 ( \mu_Y^2 +20 \mN_\psi-2 \mN_\Phi-15) \right. \nn \\
&+& \left. \pi ^2 (4 \mu_Y^2 +(2 \mN_\psi+\mN_\Phi-2) (2 \mN_\psi+ \mN_\Phi)+29)
\right] \,,
\eea
while for the fermion fields they take the form
\bea
c_{1 \, L_1 L_2}^\psi &=& \frac{1}{32}   \left[ - 2 \,  \delta_{L_1 L_2}   +   \mu^{(0)}_{L_1 L_2}  \right]  ,  \nn  \\
\tilde c_{2 \, \IR \, L_1 L_2}^\psi &=& - \frac{1}{192 \pi^2}  \left[  \left( -14 + 4 \mN_\psi + 2 \mN_\Phi \right) \delta_{L_1 L_2} + 12 \mu^{(0)}_{L_1 L_2}  + 3 \mu^{(2)}_{L_1 L_2}    \right]    \,,  \nn \\
c_{2 \,  L_1 L_2}^\psi &=&  - \frac{\tilde c_{2 \, \IR \, L_1 L_2}^\psi }{2} \left( \frac{1}{ \omega} + 2 \log \frac{\gYM^2 N}{\mu_{\rm{IR}}} \right) +   
\frac{1}{576 \pi^2} \left[ - \left( 64 + 4 \mN_\psi + 5 \mN_\Phi - \frac{9}{2} \pi^2 \right) \delta_{L_1 L_2}  \right.  \nn \\
&+& \left. 18 \, \mu^{(0)}_{L_1 L_2} + \frac{9}{2} \left( - \mu^{(2)}_{L_1 L_2}  +  \mu^{(3)}_{L_1 L_2}  + \frac{\pi^2}{8} (\mu^{(1)}_{L_1 L_2}  + \mu^{(4)}_{L_1 L_2} ) \right)
\right]. \,
\eea
The IR singularity is described in dimensional regularisation by a single pole in $\omega = d - 3$.
Similarly, for the scalar 2-point function we obtain
\bea
c_{1 \, M_1 M_2}^\Phi &=&   \frac{1}{16}  \left[4 \,   \delta_{M_1 M_2} +  \mu^{(0)}_{M_1 M_2}  \right]  , \nn \\
\tilde c_{2 \, \UV \, M_1 M_2}^\Phi &=&   
-  \frac{1}{32 \pi^2} \bigg[ (1 - 2 \mN_\psi -  \mN_\Phi) \delta_{M_1 M_2} + 2 \mu^{(0)}_{M_1 M_2} + \frac{1}{2} \mu^{(5)}_{M_1 M_2} + \mu^{(6)}_{M_1 M_2} 
- \frac{1}{72}  \lambda^{(0)}_{M_1 M_2} %
 \bigg]  
, \nn \\
\tilde c_{2 \, \IR \, M_1 M_2}^\Phi &=&  - \frac{1}{96 \pi^2}  \left[  \left( 19 - 2 \mN_\psi - \mN_\Phi \right)  \delta_{M_1 M_2}  + 3 \, \mu^{(0)}_{M_1 M_2}  \right]  , \nn 
\eea
\bea
 c_{2 \, M_1 M_2}^\Phi &=& - \frac{\tilde c_{2 \, \IR \, M_1 M_2}^\Phi }{2} \left( \frac{1}{ \omega} + 2 \log \frac{\gYM^2 N}{\mu_{\IR}} \right)    -  \tilde c_{2 \, \UV \, M_1 M_2}^\Phi  \log \frac{\gYM^2 N}{\mu_{\UV}} \nn \\
&-&  \frac{1  }{192 \pi^2} \left[ \left( 2 - \frac{3}{N^2} \right) \lambda^{(1)}_{M_1 M_2 M_3 M_4} + \left( 1 + \frac{1}{N^2} \right) \lambda^{(2)}_{M_1 M_2 M_3 M_4}  	\right]  \left(  \delta_{M_3 M_4} + \frac{1}{8} \mu^{(0)}_{M_3 M_4}  \right) \frac{1}{\bar \omega} \nn \\
&+& \frac{1}{192 \pi^2}  \left[  
  \frac{1}{6}  \left( 662 - 16 \mN_\psi   + 16 \mN_\Phi + 27 \pi^2    \right)   \delta_{M_1 M_2}    +  3 \left( 14 + \frac{3 \pi^2}{4} \right)  \mu^{(0)}_{M_1 M_2}  \right.  \nn \\
&+& \left. 3 \left( \frac{1}{8} (-12 + \pi^2) \mu^{(5)}_{M_1 M_2}  - \mu^{(6)}_{M_1 M_2}   + \frac{\pi^2}{4} \mu^{(7)}_{M_1 M_2}\right)  
+  \frac{1 }{8} \lambda^{(0)}_{M_1 M_2}   \right] \,.
\eea
 Notice that we have absorbed the $\gamma_E -  \log 4 \pi$ term in the $1/\omega$ pole and we have introduced the UV and IR scales $\mu_\UV$ and $\mu_\IR$. 
 
 As discussed before, the 2-point function of the scalar field is the only one affected by a UV divergence. This has been removed in the $\overline{\rm MS}$ renormalisation scheme by a mass counterterm $ \delta m^2_{M_1 M_2} \tr \, \Phi_{M_1} \Phi_{M_2}/2$, where
 \bea
\label{eq.massct}
\delta m^2_{M_1 M_2} &=&  \frac{(\gYM^2 \, N)^2 }{64 \pi^2} \bigg[ (1 - 2 \mN_\psi -  \mN_\Phi) \delta_{M_1 M_2} + 2 \mu^{(0)}_{M_1 M_2} + \frac{1}{2} \mu^{(5)}_{M_1 M_2} + \mu^{(6)}_{M_1 M_2} \nn \\
&-& \frac{1}{72} \lambda^{(0)}_{M_1 M_2} + \frac{1}{24}\left[ \left( 2 - \frac{3}{N^2} \right) \lambda^{(1)}_{M_1 M_2 M_3 M_4} + \left( 1 + \frac{1}{N^2} \right) \lambda^{(2)}_{M_1 M_2 M_3 M_4}  	\right] \left( 8 \, \delta_{M_3 M_4} + \mu^{(0)}_{M_3 M_4}  \right)
 \bigg]  \frac{1}{ \epsilon}  \,, \nn \\
\eea
with $\epsilon = 3 -d$.

\subsection{Generalised conformal structure constants: Results in $d=4$}

In this section we offer more details about the renormalisation of the 2-point functions of all the fields in $d=4$ dimensions.
Differently from the $d=3$ case, in $d=4$ dimensions the UV singularities already appear at 1-loop level through a $1/(d-4)$ pole in the $G_1$ function, and at 2-loop as single and double poles. On the other hand, the IR singularities do not affect the 2-point functions. These divergences can be absorbed, as usual, in the redefinition of the fields $\varphi \rightarrow Z^{1/2} \varphi$, through the wave-function renormalisation constants, and of the couplings. 
In particular, the UV divergences of the 2-point functions are removed by counterterms extracted from the kinetic part of the action. These are given by
\bea
\label{CT}
\delta \Pi^{ab}_{A \, ij}(q) &=& - \frac{1}{\gYM^2}\left(Z_A Z_g^{ -1} -1 \right) \delta^{ab} q^2 \pi_{ij} (q)  \,, \nn \\
\delta \Pi^{ab}_{c}(q) &=& - \frac{1}{\gYM^2}\left(Z_c Z_g^{-1} -1 \right) \delta^{ab} q^2 \,, \nn \\
\delta {\Pi_{\psi}}^{ab}_{L_1 L_2}(q) &=&   \frac{i}{\gYM^2}\left({Z_\psi}_{L_1 L_2} Z_g^{ -1} - \delta_{L_1 L_2} \right) \delta^{ab} q \!\!\!/ \,, \nn \\
\delta {\Pi_{\Phi}}^{ab}_{M_1 M_2}(q) &=& - \frac{1}{\gYM^2}\left({Z_\Phi}_{M_1 M_2} Z_g^{ -1} - \delta_{M_1 M_2} \right) \delta^{ab} q^2 \,,
\eea
with $Z_A$, $Z_c$, $Z_\Psi$ and $Z_\Phi$ being the wave-function renormalisation constants of the gauge, ghost, fermion and scalar fields respectively, while $Z_g$ renormalises the gauge coupling $\gYM^2$. 
The renormalisation factors are characterised by the usual pole expansion in $d = 4 - \epsilon$
\bea
\label{RenConst}
Z = 1 + \gYM^2 \frac{Z^{(1)}}{\epsilon} + \gYM^4 \left( \frac{Z^{(2)}}{\epsilon^2} + \frac{Z^{(12)}}{\epsilon} \right) + \ldots \, 
\eea
where we have omitted a possible flavour structure. Notice that the indices (12) in $Z^{(12)}$ refer to the order of the pole (order 1) and of the perturbative expansion (2). The finite parts of the corresponding counterterms will be labelled the same way (as $\delta^{(12)}$).
The $\delta \Pi^{ab}_{c}(q)$ represents the counterterm of the self-energy of the ghost field which, at 1-loop level and in $d$ dimensions, is given by
\bea
\Pi^{ab}_{c}(q) =  \delta^{ab}  \frac{N}{2}      q^{d - 2} \frac{G_1}{(4 \pi)^{d/2}} \,.
\eea
The 1-loop renormalisation of the 2-point function of the ghost fields is in this case necessary, since it contributes to the perturbative expansion of the correlator of the gauge field at 2-loop order. \\
From the structure of the singularity of the 2-point functions at 1-loop order one can easily extract the corresponding counterterms, which are given by
\bea
\label{CT1L}
\delta_A^{(1)} &=& Z^{(1)}_A - Z^{(1)}_g = - \frac{1}{48 \pi^2} \left( -10 + 4 \mN_\psi + \mN_\Phi \right) \,, \nn \\
\delta_c^{(1)} &=& Z^{(1)}_c - Z^{(1)}_g = \frac{1}{16 \pi^2}  \,, \nn \\
\delta_{\psi \, L_1 L_2}^{(1)} &=& Z^{(1)}_{\psi \, L_1 L_2} - Z^{(1)}_g \delta_{L_1 L_2} = \frac{1}{32 \pi^2} \left( -4 \delta_{L_1 L_2}  + \mu^{(0)}_{L_1 L_2}  \right) \,, \nn \\
\delta_{\Phi \, M_1 M_2}^{(1)} &=& Z^{(1)}_{\Phi \, M_1 M_2} - Z^{(1)}_g \delta_{M_1 M_2}= \frac{1}{16 \pi^2} \left(  4 \delta_{M_1 M_2} +  \mu^{(0)}_{M_1 M_2} \right) \,.
\eea
The previous relations are not sufficient to completely determine the five 1-loop renormalisation constants $Z^{(1)}$ appearing in Eq.(\ref{CT}). 
In order to close the set of equations, the analysis of the UV divergence of one of the 3-point functions involving one gauge field is still necessary.
Furthermore, the knowledge of the 1-loop counterterms of the vertices is also required by the renormalisation of the 2-loop 2-point functions of the elementary fields. Indeed, these counterterms appear in the perturbative expansion as vertex insertions in diagrams of a 1-loop topology. \\
We have explicitly computed the divergence of the fermion-gauge boson vertex at 1-loop level from which the corresponding 1-loop counterterm, proportional to the renormalisation constant $Z_{\psi} Z_A^{1/2} Z_g^{-1} -1$, is identified and it is given by the expression
\bea
\delta^{(1)}_{A \bar{\psi} \psi \, L_1 L_2 } =   \left( \frac{Z_A^{(1)}}{2} - Z_g^{(1)} \right)\delta_{L_1 L_2}+Z^{(1)}_{\psi \, L_1 L_2} =   \frac{1}{32 \pi^2} \left( - 8 \, \delta_{L_1 L_2}  +  \mu^{(0)}_{L_1 L_2} \right) \,.
\eea
The counterterms on the other 3-point vertices, with the only exception of the Yukawa coupling, are related by gauge invariance to the fermion-gauge boson correlator and to the 2-point functions, and do not need to be computed independently. They are given by
\bea
\label{delta1L}
\delta^{(1)}_{A A A} &=& \frac{3}{2} Z_A^{(1)} - Z_g^{(1)} = - \frac{1}{48 \pi^2}\left(-4 + N_\Phi + 4N_\psi \right) \,, \nn \\
\delta^{(1)}_{A \bar c c} &=& \frac{1}{2} Z_A^{(1)} + Z_c^{(1)} - Z_g^{(1)} = - \frac{1}{16 \pi^2} \,, \nn \\
\delta^{(1)}_{A \Phi \Phi \, M_1 M_2 } &=& \left( \frac{Z_A^{(1)}}{2} - Z_g^{(1)} \right)\delta_{M_1 M_2}+Z^{(1)}_{\Phi \, M_1 M_2} = \frac{1}{16 \pi^2}\left(2 \, \delta_{M_1 M_2} +  \mu^{(0)}_{M_1 M_2} \right) \,, 
\eea
which correspond, respectively, to the 3-gauge, the ghost-gauge and the scalar-gauge boson vertices. As stated above, the renormalisation of the 1-loop Yukawa vertex requires an independent computation from which we extract the counterterm
\bea
\delta^{(1)}_{\Phi \bar{\psi} \psi \, M L_1 L_2 } &=& - \frac{1}{32 \pi^2} \left( 12 \, \mu_{M L_1 L_2} + \mu_{M L_1 L_3} \mu_{M_1 L_3 L_4} \mu_{M L_4 L_2}  \right) \,.
\eea
As already noticed above, the 1-loop counterterms appear in the 2-loop perturbative expansion as propagator and vertex insertions in  diagrams with a 1-loop topology. They remove the momentum-dependent single-pole singularities of the form $1/\epsilon \log p$, which could not be absorbed by the renormalisation constants characterised by the structure given in Eq.(\ref{RenConst}). 
All the remaining divergences, $\epsilon^{-1}$ and $\epsilon^{-2}$, are proportional to constant coefficients and can be absorbed, respectively, in the renormalisation constants $Z^{(12)}$ and $Z^{(2)}$ of Eq.(\ref{RenConst}). 
The 2-loop counterterms can be found in appendix \ref{gauges}. This complete the renormalisation program of the 2-loop self-energies of the fundamental fields. 

In $d=4$ dimensions, the renormalised 2-point functions exhibit the following perturbative structure
\bea
\langle A^a_i(q) A^b_j(-q) \rangle &=&  g \, q^{-2}  \, \delta^{ab} \sum_{n = 0}^{\infty} \sum_{k = 0}^{n} \gth^{n} \, c^{A, (n,k)}_{ i j}\log^k \frac{q}{\mu_\UV} \,, \nn \\
\langle \bar \psi^a_{L_1}(q) \psi^b_{L_2}(-q) \rangle &=&  -i g \, q\!\!\!/ \, q^{-2} \, \delta^{ab}  \sum_{n = 0}^{\infty} \sum_{k = 0}^{n} \gth^{n} \, c^{\psi, (n,k)}_{ L_1 L_2}\log^k \frac{q}{\mu_\UV}  , \nn \\
\langle \Phi^a_{M_1}(q) \Phi^b_{M_2}(-q) \rangle &=&  g \, q^{-2}  \, \delta^{ab}  \sum_{n = 0}^{\infty} \sum_{k = 0}^{n} \gth^{n} \, c^{\Phi, (n,k)}_{ M_1 M_2}\log^k \frac{q}{\mu_\UV}, 
\eea
where $n$ counts the perturbative order and $k$ the logarithm power.
The first coefficient, corresponding to $n=0$, represents the identity matrix in the respective space, $c^{\varphi, (0,0)}_{lm} = \delta_{lm}$. The coefficients at 1-loop are given by
\bea
& c_{ij}^{A, (1,0)} = \frac{1}{144 \pi^2} (  31 - 10 \mN_\psi - 4 \mN_\Phi ) \, \pi_{ij} \,, \qquad   c_{ij}^{A, (1,1)} &=  - \delta_A^{(1)} \, \pi_{ij} \,, \nn \\
& c_{ L_1 L_2}^{\psi, (1,0)} =  - \frac{1}{32 \pi^2} ( 2 \delta_{L_1 L_2} - \mu^{(0)}_{L_1 L_2} ) \,,  \qquad c_{L_1 L_2}^{\psi, (1,1)} &=  - \delta_{\psi \, L_1 L_2}^{(1)}\,, \nn \\
& c_{M_1 M_2}^{\Phi, (1,0)} =  \frac{1}{16\pi^2} ( 4 \delta_{M_1 M_2} +  \mu^{(0)}_{M_1 M_2} ) \,,  \qquad c_{M_1 M_2}^{\Phi, (1,1)} &= - \delta_{\Phi \, M_1 M_2}^{(1)} \,,
\eea
with $\delta_\varphi^{(1)}$ defined in Eqs.(\ref{delta1L}). At 2-loop, in the gauge sector we have 
\bea
c_{ij}^{A, (2,0)}  &=&  \frac{1}{73728 \pi ^4} \left(-4444 \mN_\psi-2113 \mN_\Phi+6490  +144 \, \zeta_3 (8 \mN_\psi + \mN_\Phi -2)  \right. \nn \\
&-& \left.   504 \, \mu_Y^2  \right) \pi_{ij} + c_{ik}^{A, (1,0)} c_{kj}^{A, (1,0)}   \,, \nn \\
c_{ij}^{A, (2,1)}  &=&  \frac{1}{9216 \pi^4} \left( (-790 + 436 \mN_\psi + 187 \mN_\Phi) + 36 \, \mu_Y^2  \right) \pi_{ij} + 2 \, c_{ik}^{A, (1,0)} c_{kj}^{A, (1,1)}  \,, \nn \\
c_{ij}^{A, (2,2)}  &=&  c_{ik}^{A, (1,1)} c_{kj}^{A, (1,1)}    -  \delta_A^{(2)} \pi_{ij} \,.
\eea
The expressions of the remaining coefficients at 2-loop are given in appendix \ref{confst}.

\subsection{Generalised conformal structure constants: Results in $d=5$}

As a last example we consider a non-renormalisable theory in $d=5$ described by the same action in Eq.(\ref{eq.action}). 
The 2-point functions of the elementary fields develop a UV divergence at 2-loop order in perturbation theory, while the 1-loop contributions remain finite.
The singularities can be removed by counterterm operators of dimension 6 which are quadratic in the fields. The corresponding action can be written in the following form
\bea
\mathcal S_{\textrm{ct}} = \frac{2}{\gYM^2} \int d^5 x \, \tr[ \frac{\delta^{(2)}_A}{4} F_{ij} \mathcal D^2 F_{ij}   +    \frac{\delta^{(2)}_{\Phi \, M_1 M_2}}{2}  (\mathcal D \Phi^{M_1}) \mathcal D^2  (\mathcal D \Phi^{M_2})    +   \delta^{(2)}_{\psi \, L_1 L_2}  \bar{\psi}^{L_1} \mathcal D^2 \slashed{\mD} \psi^{L_2}  ]  \frac{1}{\epsilon}
\eea
where $\mathcal D^2 = \mathcal D_i \mathcal D_i$ is required by gauge invariance even though we only needed the $\Box$ to renormalise the 2-point functions. 
The coefficients $\delta^{(2)}_A, \delta^{(2)}_{\Phi \, M_1 M_2}$ and $\delta^{(2)}_{\psi \, L_1 L_2}$ are determined by the UV finiteness condition of the 2-loop 2-point functions 
\bea
\label{delta5L}
\delta^{(2)}_A &=& \gYM^4  \frac{N^2}{105 \times 2^{10} \pi^4} \left[ -152 + 6 \mN_\psi + 10 \mN_\Phi + 7 \mu_Y^2  \right] \,, \nn \\
\delta^{(2)}_{\psi \, L_1 L_2} &=& \gYM^4  \frac{N^2}{215040 \pi^4} \left[  \left( -390 + 36 \mN_\psi + 6 \mN_\Phi \right) \delta_{L_1 L_2}   + 
  52 \mu^{(0)}_{L_1 L_2}  - 3 \mu^{(2)}_{L_1 L_2} - 2 \mu^{(3)}_{L_1 L_2}   \right]   \,, \nn \\
\delta^{(2)}_{\Phi \, M_1 M_2} &=& \gYM^4 \frac{N^2}{215040 \pi^4} \left[ (-344 + 144 \mN_\psi + 24 \mN_\Phi) \delta_{M_1 M_2}  +66 \mu^{(0)}_{M_1 M_2} + \mu^{(5)}_{M_1 M_2}  - 6 \mu^{(6)}_{M_1 M_2} \nn \right. \\ 
&-& \left. \frac{1}{36} \lambda^{(0)}_{M_1 M_2} \right]  \,.
\eea
In order to highlight the generalised conformal structure of the 2-point functions we introduce the dimensionless coupling $\gth = \gYM^2 N \, q$. 
According to this definition, the coupling coefficients at 1- and 2-loop order are given by
\bea
c_{1}^{A} &=& \frac{1}{1024 \pi}  \left[ -13 + 6 \mN_\psi  +  \mN_\Phi   \right]  \,, \nn \\
\tilde c_{2 \, \UV}^{A} &=& \frac{1}{105 \times 2^{10} \pi^4} \left[ -152 + 6 \mN_\psi + 10 \mN_\Phi + 7 \mu_Y^2  \right] \,, \nn \\
c_{2}^{A} &=& - \tilde c_{2 \, \UV}^{A} \, \log \left( \gYM^2 N \mu_{\UV} \right) +  \frac{1}{3675 \times 2^{13} \pi^4} \left[  62593 + 12016 \mN_\psi  - 4135 \mN_\Phi  - 
  4018  \mu_Y^2 \right]    \nn \\
  &+&     \frac{1}{2^{20} \pi^2} \left[   8 (-3 - 12 \mN_\psi + \mN_\Phi) + (-13 + 6 \mN_\psi + \mN_\Phi)^2 + 
  16  \mu_Y^2  \right]\,,
\eea
where we have used the same notation introduced in Eq.(\ref{eq.3Dconform}) for the $d=3$ case. Similar expressions are derived for the coefficients related to the scalars and the fermions, and can be found in appendix \ref{sfermion}

\section{The $\< TT \>$ correlation function and the $A$ and $B$ form factors}
\label{TTsection}

In this section we move to discuss the structure of the 1- and 2-loop contributions to the 2-point function of the energy-momentum tensor, presenting the general expressions in $d$ dimensions.

We start by recalling that the the energy-momentum tensor of the model is defined as
\bea
T_{ij} = \frac{2}{\rg} \frac{\delta S}{\delta g^{ij}} \bigg|_{g_{ij} = \delta_{ij}} = T^A_{ij} + T^{g.f.}_{ij} + T^{gh}_{ij} + T^{\psi}_{ij} + T^{\Phi}_{ij} + T^{Y}_{ij},
\eea
where the different terms denote, respectively, the contribution of the gauge fields, the gauge-fixing, the ghost sector, the fermions, the scalars and the Yukawa interactions. These are explicitly given by  
\bea
T^{A}_{ij} &=& \frac{2}{\gYM^2} \tr \left[ F_{ik} F_{jk} - \delta_{ij} \frac{1}{4} F_{kl} F_{kl} \right] , \nn \\
T^{g.f.}_{ij} &=& \frac{2}{ \gYM^2} \frac{1}{\xi} \tr \left[ A_i \partial_j \left( \partial_k A_k\right)  + A_j \partial_i \left( \partial_k A_k\right) - \delta_{ij} \left( A_k\ \partial_k \partial_l A_l + \frac{1}{2} (\partial_k A_k) (\partial_l A_l) \right)\right] , \nn \\
T^{gh}_{ij} &=& \frac{2}{\gYM^2} \tr \left[ \partial_i \bar{c} \mD_j c + \partial_j \bar{c} \mD_i c - \delta_{ij} \partial_k \bar{c} \mD_k c\right] , \nn \\
T^{\psi}_{ij} &=& \frac{2}{\gYM^2} \tr \left[ \frac{1}{2} \bar{\psi}^L \gamma_{(i} \oa{\mD}_{j)} \psi^L - \delta_{ij} \frac{1}{2} \bar{\psi}^L \gamma_k \oa{\mD}_k \psi^L \right] , \nn \\
T^{\Phi}_{ij} &=& \frac{2}{\gYM^2} \tr \left[ \mD_i \Phi^M \mD_j \Phi^M - \delta_{ij}  \frac{1}{2} (\mD \Phi^M)^2   +  \xi_M \left( \delta_{ij} \partial^2 - \partial_i \partial_j\right) (\Phi^M)^2\right]  \nn \\
&-&  \frac{2}{\gYM^2} \delta_{ij} \left[  \frac{1}{4!}  \lambda^{(1)}_{M_1 M_2 M_3 M_4}  \tr \, \Phi^{M_1} \Phi^{M_2} \Phi^{M_3} \Phi^{M_4} +  \frac{1}{4! \, N}  \lambda^{(2)}_{M_1 M_2 M_3 M_4}  \tr \, \Phi^{M_1} \Phi^{M_2} \, \, \tr \, \Phi^{M_3} \Phi^{M_4}   \right] , \nn \\
T^{Y}_{ij} &=& \frac{2}{\gYM^2} \tr \left[ - \delta_{ij} \, \, \sqrt{2} \mu_{M L_1 L_2} \Phi^M \bar{\psi}^{L_1} \psi^{L_2} \right] .
\eea
In particular, $T^\Phi_{ij}$ represents the energy-momentum tensor for a generic non-minimal scalar $\Phi^M$, $(M = 1, \ldots, \mN_\Phi)$, which reduces to the minimal case if $\xi_M = 0$ or to the conformally coupled case if $\xi_M = (d-2)/(4(d-1))$. For instance, in three dimensions the conformal scalar is characterised by $\xi_M = 1/8$. 

In \cite{Bzowski:2011ab}, the cancellation between the gauge-fixing and the ghost contributions in correlation functions of the energy-momentum tensor has been proven on general grounds.
As such, it is sufficient to consider only $T^A_{ij}$ in the gauge sector. We have explicitly checked that this property is actually realised in the $\<TT\>$ correlator up to 2-loop order in perturbation theory.

General covariance fixes the structure of the 2-point function of the energy-momentum tensor in the following form 
\bea
\<\!\< T_{ij}(q) T_{kl}(- q)\>\!\> = A(q) \Pi_{ijkl}^{(d)} + B(q) \pi_{ij} \pi_{kl},
\eea
where the transverse and transverse traceless projection tensors are defined respectively as
\bea
\label{eq:tensorbasis}
\pi_{ij} = \delta_{ij} - \frac{q_i q_j}{q^2} \,, \qquad \Pi_{ijkl}^{(d)} = \frac{1}{2}\left(\pi_{ik} \pi_{jl} + \pi_{il} \pi_{jk} - \frac{2}{d-1} \pi_{ij} \pi_{kl} \right) \,.
\eea
The double bracket notation is used to remove the momentum conserving delta function, {\it i.e.},
\bea
\< T_{ij}(\vec{q}_1) T_{kl}(\vec{q}_2)\> = (2\pi)^3 \delta(\vec{q}_1 + \vec{q}_2) \<\!\< T_{ij}(q_1) T_{kl}(- q_1)\>\!\> \,.
\eea
and $q_1$ is the magnitude of $\vec{q}_1$.
In the following sections we provide the contribution to the $\< TT \>$ correlation function, namely to the $A$ and $B$ coefficients, from each of the individual sector of the model, the gauge, the fermion, the scalar and the Yukawa one. We present the results in arbitrary dimensions and we finally specify them to the particular cases of $d = 3,4,5$.

\subsection{Form factors:  Results in arbitrary dimensions}

\begin{figure}[h]
\centering
\includegraphics[scale=1]{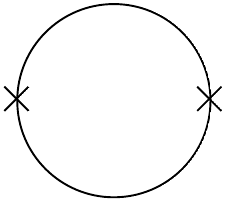}
\caption{1-loop contribution to the $\< TT \>$. \label{Fig.TT1L}}
\end{figure}

The only topology appearing at 1-loop order is the one depicted in Fig.~\ref{Fig.TT1L}. From an explicit computation we obtain
\bea
A^{(1)}_A &=&  d(G) \frac{2 d^2 - 3d - 8}{4(d^2-1)} \frac{q^d}{(4 \pi)^{d/2}} G_1 , \\
B^{(1)}_A &=&  d(G) \frac{(d-4)^2 (d-2)}{8(d-1)^2} \frac{q^d}{(4 \pi)^{d/2}} G_1 , \\
A^{(1)}_\psi &=& \mN_\psi \, d(G) \frac{1}{4(d+1)} \frac{q^d}{(4 \pi)^{d/2}} G_1 , \\
B^{(1)}_\psi &=& 0 , \\
A^{(1)}_\Phi &=& \mN_\Phi \, d(G) \frac{1}{4(d^2-1)} \frac{q^d}{(4 \pi)^{d/2}} G_1 , \\
B^{(1)}_\Phi &=&  \sum_{M = 1}^{\mN_\Phi} \, 2 d(G)  \left[ \xi_M   - \frac{d - 2}{4(d-1)} \right]^2 \frac{q^d}{(4 \pi)^{d/2}} G_1 , 
\eea
where $d(G)$ is the dimension of the adjoint representation, namely $d(G) = N^2 - 1$, while $G_1$ is the loop function defined in Eq.(\ref{Eq.G1}). Notice that, at 1-loop order, neither the Yukawa nor the quartic-scalar interactions contribute to the $A$ and $B$ coefficients. 
The $B$ coefficient describes the departure from conformality. 
In particular, it identically vanishes in the fermion sector in arbitrary dimensions while for the gauge field it is only true in $d = 4$. On the other hand, the scalars need to be conformally coupled, $\xi_M = (d-2)/(4(d-1))$.

\begin{figure}[h]
\centering
\includegraphics[scale=1]{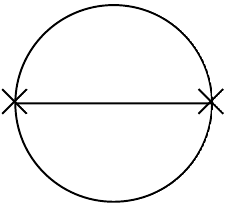} \hspace{0.5cm}
\includegraphics[scale=1]{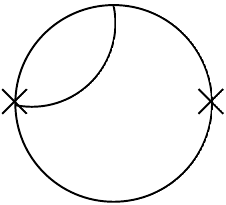} \hspace{0.5cm}
\includegraphics[scale=1]{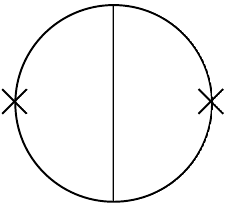} \hspace{0.5cm}
\includegraphics[scale=1]{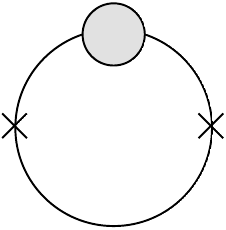} \hspace{0.5cm}
\includegraphics[scale=1]{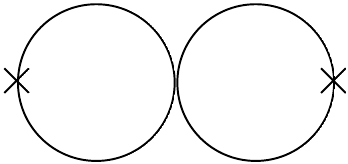}
\caption{2-loop contribution to the $\< TT \>$. The blob in the fourth diagram represents an insertion of a 1-loop self-energy. \label{Fig.TT2L}}
\end{figure}

At 2-loop order, the topologies of the diagrams appearing in perturbative expansion of the $\<TT\>$ are shown in Fig.~\ref{Fig.TT2L}. 
The 2-loop corrections are suppressed, with respect to the leading order, by $\gYM^2 \, N$ and can be organised, as usual, as the sum of different contributions: the gauge, the fermion and the scalar sectors, and are expressed in terms of the two loop functions $G_1$ and $G_2$ given in Eq.(\ref{Eq.G1}) and (\ref{Eq.G2}). 

In the gauge sector we obtain
\bea
A^{(2)}_A &=& \gth  \frac{2   d(G)}{3 (d-4)^2 (d-2) (d^2-1)} \left[  3 (d-4) \left( d^4-8 d^3+16 d^2+20 d-68 \right) G_1^2  \right.  \nn \\
    &+& \left.    \left( -13 d^5+129 d^4-462 d^3+572 d^2+376 d-1088 \right) G_2\right] \frac{q^{d}}{(4\pi)^d}, \\
B^{(2)}_A &=&  - \gth \frac{    \, d(G) }{24 (d-1)^2 } \left[  3 (d-4) ( d^3 - 16 d^2 + 68 d -88 ) G_1^2   + 16 (4 d^3 - 33 d^2 + 94 d -   92) G_2\right]\frac{q^{d}}{(4\pi)^d} ,\nn \\
\eea 
while the contribution of the fermions is
\bea
A^{(2)}_\psi &=& \gth \frac{ \mN_\psi  \, d(G) }{12 (d-4)^2 (d-2) (d^2-1) } \left[ 9 (d-4) \left( 3 d^3-23 d^2+54 d-32 \right) G_1^2 \right. \nn \\
&+& \left.   4 \left( -10 d^5+95 d^4-289 d^3+182 d^2+464 d-496 \right) G_2^2    \right] \frac{q^{d}}{(4\pi)^d} , \\
B^{(2)}_\psi &=& -\gth  \frac{\mN_\psi   \, d(G)  }{3 (d-1)^2 }  \left[ (d-4) (d-2)^2  G_2  \right]  \frac{q^{d}}{(4\pi)^d} .  
\eea
The results for the scalar contributions are
\bea
A^{(2)}_\Phi &=& \gth  \frac{2 \, \mN_\Phi  \, d(G)   }{3 (d-4)^2 (d-2) (d^2-1)} \left[ 3 (d-4) \left( d^2-7 d+13 \right) G_1^2 \right. \nn \\
&+& \left.  \left(176 - 30 d - 73 d^2 + 33 d^3 - 4 d^4  \right) G_2 \right] \frac{q^{d}}{(4\pi)^d}, 
\eea
\bea
B^{(2)}_\Phi &=& \gth \sum_{M=1}^{\mN_\Phi}   \frac{ d(G)  }{24 (d-4)^2  (d-1)^2 } \left[ 
- 3 (d-4) (d - 2  -  4 ( d -1 ) \xi_M ) \left(  16 (-9 + 2 \xi_M) + d (110 - 44 \xi_M   \right. \right. \nn \\
&+& \left. \left.  d (-27 + 2 d + 12 \xi_M )) \right)  G_1^2  + 16 \left(-272 + 370 d - 189 d^2 + 44 d^3 - 4 d^4 \right. \right.  \nn \\
&+& \left. \left.  24 (d  -3 ) (d -1 ) (8 + (-5 + d) d) \xi_M -  12 (d -2 ) (d -1 )^2 ( 3 d -8 ) \xi_M^2 \right)  G_2
\right] \frac{q^{d}}{(4\pi)^d} \nn \\
&-& \gth \sum_{M_1, M_2=1}^{\mN_\Phi} \frac{d(G)}{6}\left[ \frac{2 N^2 - 3}{ N^2}  \lambda^{(1)}_{M_1 M_1 M_2 M_2}   + \frac{1+ N^2}{N^2} \lambda^{(2)}_{M_1 M_1 M_2 M_2}  \right] \nn \\
&& \qquad \qquad \times \left(\xi_{M_1} - \frac{d-2}{4(d-1)} \right) \left(\xi_{M_2} - \frac{d-2}{4(d-1)} \right) G_1^2  \frac{q^{d}}{(4\pi)^d} \,.
\eea
Notice that the quartic scalar contribution only originates from the last diagram of Fig.~\ref{Fig.TT2L} which is simply the product of the two 1-loop topology graphs.
This term is identically vanishing if at least one of the two scalars running in the two loops is conformally coupled. 

Finally we present the contribution of the Yukawa interactions
\bea
A^{(2)}_Y &=& \gth \frac{ \mu_{Y}^2  \, d(G)}{24 (d - 4) (d - 2) (d -1)}  \left[ -3 (d -4) G_1^2 + 4 (-2 + (d -2) d) G_2 \right] \frac{q^{d}}{(4\pi)^d} , \\
B^{(2)}_Y &=& \gth \sum_{M=1}^{\mN_\Phi}  \frac{ d(G) \,  \mu_{M L_1 L_2} \, \mu_{M L_2 L_1}}{6 (d-4) (d -1)^2}  \left[ 
 8 + d (-3 + (d -3 ) d) - 12  d (d -3 ) ( d -1 )  \xi_M  \right. \nn \\
 && \left. +  12 (d -1)^2 (3d -8 ) \xi_M^2
\right] G_2 \frac{q^{d}}{(4\pi)^d},   
\eea
where we have defined $\mu_{Y}^2 = \mu_{M L_1 L_2} \, \mu_{M L_2 L_1}$ as the square of the Yukawa coupling. The sum over all the three flavour indices, where not explicitly stated, is always implicitly understood. 
In particular, notice that in $B_Y^{(2)}$ the sum of the scalar flavour has been shown explicitly because the square of the Yukawa coupling is weighted by $\xi_M$ and $\xi_M^2$.
\section{The form factors $A$ and $B$ in $d=3,4,$ and $5$ dimensions and renormalisation of the $TT$ }
In this section we will discuss the structure of the correlator in various dimensions, focusing on  
the $d=3,4$ and 5 cases.
\label{renTT}

\subsection{ Form factors: renormalised results in $d=3$}

In $d=3$ dimensions the coefficients defined above take the form 
\bea
A^{(1)}_A &=& B^{(1)}_A = \frac{d(G)}{256}   q^3 , \\
A^{(1)}_\psi &=& \frac{d(G)}{128} \mN_\psi q^3, \qquad B^{(1)}_\psi = 0, \\
A^{(1)}_\Phi &=&  \frac{d(G)}{256} \mN_\Phi  q^3 \,, \qquad  B^{(1)}_\Phi =   \frac{ d(G) }{256}  \sum_{M = 1}^{\mN_\Phi}   \left( 1 - 8 \xi_M \right)^2 q^3
\eea
and thus in total we have
\bea
\label{Eq.AB1Loop}
A^{(1)} = \frac{d(G)}{256} \mN_{(A)}  q^3, \qquad B^{(1)} = \frac{d(G)}{256} \mN_{(B)}  q^3,
\eea
with
\bea
\mN_{(A)} = 1 + 2 \mN_\psi + \mN_\Phi , \qquad \mN_{(B)} = 1 +  \sum_{M = 1}^{\mN_\Phi}   \left( 1 - 8 \xi_M \right)^2 .
\eea
Notice that in $d=3$ dimensions  the $\<TT\>$ is both UV and IR finite. 
In the large $N$ limit we recover the results obtained in \cite{McFadden:2009fg,McFadden:2010na}.

Concerning the 2-loop results,
it is possible to recognise, simply by naive dimensional analysis, that the $\<TT\>$ may develop UV and IR divergences in $d=3$ dimensions. Indeed, while $G_1$ is finite, the loop function $G_2$ has a single pole in $\epsilon$, for $d=3- \epsilon$
\bea
G_2 \underset{d \to 3 - \epsilon}{=} 2 \pi \left( \frac{1}{\epsilon} + 3 - \gamma_E \right) + O(\epsilon) ,
\eea 
where $\gamma_E$ is Euler-Mascheroni constant. 

By a closer inspection of every contribution in the diagrammatic expansion of the $\<TT\>$, we find that all the topologies give rise to UV divergences, with the only exception of the last one depicted in Fig.~\ref{Fig.TT2L} which is indeed finite. 
For the UV divergence of the $\<TT\>$ at 2-loop order we find
\bea
A^{(2)} |_{\mathrm{UV \, sing.}} =  - 2 B^{(2)}  |_{\mathrm{UV \, sing.}} =  \gth  \, q^3 \frac{ d(G)}{192 \pi ^2} \left[    (2  - \mN_\psi - 2 \mN_\Phi)  - \frac{1}{2} \mu^2_Y \right]  \frac{1}{\bar \epsilon},
\eea
which gives
\bea
\<\!\< T_{ij} (q) T_{kl} (- q) \>\!\> |_{\mathrm{UV \, sing.}} = \gth   \, q^3 \frac{  d(G) }{192 \pi ^2}\left[   (2  - \mN_\psi - 2 \mN_\Phi)  - \frac{1}{2} \mu^2_Y \right]  \frac{1}{\bar \epsilon} \left( \Pi_{ijkl} - \frac{1}{2} \pi_{ij} \pi_{kl} \right) .
\eea
This divergence can be removed by a suitable counterterm defined as the double variation, with respect to the metric tensor, of $\rg \, R$, namely
\bea
\mathrm{F.T.} \left[ \frac{\delta^2}{\delta g^{ij} \delta g^{kl}} \int d^3 x \, \delta_{CT} \, \rg \, R  \right] = - \delta_{CT} \frac{q^3}{2} \left( \Pi_{ijkl}^{(3)} - \frac{1}{2} \pi_{ij} \pi_{kl} \right) ,
\eea
where F.T. denotes the Fourier transform and the coefficient $\delta_{CT}$, in the $\overline{\rm MS}$ scheme, is given by,
\bea
\delta_{CT} =  \gth  \frac{ d(G) }{96 \pi ^2} \left[    (2  - \mN_\psi - 2 \mN_\Phi)  - \frac{1}{2} \mu^2_Y \right] \frac{1}{\bar \epsilon} \,.
\eea

The IR divergent contribution emerges only in the scalar and Yukawa sectors from the fourth topology in Fig.~\ref{Fig.TT2L}, characterised by an insertion of the scalar 1-loop self-energy. All the other diagrams have enough integration momenta in the numerator to avoid any IR singular behaviour. In particular, the IR singularity arises only from the improvement term in the scalar energy-momentum tensor and, therefore, is proportional to $\xi_M^2$. Indeed the first and the last topologies in Fig.~\ref{Fig.TT2L} do not have enough propagators to develop an IR divergence in $d=3$: when the two integration momenta ${k}_1$ and ${k}_2$ go to zero, such that ${k}_1 \sim {k}_2 \to 0$, their denominators behave at most as ${k}^4$ while the numerators goes a $k^6$. For the diagrams with the remaining topologies in Fig.~\ref{Fig.TT2L}, power counting suggests that there are possible IR logarithmic singularities, but
in all cases these are avoided because the energy-momentum tensor provides an additional integration momentum in the numerator of the Feynman diagram. The only exception to that is if we consider  the $\xi$-dependent part of the energy-momentum tensor which does not depend on any of the momenta of the two internal lines but only the external momentum $q$, thus allowing the IR singularity to appear. 
We find
\bea
\label{Eq.IRint}
\<\!\< T_{ij} (q) T_{kl} (- q) \>\!\> |_{\mathrm{IR \, sing.}} = \<\!\< T^{\Phi,\xi}_{ij} (q) T^{\Phi, \xi}_{kl} (- q) \>\!\> = \sum_{M = 1}^{\mN_\Phi} 4 \, \gYM^2  \, \xi_M^2 q^4 \pi_{ij} \pi_{kl} \int \frac{d^d {k}}{(2 \pi)^d} \frac{{\Pi_{\Phi}^{(1)}}^{aa}_{MM}({k})}{{k}^4 ({k} + q)^2}, \nn\\
\eea
where $T^{\Phi,\xi}_{ij}$ is the $\xi$-dependent part of the scalar energy-momentum tensor and ${\Pi_{\Phi}^{(1)}}^{ab}_{M_1 M_2}({k})$ is the 1-loop self-energy of a massless scalar field given in Eq.~(\ref{PiScalar1Loop}).
By naive power counting arguments we find that this contribution is logarithmic IR divergent but UV finite. In dimensional regularisation with $d=3 + \omega$, with $\omega >0$,  we obtain
\bea
\<\!\< T_{ij} (q) T_{kl} (- q) \>\!\> |_{\mathrm{IR \, sing.}} =  \gth \, q^3 \frac{d(G)}{8 \pi^2} \, \Sigma_\Phi^{IR} \left[ \frac{1}{\bar{\omega}} + 1 + 2 \log \frac{q}{\mu_{\rm{IR}}}\right]  \pi_{ij} \pi_{kl}  ,
\eea
where, as usual, $1/\bar{\omega} = 1/\omega + \gamma_E - \log4\pi$ and
\bea
\Sigma_\Phi^{IR} =  \sum_{M = 1}^{\mN_\Phi} \xi_M^2 \left( 2 +  \frac{1}{2} \, \mu_{M L_1 L_2} \, \mu_{M L_2 L_1}\right)
\eea
represents the contribution of all the scalars to the IR divergence. 

The same result can be obtained using a mass regulator which amounts to replace ${k}^2$ with ${k^2} + m^2$ in the integral in Eq.~(\ref{Eq.IRint}). In this case the singularity appears as a $\log m$, in the form 
\bea
\<\!\< T_{ij} (q) T_{kl} (- q) \>\!\> |_{\mathrm{IR \, sing.}} &=& \gth \, d(G) \frac{q^3}{8 \pi^2} \, \Sigma_\Phi^{IR}          
\frac{2 \, \mathrm{arcsinh}(\sqrt{q^2/m^2})}{\sqrt{1 + m^2/q^2}}    \pi_{ij} \pi_{kl} \nn \\
&\underset{m^2 \to 0}{\sim}& \gth \, d(G)  \frac{q^3}{8 \pi^2} \, \Sigma_\Phi^{IR}         
\left(  2 \log \frac{2 q}{m} \right)   \pi_{ij} \pi_{kl} .
\eea

The two renormalised form factors in $d=3$ are
\bea
A^{(2)} &=& \gth \, d(G)  \frac{q^3}{1536 \pi^2}  \left[ 40 + 52 \mN_\psi -8 \mN_\Phi + 2 \mu_Y^2 -3 \pi ^2 (2 + 3 \mN_\psi + 2 \mN_\Phi + \frac{1}{2} \mu_Y^2) \right. \nn \\
&+& \left.  16 (-2 + \mN_\psi+2 \mN_\Phi +\frac{1}{2} \mu_Y^2) \log \frac{q}{\mu_{\rm{UV}}}   \right] , \\
B^{(2)} &=&  - \gth \, d(G)   \frac{q^3}{6144 \pi^2}  \left[  16 - 56 \mN_\psi  - 4 \mu_Y^2 + 3 \pi ^2 
 + 32 (-2 + \mN_\psi+2 \mN_\Phi + \frac{1}{2} \mu_Y^2) \log \frac{q}{\mu_{\rm{UV}}} \right. \nn \\
&+& \left.  \sum_{M = 1}^{\mN_\Phi} 3 (8 \xi_M -1) \left(8 \left(\pi^2-16\right) \xi_M -3 \pi ^2+112 + 2 \mu_{M L_1 L_2} \, \mu_{M L_2 L_1}\right) \right] \nn \\
&+& \gth \, d(G)   \frac{q^3}{8 \pi^2} \Sigma_\Phi^{IR}  \left[ \frac{1}{\bar \omega}  + 1 + 2 \log \frac{q}{\mu_{\rm{IR}}} \right] \nn \\
&-& \sum_{M_1, M_2=1}^{\mN_\Phi} \gth \frac{q^3}{24576}  d(G)\left[ \frac{2 N^2 - 3}{ N^2}  \lambda^{(1)}_{M_1 M_1 M_2 M_2}   + \frac{1+ N^2}{N^2} \lambda^{(2)}_{M_1 M_1 M_2 M_2}  \right]  \left(8 \xi_{M_1} - 1 \right) \left(8 \xi_{M_2} - 1 \right) , \nn \\
\label{Eq.B2Loop}
\eea 
where $\mu_{\rm{UV}}$ is the renormalisation scale. In the $B^{(2)}$ form factor we have isolated the contribution affected by the IR divergence which is proportional, as we have already discussed, to $\xi_M^2$. The singularity has been regularised in dimensional regularisation with $d = 3 + \omega$, with $\omega > 0$, and it is characterised by the IR scale $\mu_{\rm{IR}}$. Notice that this term is absent in the minimal coupled scalar case, where $\xi_M=0$.\\

\subsection{Form factors: renormalised results in $d=4$}

In $d = 4$ $\<TT\>$  develops a UV singularity in both the coefficients $A$ and $B$ and a renormalisation of the correlator is necessary already at 1-loop level. For the conformal fields, namely the gauge field, the fermion field and the conformally coupled scalar, the divergence can be removed by the square of the 4-dimensional Weyl tensor $F = R_{ijkl} R^{ijkl} - 2 R_{ij} R^{ij} + 1/3 R^2$, while a non-conformally coupled scalar requires an extra $R^2$. In particular,
the second order variation with respect to the metric tensor gives
\bea
\textrm{F.T.} \left[ \frac{\delta^2}{\delta g^{ij} \delta g^{kl}} \int d^4 x \, \sqrt{g} \left( \delta_{CT} \, F + \delta_{CT}' \,  R^2 \right) \right] =  \delta_{CT} \, q^4 \,  \Pi_{ijkl}^{(4)}  + 2 \, \delta_{CT}' \, q^4 \, \pi_{ij} \pi_{kl},
\eea
where F.T. denotes  Fourier transform and the coefficients of the counterterms in the $\overline{\rm MS}$ scheme at 1-loop order are
\bea
\delta_{CT}^{(1)} &=& - \frac{1}{q^4} A^{(1)}|_{\mathrm{UV \, sing.}}  = - d(G) \left( \frac{1 }{40 \pi^2} + \frac{\mN_\psi}{160 \pi^2} + \frac{\mN_\Phi}{480 \pi^2}  \right) \frac{1}{\epsilon} \,, \nn \\  
\delta_{CT}^{(1) '} &=& - \frac{1}{2 q^4}   B^{(1)}  \bigg|_{\mathrm{UV \, sing.}}  =  -  \sum_{M = 1}^{\mN_\Phi} \, d(G) \frac{1}{288 \pi^2} (1-6 \xi_M)^2 \frac{1}{\epsilon}  \,.
\eea
The renormalised results are
\bea
A^{(1)}_A &=& d(G) \frac{q^4 }{800 \pi^2} \left( 9 - 20 \log \frac{q}{\mu_\UV} \right), \nn \\
 B^{(1)}_A &=& - d(G) \frac{q^4 }{360 \pi^2}   \,, \nn \\
A^{(1)}_\psi &=& \mN_\psi \, d(G)  \frac{q^4 }{800 \pi^2} \left( 6 - 5 \log \frac{q}{\mu_\UV} \right)  \,, \nn \\
B^{(1)}_\psi &=& - \mN_\psi \, d(G)  \frac{q^4 }{1440 \pi^2}  \,, \nn \\
A^{(1)}_\Phi &=& \mN_\Phi \, d(G)  \frac{q^4 }{7200 \pi^2} \left( 23 - 15 \log \frac{q}{\mu_\UV} \right)  \,, \nn \\
B^{(1)}_\Phi &=&  \sum_{M = 1}^{\mN_\Phi} \, d(G)  \frac{q^4 }{432 \pi^2} (1 - 6 \xi_M) \left( 2 - 18 \xi_M - 3 (1 - 6 \xi_M) \log \frac{q}{\mu_\UV} \right)  -  \mN_\Phi \, d(G)  \frac{q^4 }{4320 \pi^2} .\nn\\
\eea
\begin{figure}
\centering
\includegraphics[scale=0.8]{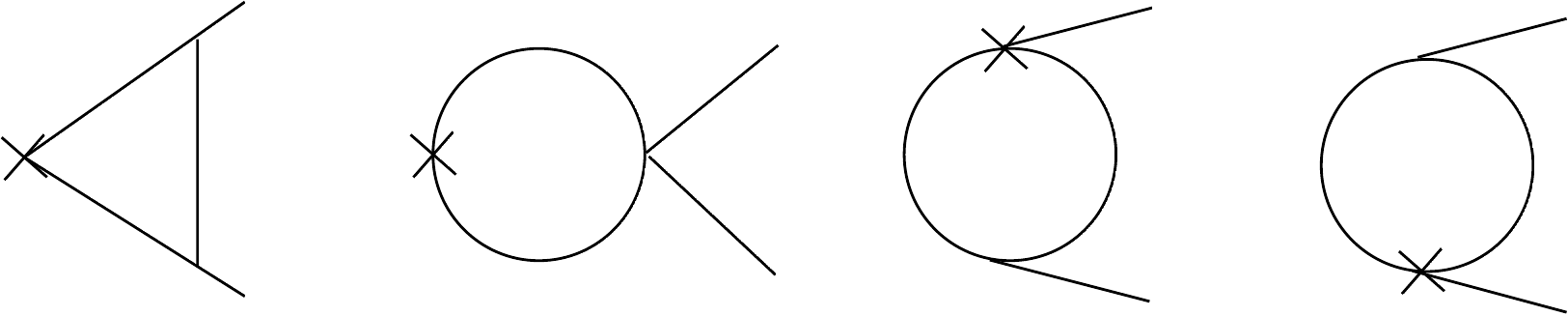} 
\caption{1-loop perturbative expansion of the $\< T \Phi \Phi \>$ correlator. \label{Fig.Tphiphi}}
\end{figure}
Notice that the $B^{(1)}$ factors in the gauge and fermion sectors and the last term in $B^{(1)}_\Phi$ are generated from the singular part of the corresponding $A^{(1)}$ coefficients due to the $d$-dependence in $\Pi_{ijkl}^{(d)}$. In particular, in $d = 4 - \epsilon$,  
\bea
\Pi_{ijkl}^{(d)} = \Pi_{ijkl}^{(4)} - \frac{\epsilon}{9} \, \pi_{ij} \pi_{kl} 
\eea
has a finite projection onto the $B^{(1)}$ coefficients. These contributions are local and they may be set to zero by 
adding a finite part in $\delta_{CT}^{(1)'}$.

Before discussing the UV behaviour of the 2-loop $\<TT\>$ correlator, we complete the renormalisation program at 1-loop order in perturbation theory by analysing the improvement term $\xi_M  R (\Phi^M)^2$. 
This is necessary for the renormalisation of the 2-point function of the energy-momentum tensor at higher orders.
The improvement term undergoes an additive renormalisation, $\xi_{M_1} \delta_{M_1 M_2} \rightarrow \xi_{M_1} \delta_{M_1 M_2} + \delta \xi_{M_1 M_2}$, when the scalar is not conformally coupled, namely, away from the $\xi_M = 1/6$ case. The definition of the $\delta \xi_{M_1 M_2}$ counterterm is encoded in the UV singularity of the $\< T \Phi \Phi \>$ correlator which we study at 1-loop order in perturbation theory. The different topologies contributing to the 3-point function are depicted in Fig.~\ref{Fig.Tphiphi} and amount to triangles and bubbles diagrams with gauge, scalar and fermion fields running in the internal lines. 
The computation is checked using the conservation Ward identity originating from the diffeomorphism invariance of the theory,
\bea
k_i \, \<\!\< T_{ij} (\bk) \Phi_{M_1}^{a} ( q_1) \Phi_{M_2}^{b} ( q_2) \>\!\> = q_{1i} \, {\Pi_{\Phi}}^{ab}_{M_1 M_2}(q_2)  +   q_{2i} \, {\Pi_{\Phi}}^{ab}_{M_1 M_2}(q_1)
\eea
where $ {\Pi_{\Phi}}^{ab}_{M_1 M_2}$ is the unrenormalised scalar field self energy defined in Eq.~(\ref{eq:selfenergyscalar}). As stated above, the $\< T \Phi \Phi \>$ develops a UV divergence which is cancelled by the wave-function renormalisation of the scalar field and by the counterterm $\delta \xi_{M_1 M_2}$. The counterterm of $\< T \Phi \Phi \>$  is extracted from the quadratic part of the renormalised energy-momentum tensor $T^{\Phi}_{ij}$ and it is given by
\bea
T^{\Phi \, \textrm{c.t.}}_{ij} &=& \delta_{\Phi \, M_1 M_2} \frac{2}{\gYM^2} \tr \left[ 
\partial_i \Phi_{M_1} \partial_j \Phi_{M_2} - \frac{\delta_{ij}}{2}  \partial_k \Phi_{M_1} \partial_k \Phi_{M_2}  + \, ( \xi_{M_1} + \delta \xi_{M_1 M_2} ) \left( \delta_{ij} \partial^2 - \partial_i \partial_j\right) \Phi_{M_1} \Phi_{M_2}  \right]  \nn \\ 
&+& \delta \xi_{M_1 M_2}  \frac{2}{\gYM^2} \tr \left[ \left( \delta_{ij} \partial^2 - \partial_i \partial_j\right) \Phi_{M_1} \Phi_{M_2} \right]  + \ldots
\eea
where the dots represent cubic and quartic terms in the fields which are unnecessary for our purpose. The counterterm $\delta_{\Phi \, M_1 M_2} $ is fixed by the renormalisation of the scalar 2-point function and it is explicitly given in Eq.~(\ref{CT1L}) at 1-loop order in perturbation theory, while $\delta \xi_{M_1 M_2}$ is determined here by the cancellation of the singularity in $\< T \Phi \Phi \>$ which is given by
\bea
\xi_{M_1 M_2}^{(1)} &=&  \frac{ \gYM^2 N}{96 \pi^2 \epsilon} \left[   (6 \, \delta_{M_1 M_2} + \mu^{(0)}_{M_1 M_2} ) (1 - 6 \xi_{M_1})  \right.  \nn \\
&-& \left.  \frac{1}{6} \left( \frac{ 2 N^2 -3}{ N^2}  \lambda^{(1)}_{M_1 M_2 M_3 M_3}  +  \frac{ 1+  N^2 }{ N^2}  \lambda^{(2)}_{M_1 M_2 M_3 M_3} \right) (1 - 6 \xi_{M_3} ) \right] \,.
\eea
Notice that there is no need of renormalisation of the improvement term if the scalars are conformally coupled as the remnant singularity of the 3-point function $\< T \Phi \Phi \>$, after the subtraction of the scalar wave-function contribution, vanishes for $\xi = 1/6$.\

Having completed the renormalisation at 1-loop order of the $\<TT\>$ and $\< T \Phi \Phi \>$ correlators we can come back to the analysis of the 2-loop 2-point function of the energy-momentum tensor in $d=4$ which also appears to be divergent in the UV.
Being the 4 dimensional theory already plagued by infinities at 1-loop level, the 2-loop perturbative expansion of the correlator is characterised by contributions of countertems inserted in the 1-loop topology diagrams, both in the $T$ vertices and in the internal propagators. With the only exception of $\delta \xi$ contribution, these counterterm insertions are proportional to the wave-function renormalisation constants of the elementary fields and exactly cancel each other. The only remaining sub-divergence is related to the non-minimal scalar coupling and can be removed by $\delta \xi$ which is extracted from the renormalisation of $\< T \Phi \Phi \>$. 
This is necessary to cancel a $1/\epsilon \log q$ singularity which otherwise could not be absorbed into a local counterterm. As such, one is left with a UV singularity entirely arising from genuine 2-loop topologies.
The divergence is removed by the same local counterterms, $F$ and $R^2$, introduced in the 1-loop analysis with coefficients given by
\bea
\delta_{CT}^{(2)} &=& - \frac{1}{q^4} A^{(2)}|_{\mathrm{UV \, sing.}}  =  - \gth \frac{ d(G)}{36864 \pi^4} \left( -64 + 28 \mN_\psi + 16 \mN_\Phi + 3  \mu_Y^2 \right)  \frac{1}{\epsilon}\,, \nn \\
\delta_{CT}^{(2)'} &=& - \frac{1}{2 q^4} B^{(2)}  \bigg|_{\mathrm{UV \, sing.}}  = 
\gth \frac{ d(G)}{4608 \pi^4}  \left[  \sum^{\mN_\Phi}_{M=1}  (1 - 6 \xi_M)^2 \left( 6 + \mu_{M L_1 L_2} \mu_{M L_2 L_1} \right)   \right. \nn \\
&-& \left. \sum^{\mN_\Phi}_{M_1, M_2=1} \frac{1}{6} \left( \frac{2N^2 - 3}{ N^2}  \lambda^{(1)}_{M_1 M_1 M_2 M_2} + \frac{1+N^2}{ N^2}  \lambda^{(2)}_{M_1 M_1 M_2 M_2} \right) (1 - 6 \xi_{M_1}) (1-6 \xi_{M_2}) \right] \frac{1}{\epsilon^2} \nn \\
&-& \gth \frac{ d(G)}{4608 \pi^4}    \sum^{\mN_\Phi}_{M=1}  (1 - 6 \xi_M)^2 \left( 4 + \frac{1}{2} \mu_{M L_1 L_2} \mu_{M L_2 L_1} \right)   \frac{1}{\epsilon}  \,.
\eea

We present below the renormalised expressions of the 2-loop contributions for the $A$ and $B$ form factors due to the gauge and fermion sectors.  They are given by 

\bea
A^{(2)}_A &=& \gth \, d(G) \frac{q^4}{17280 \pi^4} \left(    -1 + 60 \log \frac{q}{\mu_\UV} - 162 \, \zeta_3 \right) \,, \nn \\
B^{(2)}_A &=&  \gth \, d(G) \frac{37 q^4}{20736 \pi^4} \,, \nn \\
A^{(2)}_\psi &=& - \gth \, \mN_\psi  \, d(G) \frac{q^4}{552960 \pi^4} \left(  -1367 + 840 \log \frac{q}{\mu_\UV} + 1296 \, \zeta_3 \right)  \,, \nn \\
B^{(2)}_\psi &=& - \gth \, \mN_\psi   \,  d(G) \frac{31 q^4}{82944 \pi^4}  \,. \nn \\
\eea
and as in the 1-loop case, the $B^{(2)}$ form  factors for the gauge fields and scalars may be set to zero by 
adding suitable finite terms in $\delta_{CT}^{(2)'}$. The remaining contributions can be found in appendix \ref{sect}.

\subsection{Form factors: renormalised results in $d=5$}

We now turn to the case of $d=5$.
At 1-loop order we obtain
\bea
A^{(1)}_A &=& - d(G) \frac{9}{4096 \pi} q^{5} \, \qquad
B^{(1)}_A = - d(G) \frac{3}{16384 \pi} q^{5} \qquad A^{(1)}_\psi = - \mN_\psi d(G) \frac{1}{3072 \pi} q^{5} \,, \nn \\
B^{(1)}_\psi &=& 0 \,, \qquad A^{(1)}_\Phi = - \mN_\Phi d(G) \frac{1}{12288 \pi} q^{5} \,, \qquad B^{(1)}_\Phi =  - \sum_{M = 1}^{\mN_\Phi} d(G) \frac{1}{16384 \pi} \left( 3 - 16 \xi_M \right)^2 q^{5} \,. \nn\\
\eea

At 2-loop order in perturbation theory a UV divergence appears in both  coefficients. The singularities can be removed, as usual, by local counterterms constructed from $R_{ijkl}$, $R_{ij}$ and $R$. The corresponding second order variation with respect to the metric tensor is 
\bea
&& \textrm{F.T.} \left[ \frac{\delta^2}{\delta g^{ij} g^{kl}} \int d^5 x \sqrt{g} \left( \delta_{CT} \,  (R_{ijkl} \Box R^{ijkl} - 2 R_{ij} \Box R^{ij} + \frac{3}{8}R \, \Box R) + \delta'_{CT} \, R \, \Box R \right) \right] \nn \\
&& =  - \delta_{CT} \, q^5 \, \Pi_{ijkl}^{(5)} - 2 \delta'_{CT} \, q^5 \, \pi_{ij} \pi_{kl},
\eea
where the counterterms are given by
\bea
\delta_{CT}^{(2)} = \frac{1}{q^5} A|_{\mathrm{UV \, sing.}}  &=& \gth \frac{ d(G)}{1935360 \pi^4} \left[ -443 - \frac{271}{2} \mN_\psi -29 \mN_\Phi + \frac{13}{4} \mu_Y^2  \right] \frac{1}{\epsilon}  \,, \nn \\
\delta_{CT}^{'(2)} = \frac{1}{2q^5} B|_{\mathrm{UV \, sing.}}  &=& \gth \frac{ d(G)}{1146880 \pi^4} \bigg[ - \frac{106}{9} - \mN_\psi   + \sum_{M=1}^{\mN_\Phi} \bigg[  \frac{2}{3} (-49 + 64 (8 - 21 \xi_M) \xi_M)     \nn \\
&+&    \frac{1}{18} (43 + 96 \xi_M (-5 + 14 \xi_M)) \mu_{M L_1 L_2} \mu_{M L_2 L_1} \bigg]   \bigg] \frac{1}{\epsilon} \,.
\eea
The renormalised results due to the gauge and fermion sectors are
\bea
A^{(2)}_A &=&  \gth \, d(G) \frac{q^5}{6502809600 \pi^4} \left(-6725288 + 628425 \pi^2 + 2976960 \log\frac{q}{\mu_\UV} \right)   \,, \nn \\
B^{(2)}_A &=&  	\gth \, d(G)  \frac{q^5}{69363302400 \pi^4} \left(-3026624 + 760725 \pi^2 + 2849280 \log\frac{q}{\mu_\UV} \right) \,, \nn \\
A^{(2)}_\psi &=& \gth \, \mN_\psi \,  d(G)  \frac{q^5}{8670412800 \pi^4} \left(-2638704 + 209475 \pi^2 + 1214080 \log\frac{q}{\mu_\UV} \right)  \,, \nn \\
B^{(2)}_\psi &=& \gth \, \mN_\psi \,  d(G)  \frac{q^5}{2167603200 \pi^4} \left(557 + 7560 \log\frac{q}{\mu_\UV} \right). 
\eea
The remaining contributions can be found in appendix \ref{yuk}.

\section{Connection with holographic cosmology}
\label{hol}

One of the motivations for the current work was the need for the $d=3$ results in the context of holographic cosmology. In particular, the form factors $A$ and $B$ of the 2-point function of the energy-momentum tensor is related to the cosmological power spectra $\mathcal{P}$ and $\mathcal{P}_T$, respectively 
\cite{McFadden:2009fg,McFadden:2010na},
\begin{equation} \label{holo_2pt}
\mathcal{P}(q) = -\frac{q^3}{16  \pi^2} \frac{1}{{\rm Im} B(q)} , \quad
\mathcal{P}_T(q) = -\frac{ 2 q^3}{\pi^2} \frac{1}{{\rm Im} A(q)},
\end{equation}
where the imaginary part is taken after the analytic continuation,
\begin{equation} \label{analytic}
q \to -i q, \quad N \to -i N, 
\end{equation}

The generalised conformal structure and the large $N$ limit imply 
\begin{equation}
A(q, N)= q^3 N^2 f^T(\gth), \qquad B(q, N)= \frac{1}{4}  q^3 N^2 f^S(\gth)
\end{equation}
This is the analogue of (\ref{corr_GCS}) for the 2-point function of the energy-momentum tensor (the factor of 1/4 in $B$ is conventional).  In particular, factor $q^3$ reflects the fact that the energy-momentum tensor has dimension 3 in three dimensions (and $(2 \Delta -d)=3$) and the factor of $N^2$ is due to the fact that we are considering the leading term in the large $N$ limit.  Under the analytic continuation (\ref{analytic})
\begin{equation}
q^3 N^2 \to - i q^3 N^2, \qquad \gth \to \gth
\end{equation} 
so for this class of theories one may readily perform the analytic continuation and (\ref{holo_2pt}) becomes
\begin{equation} \label{holo_2pt_gcs}
\mathcal{P}(q) = \frac{q^3}{4 \pi^2 N^2 f^S(\gth)} , \quad
\mathcal{P}_T(q) = \frac{2 q^3}{\pi^2 N^2 f^T(\gth)}. 
\end{equation}
We have thus now arrived in a relation between cosmological observables and correlators of standard QFT.

In perturbation theory, the functions $f^{(S,T)}(\gth)$ can be expanded as
\bea
f(\gth) = f_0 \left( 1 -  f_1\,  \gth \log \gth  +  f_2 \, \gth  + O(\gth^2)  \right) \,.
\eea 
where we use the conventions (names of coefficients and relative signs)  of \cite{Easther:2011wh}. 
The leading order contribution, $f_0$, can be extracted from the 1-loop computation of the $\<TT\>$ and, in particular, from (\ref{Eq.AB1Loop}) thus obtaining
\bea
f_{0}^S = \frac{1}{64} \mN_{(B)} \,, \qquad  f_{0}^T = \frac{1}{256} \mN_{(A)} \,.
\eea
The coefficient $f_1$ of the logarithm term is computed from the 2-loop corrections given in (\ref{Eq.B2Loop}). Using the definition of the effective coupling we can exploit the relation
\bea
\gth \, \log (q / \mu) = - \gth \, \log \gth + \gth \, \log (\gYM^2 N/\mu), 
\eea
which can be used to recast the coefficients $A^{(2)}$ and $B^{(2)}$ in the form
\bea
A^{(2)} &=& - \gth \, N^2   \frac{q^3}{96 \pi^2}     (-2 + \mN_\psi+2 \mN_\Phi +\frac{1}{2} \mu_Y^2) \log \gth   + \ldots  \,, \nn \\
B^{(2)} &=&  \gth \, N^2 \,  q^3 \left[ \frac{1}{192 \pi^2}  (-2 + \mN_\psi + 2 \mN_\Phi + \frac{1}{2} \mu_Y^2)  -    \frac{1}{4 \pi^2} \Sigma_\Phi^{IR}  \right]     \log \gth     + \ldots \,.
\eea
Notice that, contrary to the 1-loop case, the two functions acquire contributions from all fields, even fermions and conformal scalars. 
Finally, the $f_1$ function is given by
\bea
f_1^S &=& - \frac{4}{3 \pi^2} \frac{1}{\mN_{(B)}} \left[ -2 + \mN_\psi + 2 \mN_\Phi + \frac{1}{2} \mu_Y^2 - 48 \sum_{M = 1}^{\mN_\Phi} \xi_M^2 \left( 2 + \frac{1}{2} \mu_{M L_1 L_2} \mu_{M L_2 L_1}\right) \right] \,, \nn \\
f_1^T &=&  \frac{8}{3 \pi^2} \frac{1}{\mN_{(A)}} \left[ -2 + \mN_\psi + 2 \mN_\Phi + \frac{1}{2} \mu_Y^2    \right] \,.
\eea
We also have
\bea
f_2^S &=& - \frac{1}{24 \, \mN_{(B)} \, \pi^2}  \bigg[  16 - 56 \mN_\psi  - 4 \mu_Y^2 + 3 \pi ^2 
 + 32 (-2 + \mN_\psi+2 \mN_\Phi + \frac{1}{2} \mu_Y^2) \log \frac{\gYM^2 \, N}{\mu}  \nn \\
&-&   3 \sum_{M =1}^{\mN_\Phi}(1 - 8 \xi_M)  \left(   112 - 3\pi^2 + 8 (\pi^2 - 16) \xi_M + 24 \mu_{M L_1 L_2} \mu_{M L_2 L_1}     \right) \bigg] \nn \\
&+&  \frac{32}{ \mN_{(B)} \, \pi^2} \sum_{M=1}^{\mN_\Phi} \xi_M^2 \left( 2  +  \frac{1}{2} \, \mu_{M L_1 L_2} \, \mu_{M L_2 L_1}\right) \left[ \frac{1}{\bar \omega}  + 1 + 2 \log \frac{\gYM^2 N}{\mu_*} \right] \nn \\
&-&   \frac{1}{48 \, \mN_{(B)}} \sum_{M_1,M_2 = 1}^{\mN_\Phi} (1 - 8 \xi_{M_1}) (1 - 8 \xi_{M_2}) \left[ \lambda^{(1)}_{M_1 M_1 M_2 M_2} + \frac{1}{2}   \lambda^{(2)}_{M_1 M_1 M_2 M_2}\right] \,, \nn \\
f_2^T &=&   \frac{1}{6 \, \mN_{(A)} \, \pi^2}  \bigg[ 40 + 52 \mN_\psi -8 \mN_\Phi + 2  \mu_Y^2 -3 \pi ^2 (2 + 3 \mN_\psi 
+  2 \mN_\Phi + \frac{1}{2}  \mu_Y^2   ) \nn \\
&+&  16 (-2 + \mN_\psi+2 \mN_\Phi  +   \frac{1}{2}  \mu_Y^2  ) \log \frac{\gYM^2 \, N}{\mu}  \bigg] \,.
\eea
(Recall that $\mu$ is a UV scale and $\mu^*$ is an IR scale). 
These results (with $\lambda^{(2)}_{M_1 M_1 M_2 M_2}=0$) were used in \cite{Afshordi:2016dvb}, where the predictions of these holographic models were compared with Planck data.

These results were instrumental in the comparison between the predictions of holographic cosmology and Planck data in   \cite{Afshordi:2016dvb}. In particular, the precise 2-loop results were needed in order to analyse whether there are models within this class that realise the best fit values obtained from the fit to data, and to check that the effective coupling constant is indeed small enough to justify the use of perturbation theory, for the momentum scales seen by Planck. It was found that gauge theory coupled to fermions only is ruled out by the data, but gauge theory coupled to sufficient number of scalars is ruled in, and it was further confirmed that this theory is indeed perturbative for almost all but the very low momenta (the theory becomes
non-perturbative in the region corresponding to CMB multipoles less the 30).

\subsection*{Holographic formulae for a model with no gauge fields}

We also quote here the results for a holographic model with no gauge fields.
 In this case the holographic coefficients read as
\bea
f_0^S &=& \frac{1}{64} \sum_{M=1}^{\mN_\Phi}(1-8 \xi_M)^2 \,, \nn \\
f_1^S &=& - \sum_{M=1}^{\mN_\Phi} \frac{2}{3 \pi^2 \mN_{(B)}} (1-48 \xi_M^2) \mu_{M L_1 L_2} \mu_{M L_2 L_1} \,, \nn \\
f_2^S &=& - \frac{1}{6 \, \mN_{(B)} \pi^2} \left( -1 + 4 \log \frac{\gYM^2 N}{\mu} \right) \mu_Y^2	\nn \\ 
&+& \frac{1}{ \mN_{(B)} \pi^2} \sum_{M=1}^{\mN_\Phi} \left[ 3(1 - 8 \xi_M) + 16 \xi_M^2  \left( \frac{1}{\omega} + 1 + 2 \log \frac{\gYM^2 N}{\mu_*} \right) \right] \mu_{M L_1 L_2} \mu_{M L_2 L_1}   \nn \\
&-&  \sum_{M_1, M_2 = 1}^{\mN_\Phi} \frac{1}{48 \mN_{(B)}} (1 - 8 \xi_{M_1})(1 - 8 \xi_{M_2}) \left[ \lambda^{(1)}_{M_1 M_1 M_2 M_2} + \frac{1}{2}   \lambda^{(2)}_{M_1 M_1 M_2 M_2}\right] \,,
\eea
 for the scalar perturbations where $\mN_{(B)} = \sum_{M=1}^{\mN_\Phi}(1-8 \xi_M)^2$ and 
\bea
f_0^T &=& \frac{1}{256} \left(  2 \mN_\psi + \mN_\Phi  \right) \,, \nn \\
f_1^T &=&  \frac{4}{3 \pi^2} \frac{1}{\mN_{(A)}}  \mu_Y^2   \,, \nn \\
f_2^T &=&   \frac{1}{6 \, \mN_{(A)} \, \pi^2}  \left[  2    -\frac{3}{2} \pi ^2 +  8 \log \frac{\gYM^2 \, N}{\mu}  \right] \mu_Y^2 \,,
\eea
for the tensor perturbations with $\mN_{(A)} = 2 \mN_\psi + \mN_\Phi$.

Notice that if we consider only scalars (or slightly more generally if we keep fermions but turn off the Yukawa couplings), then $f_1^S=f_1^T=f_2^S=0$, and there are also no infinities. Ordinarily, $f_1$ is computable in perturbation theory but $f_2$ is ambiguous due to UV and/or IR divergences. In this case $f_1=0$ and $f_2$ is unambiguous. In particular, the 2-loop result is UV finite because it is the square of an 1-loop diagram, and odd loops in odd dimensions are finite.   If we keep only a single non-minimal scalar then the non-zero coefficients are
\begin{equation}
f_0^S=\frac{1}{64}(1-8\xi)^2, \qquad  f_2^S=-\frac{1}{48}(\lambda^{(1)} + \frac{1}{2} \lambda^{(2)}), \qquad 
f_0^T = \frac{1}{256}
\end{equation} 
It turns that this model still provides a good fit to Planck data, though now the model becomes non-perturbative for a large portion of the Planck data.  The non-perturbative evaluation of the 2-point function of the energy-momentum for this theory using lattice method is 
currently in progress (see \cite{Lee:2019zml} for preliminary results).

\section{Irrelevant deformation: $\Phi^6$ coupling}\label{rgflows}
In the Wilsonian approach to renormalisation, operators are classified as irrelevant, (exactly) marginal and relevant
depending on their effect under renormalisation group flow. Irrelevant operators modify the UV of the theory but are irrelevant in the IR, and vice versa for relevant ones. One may wonder whether there is a similar classification holds relative to the generalised conformal structure. 

This question is also relevant in the context of holographic cosmology, where inverse RG flow is connected with time evolution.
In this context if we want to exit from the non-geometric phase we would need to change the UV of the theory. In this section 
we will discuss the impact of a $\Phi^6$ operator to the theory defined by Eq.(\ref{eq.action}).  While this operator is marginal in the usual sense, it is irrelevant from the perspective of the generalised conformal structure.

In particular we consider its leading contribution to the $\langle TT \rangle$ and the 2-point functions of the elementary fields, focusing on the $3$ dimensional case, and we compute its effect on the renormalisation group running of the coupling constants. 
The new action is defined by
\bea
\label{eq.actionPhi6}
S_{new} = S+ \frac{2}{\gYM^2} \int d^3 x  \frac{1}{6!} c_{M_1 \cdots M_6} \tr[ \Phi^{M_1} \cdots \Phi^{M_6} ] \, ,
\eea
where $S$ is the action Eq.(\ref{eq.action}). 
The sum over the flavour indices $M$ is implicitly understood. 
The coupling $c$ are symmetric in the flavour indices and thus selects the following gauge structure
\bea
\textrm{Str}  T^{a_1} T^{a_2} T^{a_3} T^{a_4} T^{a_5} T^{a_6} &=& \frac{1}{6!} \sum_{\pi} \tr \, T^{a_{\pi(1)}} T^{a_{\pi(2)}}  T^{a_{\pi(3)}} T^{a_{\pi(4)}} T^{a_{\pi(5)}} T^{a_{\pi(6)}}\, . 
\eea
The coefficients $c$ are completely symmetric in flavour space and are dimensionful with mass dimension -2. 
We would like to  understand the effect of the new term in perturbation theory where $\lambda$ is a small parameter
and also perturbatively in $c$, where $c$ denotes any of the components of  $c_{M_1 \cdots M_6}$. All factors of $g_{YM}^2$ may be converted into $\lambda$ and then on dimensional grounds any factor of $c$ will appear in the dimensionless combination $c q^2$. It follows that if we want to treat $c$ perturbative we need
\begin{equation} \label{limit}
 c q^2 \ll 1 \quad \Rightarrow \quad q^2 \ll \frac{1}{c}\, , 
 \end{equation}
In other words, this is a low energy limit relative to the new scale introduced by $c$. For perturbation theory to be valid we also need 
\begin{equation} \label{pert}
\lambda \ll1  \quad \Rightarrow \quad g_{YM}^2 N \ll q\, .
\end{equation} 
 Altogether we will be working in the range
\begin{equation} \label{range}
g_{YM}^2 N \ll q \ll \frac{1}{\sqrt{c}}\, ,
\end{equation}
Note that (\ref{limit}) implies 
\begin{equation} \label{c-perb}
c (g_{YM}^2 N)^2 \ll \lambda^2\, ,
\end{equation}
which upon use of (\ref{pert}) implies $c (g_{YM}^2 N)^2 \ll 1$, or 
\begin{equation} \label{limit2}
c \lambda^2 q^2 \ll 1.
\end{equation}
We will use this equation below.
 
 Using power counting (see section \ref{sec:prelim}, Eqs. (\ref{eq.superficialD}) and (\ref{eq.loop})) one may identify the relevant diagrams that are linear in $c$ and require renormalisation.
Up to 2-loops, the new $\Phi^6$ interaction does not provide any new contribution to the 2-point functions of the elementary fields. 
One could write down diagrams, which are linear in $c$, and contribute to the $\langle \Phi \Phi \rangle$ correlator, but they
are proportional to the square of 1-loop massless tadpoles and as such they vanish.
One may check that the first time the $\Phi^6$ coupling contributes at 2-loops is in the 4-point vertex of the scalar fields, thus potentially affecting the running of the quartic couplings $\lambda$. 
The RG behaviour of the gauge and Yukawa couplings remains unchanged at 2-loop as the UV divergent corrections induced by $\Phi^6$ are introduced only at higher orders. 

\begin{figure}[h]
\centering
\includegraphics[scale=1]{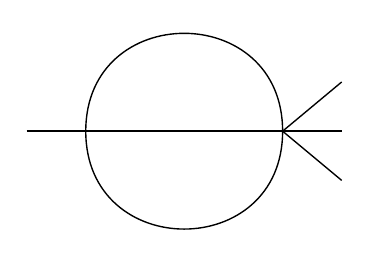} \hspace{1cm}
\includegraphics[scale=1]{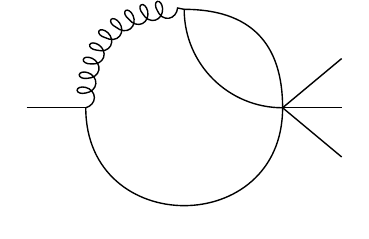} 
\caption{2-loop corrections to the quartic scalar vertex linear in the $c$ coefficient. \label{Fig.phi42Loop}}
\end{figure}

Here we focus on the 2-loop corrections to the 4-point function of the scalar fields which represent the first source of UV divergences proportional to the $c$ parameter \footnote{There are non-vanishing 1- and 2-loop 4-point diagrams constructed from vertices coming from (\ref{eq.action}) only but these diagrams are 
finite (reflecting the fact that the theory is super-renormalisable) and they will not be discussed here.}.
We depict in Fig.~\ref{Fig.phi42Loop} the relevant diagrams. The second one trivially vanishes due to the contraction of the antisymmetric $f^{abc}$, arising from the internal gauge-scalar vertex, with the fully symmetric gauge structure from the $\Phi^6$ coupling.
(There are additional diagrams but they all contain 1-loop massless tadpoles as such they vanish). 

The 4-point function is given by
\bea
\label{eq.g4}
G^{(2)}_4(q_1,q_2,q_3,q_4) &=& \frac{2}{3}  \gYM^2  \frac{G_2}{(4 \pi)^d}  \bigg[ C^{a_1 a_2 a_3 a_4}_{M_1 M_2 M_3 M_4}  \, q_1^{2d-6}   + C^{a_2 a_1 a_3 a_4}_{M_2 M_1 M_3 M_4}  \, q_2^{2d-6}  + C^{a_3 a_1 a_2a_4}_{M_3 M_1 M_2 M_4}  \, q_3^{2d-6}  \nn \\
&+&  C^{a_4 a_1 a_2 a_3}_{M_4 M_1 M_2 M_3}  \, q_4^{2d-6}    \bigg] + \ldots
\eea
where the dots represent $c$-independent terms and $\sum \vec{q}_i = 0$. 
The coupling $C$ is given by the sum of the two contributions proportional to $\lambda^{(1)}$ and $\lambda^{(2)}$, namely,
\bea
\label{eq.Cstructure}
C^{a_1 a_2 a_3 a_4}_{M_1 M_2 M_3 M_4} = C_{(1)}^{a_1 a_2 a_3 a_4} (  \lambda^{(1)} \cdot c )_{M_1 M_2 M_3 M_4} + \frac{1}{N} C_{(2)}^{a_1 a_2 a_3 a_4} (  \lambda^{(2)} \cdot c )_{M_1 M_2 M_3 M_4} \,,
\eea
where the coefficients $( \lambda^{(n)} \cdot c)_{M_1 M_2 M_3 M_4} = \lambda^{(n)}_{M_1 M_5 M_6 M_7} c_{M_5 M_6 M_7 M_2 M_3 M_4}$ while $C_{(n)}$ are the gauge contractions given by
\bea
\label{eq.Ctensors}
C_{(1)}^{a_1 a_2 a_3 a_4} &=& \textrm{Str} [ T^{a_1} T^{a_5} T^{a_6} T^{a_7}] \, \textrm{Str} [T^{a_5} T^{a_6} T^{a_7} T^{a_2} T^{a_3} T^{a_4}]  \nn \\
 &=& C^{1}_{(1)} \, \textrm{Str}_{(1,3)} [ T^{a_1} T^{( a_2} T^{a_3} T^{a_4)}] + C^{2}_{(1)} \, \textrm{Str}[ T^{a_1} T^{a_2}][ T^{a_3} T^{a_4}]    \,, \nn \\
C_{(2)}^{a_1 a_2 a_3 a_4} &=& \textrm{Str}[ T^{a_1} T^{a_2}][ T^{a_3} T^{a_4}] \, \textrm{Str} [T^{a_5} T^{a_6} T^{a_7} T^{a_2} T^{a_3} T^{a_4}]  \nn \\
&=& C^{1}_{(2)} \, \textrm{Str}_{(1,3)} [ T^{a_1} T^{( a_2} T^{a_3} T^{a_4)}]  + C^{2}_{(2)} \, \textrm{Str}[ T^{a_1} T^{a_2}][ T^{a_3} T^{a_4}]   \,.
\eea
Notice that, due to the contraction of $\lambda$ and $c$, the 4-point function $G^{(2)}_4$ is not completely symmetric in the flavour and gauge indices, separately, but it is, obviously, still symmetric under any exchange of any $(a_i, M_i)$ pair. 

The gauge factors are $C^{1}_{(1)} = (N^4-5 N^2+60)/(160 N^2)$, $C^{2}_{(1)} = (2 N^2 -15)/(160 N)$, $C^{1}_{(2)} = (2 N^2 -5)/(40 N)$ and $C^{2}_{(2)} = 1/40$. 
The first gauge structure appearing in the decomposition of the two terms of Eq.(\ref{eq.Ctensors}) is given by the symmetrisation of the double $\tr T^a T^b \, \tr T^c T^d$ over all the permutations of the four gauge indices $a_{1},\ldots, a_{4}$ while the last one is obtained from the symmetrisation of a single trace over the last three indices $a_{2},\ldots, a_{4}$
\bea
\textrm{Str}_{(1,3)} \, T^{a_1} T^{(a_2} T^{a_3} T^{a_4)} &=& \frac{1}{3!} \sum_{\pi} \tr \, T^{a_1} T^{a_{\pi(2)}} T^{a_{\pi(3)}} T^{a_{\pi(4)}} \,.
\eea
As such, the first gauge structure projects the 4-point function on a $\tr [ \Phi^2]^2$ operator, thus introducing an operator mixing with $\tr [\Phi^4]$ under renormalisation.

The UV divergence in Eq.(\ref{eq.g4}) can be removed with the counterterm obtained, as usual, by a rescaling of the fields and the quartic coupling constants. 
The counterterm action involved in the renormalisation of Eq.(\ref{eq.g4}) is
\bea
S_\textrm{c.t} = \frac{2}{\gYM^2} \int d^d x  \frac{\delta \lambda^{(1)}}{4!}  \tr[ \Phi^4] + \frac{1}{N} \frac{\delta \lambda^{(2)}}{4!}  \tr[ \Phi^2]^2 
\eea
where we have used that at  2-loops, $\delta Z_\Phi = \delta Z \gYM = 0$.  This follows from the absence of UV divergences in the theory other than the ones cancelled by the mass counterterm, and in particular the finiteness of the $q^2$ dependent part of scalar propagator. 
In $d = 3-\epsilon$ we find
\bea
\label{eq.lambdact}
\delta \lambda^{(1)}_{M_1 M_2 M_3 M_4} &=&   \frac{\gYM^4}{24 \pi^2} \bigg[ C_{(1)}^{1} (\lambda^{(1)} \cdot c)_{M_1 M_2 M_3 M_4} + \frac{1}{N} C_{(2)}^{1} (\lambda^{(2)} \cdot c)_{M_1 M_2 M_3 M_4}  \bigg] \frac{1}{\epsilon}  \,, \nn \\
\frac{1}{N} \delta \lambda^{(2)}_{M_1 M_2 M_3 M_4} &=&   \frac{\gYM^4}{24 \pi^2} \bigg[ C_{(1)}^{2} (\lambda^{(1)} \cdot c)_{M_1 M_2 M_3 M_4} + \frac{1}{N} C_{(2)}^{2} (\lambda^{(2)} \cdot c)_{M_1 M_2 M_3 M_4}  \bigg] \frac{1}{\epsilon}  \,.
\eea
From the counterterms given in Eq.(\ref{eq.lambdact}) we can extract the $\beta$ functions, $\beta_\lambda = \mu \partial \lambda/\partial \mu$, controlling the running of the quartic scalar couplings
\bea
\beta_{\lambda^{(1)}_{M_1 M_2 M_3 M_4}} &=&  \frac{\gYM^4}{12 \pi^2} \bigg[ C_{(1)}^{1} (\lambda^{(1)} \cdot c)_{M_1 M_2 M_3 M_4} + \frac{1}{N} C_{(2)}^{1} (\lambda^{(2)} \cdot c)_{M_1 M_2 M_3 M_4}  \bigg] \,,   \\
\frac{1}{N} \beta_{\lambda^{(2)}_{M_1 M_2 M_3 M_4}} &=&  \frac{\gYM^4}{12 \pi^2} \bigg[ C_{(1)}^{2} (\lambda^{(1)} \cdot c)_{M_1 M_2 M_3 M_4} + \frac{1}{N} C_{(2)}^{2} (\lambda^{(2)} \cdot c)_{M_1 M_2 M_3 M_4}  \bigg]   \,.
\eea
In order to highlight the behaviour of the scalar coupling with the renormalisation scale $\mu$, we can solve the RG equation in the simple case where both $\lambda$ and $c$ are proportional to the identity matrix in the flavour space. In the large $N$ limit, the running of the quartic couplings is driven by $\lambda^{(1)}$ and two $\beta$ functions simplify to 
\bea
\beta_{\lambda^{(1)}} &\simeq & \frac{\gYM^4}{12 \pi^2}  C_{(1)}^{1} \lambda^{(1)} c      \simeq    \frac{(\gYM^2 \, N)^2}{1920 \pi^2}   \lambda^{(1)} c       \,, \nn \\
\beta_{\lambda^{(2)}} &\simeq & \frac{\gYM^4}{12 \pi^2}  N \, C_{(1)}^{2} \lambda^{(1)} c      \simeq    \frac{(\gYM^2 \, N)^2}{960 \pi^2}   \lambda^{(1)} c    \,.
\eea
The corresponding RG solutions are
\bea
\lambda^{(1)}(\mu) & \simeq &  \lambda^{(1)}(\mu_0) \left( \frac{\mu}{\mu_0} \right)^k \,, \nn \\
\lambda^{(2)}(\mu) & \simeq &  \lambda^{(2)}(\mu_0)  + 2 \lambda^{(1)}(\mu_0) \left[ \left( \frac{\mu}{\mu_0} \right)^k - 1 \right] \,,\qquad k =  \frac{(\gYM^2 \, N)^2}{1920 \pi^2} c.
\eea
Notice that in the regime (\ref{limit}), $k \ll 1$.

\begin{figure}[h]
\centering
\includegraphics[scale=0.5]{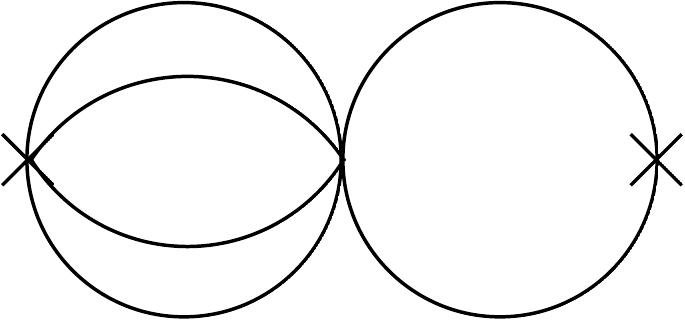} \hspace{1cm}
\includegraphics[scale=0.5]{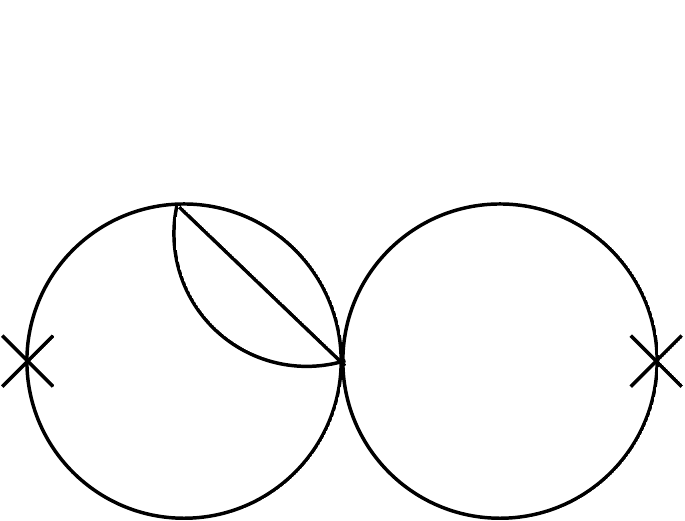} \hspace{1cm}
\includegraphics[scale=0.5]{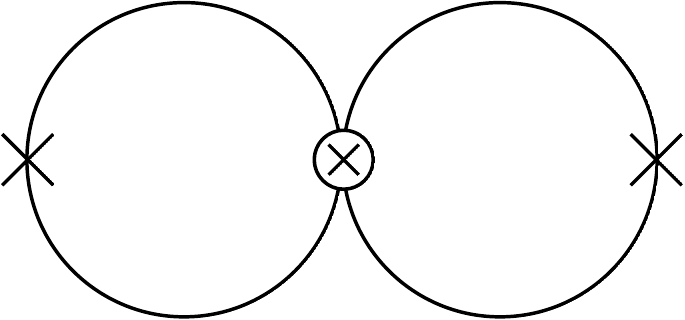} 
\caption{Leading contributions to the $\< TT \>$ from the $\Phi^6$ interaction. \label{Fig.TTphi6}}
\end{figure}

In the following we present the analysis of the leading contribution in the limit (\ref{limit}) ({\it i.e.} linear in $c$) of the $\Phi^6$ operator to the $\langle TT \rangle$ correlation function. 
The contributions to the energy-momentum tensor from the quartic couplings and the $\Phi^6$ term in the action of Eq.(\ref{eq.actionPhi6}) are
\bea
T_{ij} =   - \frac{2}{\gYM^2} \delta_{ij}  \bigg[ \frac{\lambda^{(1)}}{4!}  \tr[ \Phi^4] + \frac{1}{N} \frac{\lambda^{(2)}}{4!}  \tr[ \Phi^2]^2 + \frac{c}{6!} \tr[ \Phi^6 ]  \bigg] \,.
\eea
The leading contributions in $c$ to $\< TT \>$ appear first at 4-loop order in perturbation theory and correspond to the diagrams depicted in Fig.~\ref{Fig.TTphi6}. 
Here, we focus only on these corrections, neglecting all the other contributions of higher order in or independent of $c$. 

The explicit results for the first two diagrams of Fig.~\ref{Fig.TTphi6} in arbitrary $d$ dimensions are
\bea
\label{eq:phi6results}
A^{(4)}_{1+2} &=& 0 , \nn \\
B^{(4)}_{1+2} &=&  \gYM^6 \, C^{a_1 a_1 a_2 a_2}_{M_1 M_1 M_2 M_2} \frac{ \Gamma \left(4-\frac{3 d}{2}\right) \Gamma \left(2-\frac{d}{2}\right) \Gamma \left(\frac{d}{2}-1\right)^6  }{\Gamma (d-2) \Gamma (2 d-4)}   \left( \xi_{M_2} - \frac{d-2}{4 (d-1)} \right)  \nn \\ 
&\times&  \frac{ (8 (2 d-5) \xi_{M_1} -3 d+8)}{3 (d-3) }  \frac{1}{(4\pi)^{2d}} \, q^{(4 d -10)} 
\eea
where the gauge factor is given by the same contraction of the $c$ and $\lambda$ couplings appearing in the first diagram of Fig.\,\ref{Fig.phi42Loop} and it is simply obtained from Eq.(\ref{eq.Cstructure}) by summing over pairs of indices. The two gauge structures must necessarily have a common origin in order to guarantee the cancellation of the UV divergence in the $\< TT \>$ by the counterterm in the third diagram of Fig.\,\ref{Fig.TTphi6}, as we will explicitly show below.
In particular, we have
\bea
C^{a_1 a_1 a_2 a_2}_{M_1 M_1 M_2 M_2} =  c_1 \, (  \lambda^{(1)} \cdot c )_{M_1 M_1 M_2 M_2} + c_2 \, (  \lambda^{(2)} \cdot c )_{M_1 M_1 M_2 M_2} \,,
\eea
with
\bea
c_1 &=&  \frac{N^5}{480} - \frac{N^3}{64} + \frac{73 N}{960} - \frac{5}{32 N} + \frac{3}{32 N^3} \,, \nn \\
c_2 &=& \frac{N^3}{96} - \frac{N}{24} + \frac{1}{16 N} - \frac{1}{32 N^3} \,.
\eea
Notice that the $B^{(4)}_{1+2}$ function vanishes identically if the scalar fields are conformally coupled in $d$ dimensions, namely, $\xi_M = (d-2)/(4(d-1))$. 

By closer inspection of the structure of the topologies in Fig.~\ref{Fig.TTphi6}, one can realise that the first diagram is finite in $d=3$ dimensions, both in the UV and in the IR, while the second one is UV divergent. This singularity appears in the $d=3$ pole in Eq.(\ref{eq:phi6results}). In dimensional regularisation, with $d = 3 - \epsilon$, one obtains
\bea
\label{eq:B12}
B^{(4)}_{1+2} &=&  \gYM^6 \, C^{a_1 a_1 a_2 a_2}_{M_1 M_1 M_2 M_2} \frac{q^2}{12 (64\pi)^2}  \bigg[ (1 - 8 \xi_{M_1}) (1-8 \xi_{M_2}) \left( \frac{1}{ \epsilon}   - 4 \log \frac{q}{\mu} +  \log 4  \right)  + 
\frac{3}{2}  \nn \\
&-&  16 \xi_{M_2} - 20 \xi_{M_1} + 192 \xi_{M_1} \xi_{M_2}  \bigg]  .
\eea
The divergence in Eq.(\ref{eq:B12}) is cancelled by the third diagram in Fig.~\ref{Fig.TTphi6} which is characterised by the insertion of the counterterms $\delta \lambda^{(1)}$ and $\delta \lambda^{(2)}$. From an explicit computation in $d=3$ dimensions, we find $A^{(4)}_3 = 0$ and
\bea
\label{eq:B3}
B^{(4)}_{3} &=& -  \gYM^6 \, C^{a_1 a_1 a_2 a_2}_{M_1 M_1 M_2 M_2} \frac{q^2}{12 (64\pi)^2}  \bigg[ (1 - 8 \xi_{M_1}) (1-8 \xi_{M_2})  \left( \frac{1}{ \epsilon}  - 2 \log \frac{q}{\mu}  + \log 4  \right)  
\nn \\
&-& 1 + 4 \xi_{M_1} + 4 \xi_{M_2}  \bigg] \,,
\eea
where we have exploited the explicit expressions of the couterterms defined in Eq.(\ref{eq.lambdact}) and re-expressed them in terms of $C^{a_1 a_1 a_2 a_2}_{M_1 M_1 M_2 M_2}$.
The complete result, given by the sum of Eq.(\ref{eq:B12}) and Eq.(\ref{eq:B3}), is clearly UV finite and in the large $N$ limit reads as
\bea
A^{(4)} &=& 0,  \label{2loopc} \\
B^{(4)} &=&     \frac{ N^2 \, q^3}{10 (1536\pi)^2} \gth^3 \left[ q^2 (\lambda^{(1)} \cdot c)_{M_1 M_1 M_2 M_2}  \right] 
(1-8 \xi_{M_2})\left[  - 2 (1-8 \xi_{M_1})   \log \frac{q}{\mu}   + \frac{1}{2} (5- 48 \xi_{M_1})  \right]  \,.\nn
\eea
The $A^{(4)}$ coefficient originating from each of the three diagrams in Fig.~\ref{Fig.TTphi6} identically vanishes due to the peculiar structure of the 2-loop corrections. 
These are all given by the product of two one-loop bubbles, each of them contains a single energy-momentum tensor and as such 
each  proportional to the transverse tensor $\pi_{ij}$ (defined in  Eq.(\ref{eq:tensorbasis})).
Therefore, the complete diagram naturally gives vanishing contributions onto the transverse and traceless part. 

The condition (\ref{limit2}) implies 
\begin{equation}
 \lambda^3 c q^2 \ll \lambda
\end{equation}
and the contribution (\ref{2loopc}) is indeed subleading to the 2-loop contribution we considered earlier in section \ref{hol}.
If the CMB scales lie within this regime then the holographic model based on (\ref{eq.actionPhi6}) would fit the data equally well as the model without the $\Phi^6$ term, but the new model would start to deviate at higher energies (later times from the bulk perspective) triggering an exit from this period. For this model to describe the right physics, the (inverse) RG flow should drive us to strong coupling at higher energies describing the transition to Einstein gravity. Analysing this interesting question is beyond the scope of this paper.

We finish this section with a few comments about a Wilsonian view of the generalised conformal structure. The coefficient of terms with {\it generalised dimension} $\Delta$ will appear in perturbation theory in the dimensionless combination $c_\Delta q^{\Delta-4}$. Therefore if $\Delta > 4$ their effect  will be washed out in the IR (relative to the scale set by $c_\Delta$) and they dominate in the UV, as such are they are the analogue of the irrelevant terms of the usual Wilsonian picture. In our case  the $\Phi^6$ operator 
has $\Delta=6$ so it is indeed irrelevant \footnote{Note that in standard Wilsonian approach (with dimensions assigned using the Gaussian UV fixed point)
the $\Phi^6$ coupling in $d=3$ is marginal.}.
In the opposite case, $\Delta < 4$, the operators dominate in the IR and are washed out in the UV. An example would be the operator $\Phi^2$ which is relevant.

\section{Conclusions and Perspectives} \label{conclusions}

We presented  in this paper the perturbative computation of the two-point function of the energy-momentum tensor  to 2-loop order in a 
class of theories that has generalised conformal structure, namely $SU(N)$ gauge theory coupled to massless fermions and scalars with Yukawa and quartic interactions, with all fields in the adjoint of $SU(N)$.
The computation was done for general $d$ using dimensional regularisation.  Generalised conformal structure implies that  
the momentum dependence of the perturbative correlators is a Laurent series in the magnitude of momenta
with coefficients that have poles as $d$ approaches integer values. When $d$ is odd the first poles appear at 2-loop order 
while for $d$ even they are already present at 1-loop. The poles may be associated with either IR and/or UV infinities.
We discussed renormalisation when $d=3, 4, 5$.  

When $d=3$ the theory is super-renormalisable. The 2-point function of the energy-momentum tensor has UV divergences 
at 2-loops, which may be cancelled using a counterterm proportional to the scalar curvature, and an IR singularity if the theory 
contains non-minimally coupled scalars. Such super-renormalisable theories are expected to be IR finite, with the Yang-Mill coupling constant acting as an IR regulator \cite{Jackiw:1980kv, Appelquist:1981vg}. 

When $d=4$ the coupling constant is dimensionless, so this theory does not have generalised conformal structure. Instead the theory is classically scale invariant. The 2-point function of the energy-momentum tensor has UV infinities both at 1- and 2-loop order, which may be cancelled using a Weyl squared and curvature squared countertrems. The renormalised 2-point function has the form dictated by scale invariance, modulo the logarithms originating from the UV subtractions. 

When $d=5$ the theory is non-renormalisable and renormalisation of elementary fields induces higher dimension terms in the action, spoiling the generalised conformal structure. The generalised conformal structure is then only present at low enough energies so that the higher dimension operators are suppressed. Using the same Lagrangian as in the $d=3$ and $d=4$ cases we find the 2-point function of the energy-momentum tensor has UV divergences at 2-loops, which may be removed using local counterterms of the schematic form of D'Alembertian operator acting on squares of curvatures. 

It would be interesting to extend the computation described here to the maximally supersymmetric theories in the corresponding dimensions.
In $d=4$ the result is well-known as the  corresponding theory is ${\cal N} =4$ SYM, but we are not aware of such
result in different dimensions. To do this computation we would need to relax the condition that the quartic self-coupling $\lambda^{(1)}$ is completely symmetric in flavour space and consider appropriate Yukawa couplings. It would also be interesting 
to analyse the lower dimensional cases $d<3$, and in particular understand the fate of the IR singularities.

One of the main motivations of this work was the application of the $d=3$ results to holographic cosmology. Indeed, the detailed 
form of the results relevant for the scalar power spectrum was already used in \cite{Afshordi:2016dvb} when analysing the fit of these models to CMB data. Here we present the derivation of this result  as well as the corresponding result for the tensor power spectrum.

Another important issue in holographic cosmology is to understand how to exit from the non-geometric phase and develop a theory of holographic reheating. A general expectation is that this should involve turning-on irrelevant operators. 
Here we made a first step in this direction by analysing the effects of turning-on a $\Phi^6$ term, to leading order at low energies. This operator, while marginal in the usual sense, is the leading irrelevant operator from the perspective 
of the generalised conformal structure, and indeed  we confirmed that its effects are washed out in the IR.
It would be interesting to further develop this model.

In terms of the complexity of correlators, theories with generalised conformal structure sit between CFTs 
and generic QFTs. Here we analysed 2-point functions, and it would be interesting to extend such analysis to 
higher point functions.

\vspace{2 cm}
\centerline{\bf Acknowledgements} 
The work of C.C. is partially suppoted by INFN under grant Iniziativa Specifica QFT-HEP.
 KS is supported in part by the Science and Technology Facilities Council (Consolidated Grant “Exploring the Limits of the Standard Model and Beyond”).
\newpage
\appendix
\section{2-loop results for counterterms, coupling coefficients and $A$, $B$ form factors in various dimensions}
\label{appendixA}
In this appendix we present the result for some of the coefficients defined in main text. We start from the 2-loop self energies in appendix \ref{selfies}, moving to the counterterms of the gauge sector in \ref{gauges} and the generalised conformal structure constants in \ref{confst} and \ref{sfermion}.  Appendices  \ref{sect} and \ref{yuk} contain  the part of the expressions of the $A$ and $B$ form factors at 2-loops not given in the main text.
\subsection{2-loop self energies}
\label{selfies}
The 2-loop corrections are expressed in terms of the coefficients $\alpha_{\psi i},\alpha_{\Phi k}$ defined in Eqs.~\eqref{pi1}, \eqref{pi2}
\bea
\alpha_{A0}&=& \frac{1}{2 (d-6) (d-4)^2 (d-1)} \nn\\
\alpha_{A 1}&=& -(d-6) (d-4) (d-2) ((d-9) d+24) \mN_\psi \nn\\
&& +(d-6) (d-4) (2 d-9) \mN_\Phi -(d-6) (d-4) ((d-4) d-2)
\eea
\bea
\alpha_{A 2}&=& 3 d^5-59 d^4+376 d^3-1084 d^2  +1648 d-1248  \nn\\
&&   +\left(6 d^5-106 d^4+796 d^3-3096 d^2+5968 d-4416\right) \mN_\psi  \nn\\
&&  +\left(3 d^4-53 d^3+366 d^2-1112 d+1200\right) \mN_\Phi 
\eea

\bea
\alpha_{\psi 0}&=&-  \frac{(d-2) }{4 (d-6) (d-4)^2} \nn\\
\alpha_{\psi 1}&=& (d-6) (d-4) (d-2)^2\nn\\
\alpha_{\psi 2}&=& -4 (d-2) (d-4)^2 \mN_\psi  -2 (d-4)^2 \mN_\Phi+2 d (d ((29-2 d) d-150)+352)-640 \nn\\
\alpha_{\psi 3}&=&(d -4) (2 + (d -4) d) \nn\\
\alpha_{\psi 4}&=&- 4 (-32 + d (29 + (d -9) d)) 
\eea

\bea
\alpha_{\Phi 0}&=&\frac{ 1}{2 (d-6) (d-4)^2}\nn\\
\alpha_{\Phi 1}&=&  -(d-6) (d-4) (3 d-14)   \nn\\
\alpha_{\Phi 2}&=&  8 (d-4) (d-2)^2  \mN_\psi  +4 (d-4) (d-2) \mN_\Phi +d (d (d (3 d-41)+94)+344)-992       \nn\\
\alpha_{\Phi 3}&=&  (d -4 ) (20 + (d -8) d)\nn\\
\alpha_{\Phi 4}&=&     2 (192 + d (-140 + (34 - 3 d) d)).  
\eea

The contraction of the quartic and Yukawa couplings, which generalise the expressions given in 
\eqref{thisone} are defined as follows
\bea
\label{refquartic}
\mu^{(2)}_{L_1 L_2} &=&\mu_{M_1 L_1 L_3} \, \mu_{M_2 L_3 L_2} \, \mu_{M_1 L_4 L_5} \, \mu_{M_2 L_5 L_4} , \nn \\
\mu^{(3)}_{L_1 L_2} &=&\mu_{M_1 L_1 L_3} \, \mu_{M_2 L_3 L_4} \, \mu_{M_2 L_4 L_5} \, \mu_{M_1 L_5 L_2} , \nn \\
\mu^{(4)}_{L_1 L_2} &=&\mu_{M_1 L_1 L_3} \, \mu_{M_1 L_3 L_4} \, \mu_{M_2 L_4 L_5} \, \mu_{M_2 L_5 L_2} , \nn \\
\mu^{(5)}_{M_1 M_2} &=& \mu_{M_1 L_1 L_2} \, \mu_{M L_2 L_3} \, \mu_{M_2 L_3 L_4} \,  \mu_{M L_4 L_1} , \nn \\
\mu^{(6)}_{M_1 M_2} &=& \mu_{M_1 L_1 L_2} \, \mu_{M L_2 L_3} \, \mu_{M L_3 L_4} \,  \mu_{M_2 L_4 L_1} , \nn \\
\mu^{(7)}_{M_1 M_2} &=& \mu_{M_1 L_1 L_2} \, \mu_{M L_2 L_1} \, \mu_{M L_3 L_4} \,  \mu_{M_2 L_4 L_3} .
\eea

\subsection{Counterterms at 2-loop order in $d=4$ }
\label{gauges}
For the counterterms of the 2-point functions of the gauge, scalar and fermion fields, extending at 2-loops the results given in \eqref{CT1L},  we obtain
\bea
\delta_A^{(12)} &=& - \frac{1}{2048 \pi^4} ( -46 + 36 \mN_\psi + 15 \mN_\Phi +  4 \mu_Y^2   ) \,, \nn \\
\delta_A^{(2)} &=& \frac{5 }{1536 \pi^4} (-10 + 4 \mN_\psi + \mN_\Phi) \,, \nn \\
\delta_{\psi \, L_1 L_2}^{(12)} &=&  \frac{1}{8192 \pi^4} ((-224 + 32 \mN_\psi + 8 \mN_\Phi) \delta_{L_1 L_2} + 40 \,  \mu^{(0)}_{L_1 L_2} +  6 \mu^{(2)}_{L_1 L_2} + \mu^{(3)}_{L_1 L_2}) \,, \nn \\
\delta_{\psi \, L_1 L_2}^{(2)} &=& -  \frac{1}{2048 \pi^4} (-48 \, \delta_{L_1 L_2} + 32 \,  \mu^{(0)}_{L_1 L_2} + 2 \mu^{(1)}_{L_1 L_2} + 2 \mu^{(2)}_{L_1 L_2} + \mu^{(3)}_{L_1 L_2} ) \,, \nn \\
\delta_{\Phi \, M_1 M_2}^{(12)} &=& \frac{1}{6144 \pi^4}   \left( 
   (190 - 40 \mN_\psi - 22 \mN_\Phi) \delta_{M_1 M_2} + 60   \mu^{(0)}_{M_1 M_2} +  3  \mu^{(5)}_{M_1 M_2} + 9  \mu^{(6)}_{M_1 M_2}      - \frac{1}{12} \lambda^{(0)}_{M_1 M_2}  \right) \,, \nn \\
\delta_{\Phi \, M_1 M_2}^{(2)} &=& - \frac{1}{512 \pi^4} ( (20 - 8 \mN_\psi - 2 \mN_\Phi) \delta_{M_1 M_2} + 4  \mu^{(0)}_{M_1 M_2} + \mu^{(5)}_{M_1 M_2} + \mu^{(6)}_{M_1 M_2}) \,,
\eea

\subsection{GCSC of scalars and fermions at 2-loop order in $d=4$}
\label{confst}
They are given by
\bea
c_{L_1 L_2}^{\psi, (2,0)}  &=& \frac{ 1}{32768 \pi ^4} ( (224 \mN_\psi +72 \mN_\Phi +768 \zeta_3 - 2208) \delta_{L_1 L_2} + 8 (127-24 \zeta_3)  \mu^{(0)}_{L_1 L_2}  \nn \\
&+& 32 \mu^{(1)}_{L_1 L_2} + 114 \mu^{(2)}_{L_1 L_2} +31 \mu^{(3)}_{L_1 L_2})  + c_{L_1 L_3}^{\psi, (1,0)} c_{L_3 L_2}^{\psi, (1,0)} \,, \nn \\
c_{L_1 L_2}^{\psi, (2,1)}  &=& -\frac{1}{4096 \pi ^4} ((32 \mN_\psi +8 \mN_\Phi -320) \delta_{L_1 L_2} + 160   \mu^{(0)}_{L_1 L_2} \nn \\
&+& 8  \mu^{(1)}_{L_1 L_2} +14 \mu^{(2)}_{L_1 L_2}+5 \mu^{(3)}_{L_1 L_2})  + 2 \, c_{L_1 L_3}^{\psi, (1,0)} c_{L_3 L_2}^{\psi, (1,1)} \,, \nn \\
c_{L_1 L_2}^{\psi, (2,2)}  &=&  c_{L_1L_3}^{\psi, (1,1)} c_{L_3 L_2}^{\psi, (1,1)}    -  \delta_{\psi \, L_1 L_2}^{(2)} \,, \nn \\
c_{M_1 M_2}^{\Phi, (2,0)}  &=& \frac{1}{24576 \pi ^4} \left(  (- 760 \mN_\psi -298 \mN_\Phi - 288 \zeta_3+3070)\delta_{M_1 M_2} +87  \mu^{(5)}_{M_1 M_2}+171  \mu^{(6)}_{M_1 M_2}   \right. \nn \\
&+& \left.  12 (91-24 \zeta_3)  \mu^{(0)}_{M_1 M_2}    - \frac{13}{12}   \lambda^{(0)}_{M_1 M_2}  \right) \nn \\
&+&  c_{M_1 M_3}^{\Phi, (1,0)} c_{M_3 M_2}^{\Phi, (1,0)} \,, \nn \\
c_{M_1 M_2}^{\Phi, (2,1)}  &=& \frac{1}{3072 \pi ^4} \left((104 \mN_\psi +38 \mN_\Phi -350) \delta_{M_1 M_2} -108  \mu^{(0)}_{M_1 M_2}  - 15  \mu^{(5)}_{M_1 M_2} \right. \nn \\
&-& \left. 21 \mu^{(6)}_{M_1 M_2}  + \frac{1}{12} \lambda^{(0)}_{M_1 M_2}  \right) + 2 \, c_{M_1 M_3}^{\Phi, (1,0)} c_{M_3 M_2}^{\Phi, (1,1)} \,, \nn \\
c_{M_1 M_2}^{\Phi, (2,2)}  &=&  c_{M_1M_3}^{\Phi, (1,1)} c_{M_3 M_2}^{\Phi, (1,1)}    -  \delta_{\Phi \, M_1 M_2}^{(2)} \,.
\eea
\subsection{GCSC of scalars and fermions in $d=5$}
\label{sfermion}

They are given by
\bea
c_{1 \, L_1 L_2}^\psi &=& \frac{1}{512 \pi}   \left[ 6 \,  \delta_{L_1 L_2}   -   \mu^{(0)}_{L_1 L_2} \right]  \,,  \nn  \\
\tilde c_{2 \, \UV \, L_1 L_2}^\psi &=&   \frac{1}{215040 \pi^4} \left[  \left( -390 + 36 \mN_\psi + 6 \mN_\Phi \right) \delta_{L_1 L_2}   + 
  52 \mu^{(0)}_{L_1 L_2}  - 3 \mu^{(2)}_{L_1 L_2} - 2 \mu^{(3)}_{L_1 L_2}   \right]   \,, \nn \\
c_{2  \, L_1 L_2}^\psi &=& - \tilde c_{2 \, \UV \, L_1 L_2}^\psi  \, \log \left( \gYM^2 N \mu_{\UV} \right)    +   \frac{1}{3675 \times 2^{18} \pi^4}  \left[ 
- 32 (   (-103190 + 4692 \mN_\psi  \right. \nn \\
&+& \left.   922 \mN_\Phi) \delta_{L_1 L_2}  +  16204 \mu^{(0)}_{L_1 L_2}  - 881 \mu^{(2)}_{L_1 L_2} - 564 \mu^{(3)}_{L_1 L_2}   )   \right. \nn \\
&+& \left. 3675 \pi^2 (  -72 \delta_{L_1 L_2} + 16 \mu^{(0)}_{L_1 L_2}   +   \mu^{(4)}_{L_1 L_2}   + \mu^{(1)}_{L_1 L_2} )   \right]   \,,
\eea
\bea
c_{1 \, M_1 M_2}^\Phi &=& -  \frac{1}{256 \pi}   \left[ 4 \,   \delta_{M_1 M_2}  +  \mu^{(0)}_{M_1 M_2}  \right]  \,, \nn \\
c_{2 \, \UV \, M_1 M_2}^\Phi &=& \frac{1}{215040 \pi^4} \left[ (-344 + 144 \mN_\psi + 24 \mN_\Phi) \delta_{M_1 M_2}  +66 \mu^{(0)}_{M_1 M_2} + \mu^{(5)}_{M_1 M_2}  - 6 \mu^{(6)}_{M_1 M_2} 
- \frac{1}{9} \lambda^{(0)}_{M_1 M_2} \right]   \,, \nn \\
c_{2 \,  M_1 M_2}^\Phi &=& - \tilde c_{2 \, \UV \, L_1 L_2}^\Phi  \, \log \left( \gYM^2 N \mu_{\UV} \right)    - \frac{1}{3675 \times 2^{13}\pi^4} 
\left[  ( -68308 + 28848 \mN_\psi + 5368 \mN_\Phi) \delta_{M_1 M_2}   \right. \nn \\
 &+&  \left.   22042 \mu^{(0)}_{M_1 M_2}  + 247 \mu^{(5)}_{M_1 M_2}  - 1762 \mu^{(6)}_{M_1 M_2} 
-\frac{247}{9} \lambda^{(0)}_{M_1 M_2}  \right] \nn \\
&+& \frac{1}{2^{17}\pi^2} \left[  28 \delta_{M_1 M_2} + 26 \mu^{(0)}_{M_1 M_2}  +  \mu^{(5)}_{M_1 M_2}  + 2 \mu^{(7)}_{M_1 M_2} \right]
\eea
where we have used the same notation introduced in Eq.(\ref{eq.3Dconform}) for the $d=3$ case.

\subsection{$A$ and $B$ form factors at 2-loop order in $d=4$}
\label{sect}
For the scalar sector we obtain
\bea
A^{(2)}_\Phi &=& - \gth \, \mN_\Phi \, d(G) \frac{q^4}{138240 \pi^4} \left(  -251 + 120 \log \frac{q}{\mu_\UV} + 108 \, \zeta_3 \right)  \,, \nn \\
B^{(2)}_\Phi &=&  \gth \sum_{M = 1}^{\mN_\Phi}   d(G) \frac{q^4}{10368 \pi^4} \bigg( 61 + 9 \xi_M (-101 + 360 \xi_M)  - 27 (1 - 6 \xi_M)^2 \zeta_3 \nn \\
&& - 27 (1 - 6 \xi_M)  (3- 20 \xi_M) \log \frac{q}{\mu_\UV} + 27 (1 - 6 \xi_M)^2 \log^2 \frac{q}{\mu_\UV}\bigg) \,, \nn \\
 &-& \gth \sum_{M_1, M_2 = 1}^{\mN_\Phi} \frac{q^4}{497664 \pi^4}  d(G) \left[ \frac{2 N^2 -3 }{N^2} \lambda^{(1)}_{M_1 M_1 M_2 M_2}  + \frac{1+ N^2}{N^2} \lambda^{(2)}_{M_1 M_1 M_2 M_2} \right]    \bigg(  
    (- 5 + 36 \xi_{M_1}) (- 5 + 36 \xi_{M_2}) \nn \\
 &&  + 6 ( (1 - 6 \xi_{M_1})  + (1 - 6 \xi_{M_2})    - 12 (1 - 6 \xi_{M_1}) (1 - 6 \xi_{M_2})    )  \log \frac{q}{\mu_\UV}  \nn \\
 && + 36 (1 - 6 \xi_{M_1})  ( 1 - 6 \xi_{M_2}) \log^2 \frac{q}{\mu_\UV} 
 \bigg) \,,
 \eea
while the Yukawa sector gives
 \bea
A^{(2)}_Y &=& \gth \, d(G) \frac{q^4}{49152 \pi^4} \left(  21 -8 \log \frac{q}{\mu_\UV}  \right)  \mu^2_Y \,, \nn \\
B^{(2)}_Y &=& \gth \sum_{M = 1}^{\mN_\Phi}    d(G) \frac{q^4}{331776 \pi^4} \bigg( 289 + 216 \xi_M (-19 + 66 \xi_M)  - 48 (1 - 6 \xi_M) (8 - 54 \xi_M) \log \frac{q}{\mu_\UV}  \nn \\
&+&  144 (1 - 6 \xi_M)^2 \,  \log^2 \frac{q}{\mu_\UV} \bigg)  \mu_{M L_1 L_2} \mu_{M L_2 L_1} \,. 
\eea

\subsection{$A$ and $B$ form factors at 2-loop order in $d=5$}
\label{yuk}
They are given by
\bea
A^{(2)}_\Phi &=& \gth \, \mN_\Phi \,    d(G)   \frac{q^5}{2167603200 \pi^4} \left(-150728 + 11025 \pi^2 + 64960 \log\frac{q}{\mu_\UV} \right)   \,, \nn \\
B^{(2)}_\Phi &=& \gth  \sum_{M=1}^{\mN_\Phi}    d(G)   \frac{q^5}{69363302400 \pi^4}   \bigg( -64 (278789 + 1152 \xi_M (-2606 + 7007 \xi_M))  \nn \\
&+& 33075 \pi^2 (-3 + 16 \xi_M) (-19 + 112 \xi_M) 
+ 161280 (49 + 64 \xi_M(-8 + 21 \xi_M))  \log\frac{q}{\mu_\UV}  \bigg)    \nn \\
& -& \gth  \sum_{M_1,M_2=1}^{\mN_\Phi}   \frac{q^5}{25165824 \pi^4}    d(G) \left[ \frac{2 N^2 -3}{ N^2} \lambda^{(1)}_{M_1 M_1 M_2 M_2} + \frac{1+N^2}{N^2} \lambda^{(2)}_{M_1 M_1 M_2 M_2}  \right]  (16 \xi_{M_1} - 3)(16 \xi_{M_2} - 3)        \,, \nn \\
A^{(2)}_Y &=& - \gth \, \mu_Y^2 \,  d(G)  \frac{q^5}{52022476800 \pi^4} \left(-392816 + 33075 \pi^2 + 174720 \log\frac{q}{\mu_\UV} \right)  \,, \nn \\
B^{(2)}_Y &=&   \gth \, \sum_{M=1}^{\mN_\Phi}     d(G)  \frac{q^5}{4335206400 \pi^4} \bigg( 72721 + 288 \xi_M (-2855 + 8036 \xi_M)  \nn \\
 &-& 840 (43 + 96 \xi_M (-5 + 14 \xi_M))  \log\frac{q}{\mu_\UV} \bigg) \mu_{M L_1 L_2} \mu_{M L_2 L_1} \,.
\eea

\section{Tensor reductions: technical details}
\label{tensrid}

The 1-loop tensor reduction of the 2-point functions is completely solved by the following decompositions
\bea
I_{i} &=&  \int \frac{d^d k }{(2 \pi)^d} \frac{k_{i}}{k^2 \, (k + p)^2  }  = p_i \, B_1 \,, \nn \\
I_{ij} &=&  \int \frac{d^d k }{(2 \pi)^d} \frac{k_{i} k_{j}}{k^2 \, (k + p)^2  }  = \delta_{ij} \, B_{00} + p_i p_j \, B_{11} \,, \nn  \\
I_{ijk} &=&  \int \frac{d^d k }{(2 \pi)^d} \frac{k_{i} k_{j} k_{k}}{k^2 \, (k + p)^2  }  = \delta_{\left\{ \right. ij} \, p_{k \left. \right\}} \, B_{001} + p_i p_j p_k \, B_{111} \,, \nn \\
I_{ijkl} &=&  \int \frac{d^d k }{(2 \pi)^d} \frac{k_{i} k_{j} k_{k} k_{l}}{k^2 \, (k + p)^2  }  = \delta_{\left\{ \right. ij} \, \delta_{k l \left. \right\}} \, B_{0000} + \delta_{\left\{ \right. ij} \, p_{k} p_{l \left. \right\}} \, B_{0011}  + p_i p_j p_k p_l \, B_{1111} \,, 
\eea
where the curly brackets denote fully symmetrisation of the indices\footnote{Notice that the symmetrisation is not weighted, therefore, as an example, $\delta_{\left\{ \right. ij} \, p_{k \left. \right\}} = \delta_{ij} p_k + \delta_{ik} p_j +\delta_{kj} p_i$.}. The scalar coefficients are then reduced by algebraic manipulations to the main integral $B_0$ and are explicitly given by
\bea
B_1 &=& -\frac{1}{2} B_0 \,, \nn \\
B_{00} &=& -\frac{ p^2}{4 (d-1)} B_0 \,, \qquad  B_{11} = \frac{ d}{4 (d-1)} B_0 \,, \nn \\
B_{001} &=& \frac{ p^2}{8 (d-1)}B_0 \,, \qquad  B_{111} = -\frac{ (d+2)}{8 (d-1)} B_0 \,, \nn \\
B_{0000} &=& \frac{ p^4}{16 \left(d^2-1\right)} B_0  \,, \qquad   B_{0011} =  - \frac{ (d+ 2) p^2}{16 \left(d^2-1\right)} B_0 \,, \qquad   B_{1111} = \frac{ \left(d^2+6 d+8\right)}{16 \left(d^2-1\right)} B_0 \,. 
\eea

As stated above, the tensor reduction of the 2-loop diagrams is much more involved. For example, among the technical complications of the reduction, the presence of two integration momenta 
provides different tensor expansions for a give rank. Indeed, even in the simplest rank-1 case we have
\bea
J_{i}^{(n)} &=&  \int \frac{d^d k_1 }{(2 \pi)^d}  \frac{d^d k_2 }{(2 \pi)^d}     \frac{k_{n \, i}}{ \textrm{den}(k_1,k_2,p; \left\{\nu_i \right\})  }  = p_i \, J_1^{(n)} \,,
\eea
with $n =1,2$ and $\textrm{den}(k_1,k_2,p; {\nu_i}) = k_1^{2 \nu_1} k_2^{2 \nu_2} (k_1 + p)^{2\nu_3} (k_2 + p)^{2\nu_4} (k_1 - k_2)^{2\nu_5}$, in which the two coefficients $J_1^{(1)}$ and $J_1^{(2)}$ may differ depending on the specific choice of the $\nu_i$ exponents. Following the same reasoning, it is not difficult to identify, for example, three different tensor expansions for the rank-2 integrals and four for the rank-3. 
The latter manifests another source of complication. Indeed, while in the tensor decomposition of $J_{ijk}^{(111)}$ or $J_{ijk}^{(222)}$ we can fully exploit the symmetrisation of the indices (as in the 1-loop case), namely
\bea
J_{ijk}^{(111)} =  \int \frac{d^d k_1 }{(2 \pi)^d}  \frac{d^d k_2 }{(2 \pi)^d}     \frac{k_{1 \, i} \, k_{1 \, j} \, k_{1 \, k}}{ \textrm{den}(k_1,k_2,p; \left\{\nu_i \right\})  }  = \delta_{\left\{ \right. ij} \, p_{k \left. \right\}} \, J_{001}^{(111)} + p_i p_j p_k \, J_{111}^{(111)} \,,
\eea
and similarly for the $(222)$, the tensor integrals $J_{ijk}^{(112)}$ and $J_{ijk}^{(122)}$ can only be expanded onto a partially symmetrised tensor basis as
\bea
J_{ijk}^{(112)} =  \int \frac{d^d k_1 }{(2 \pi)^d}  \frac{d^d k_2 }{(2 \pi)^d}     \frac{k_{1 \, i} \, k_{1 \, j} \, k_{2 \, k}}{ \textrm{den}(k_1,k_2,p; \left\{\nu_i \right\})  }  = 
\delta_{  ij} \, p_{k } \, J_{001}^{(112)}  + ( \delta_{  ik} \, p_{j } + \delta_{jk} \, p_{i} ) J_{010}^{(112)}  + p_i p_j p_k \, J_{111}^{(112)} \,, \nn \\
\eea
where an extra scalar coefficient appears with respect to the $J_{ijk}^{(111)}$ and $J_{ijk}^{(222)}$ cases. \\

Here we present the tensor decomposition of the 2-loop integrals needed in the present work.
For the rank-1 and rank-2 tensor integrals, the tensorial structure is simply given by
\bea
\int \frac{d^d k_1 }{(2 \pi)^d}  \frac{d^d k_2 }{(2 \pi)^d}     \frac{(k_{\alpha})_i}{ \textrm{den}(k_1,k_2,p; \left\{\nu \right\})  }  &=& p_i \, {\mathcal C}_1^{(\alpha)} \,, \nn \\
\int \frac{d^d k_1 }{(2 \pi)^d}  \frac{d^d k_2 }{(2 \pi)^d}     \frac{(k_{\alpha})_i (k_{\beta})_j}{ \textrm{den}(k_1,k_2,p; \left\{\nu \right\})  }  &=& \delta_{ij} \, {\mathcal C}_{00}^{(\alpha\beta)} + p_i p_j \, {\mathcal C}_{11}^{(\alpha\beta)} \,,
\eea
where the coefficients $\mathcal C$ depend on the topology of the integrals, namely on the form of the denominators.  

For higher ranks, the symmetries of the integral must be explicitly exploited. For the rank-3 we get
\bea
\int \frac{d^d k_1 }{(2 \pi)^d}  \frac{d^d k_2 }{(2 \pi)^d}     \frac{(k_{\alpha})_i (k_{\beta})_j (k_{\gamma})_l}{ \textrm{den}(k_1,k_2,p; \left\{\nu \right\})  }  &=& \sum_n  [S]_{ijl; n}^{(\alpha\beta\gamma)}(1\delta,1p) \,  {\mathcal C}_{001;n}^{(\alpha\beta\gamma)} + p_i p_j p_l \, {\mathcal C}_{111}^{(\alpha\beta\gamma)} \,,
\eea
where $n$ runs over all the tensors of the basis $[S]_{ijl}^{(\alpha\beta\gamma)}(1\delta,1p)$ which is built with one Kronecker $\delta$ and one momentum $p$. 
The symmetries of the tensors in the Lorentz indices $i,j,l$ are defined by the combination of $(\alpha\beta\gamma)$. For instance, $[S]_{ijl}^{(111)}(1\delta,1p)$ and $[S]_{ijl}^{(222)}(1\delta,1p)$ contain tensor structures that are fully symmetric under the indices $i,j,l$,
\bea
[S]_{ijl}^{(111)}(1\delta,1p) = [S]_{ijl}^{(222)}(1\delta,1p) =  \left\{ \delta_{ij} p_l + \delta_{il} p_j +\delta_{lj} p_i \right\} \,,
\eea
while $[S]_{ijl}^{(112)}(1\delta,1p)$ and $[S]_{ijl}^{(221)}(1\delta,1p)$ denote tensors that are symmetric only under exchange of the first two indices $i,j$,
\bea
[S]_{ijl}^{(112)}(1\delta,1p) = [S]_{ijl}^{(221)}(1\delta,1p) =  \left\{ \delta_{ij} p_l \,, \delta_{il} p_j +\delta_{lj} p_i \right\} \,.
\eea
The integrals with higher ranks, from 4 to 6, are given by
\bea
\int \frac{d^d k_1 }{(2 \pi)^d}  \frac{d^d k_2 }{(2 \pi)^d}     \frac{(k_{\alpha})_i (k_{\beta})_j (k_{\gamma})_l (k_{\eta})_m}{ \textrm{den}(k_1,k_2,p; \left\{\nu \right\})  }  &=&
 \sum_n  [S]_{ijlm; n}^{(\alpha\beta\gamma\eta)}(2\delta,0p) \,  {\mathcal C}_{0000;n}^{(\alpha\beta\gamma\eta)}  \nn \\ 
&& \hspace{-4cm} + \sum_n  [S]_{ijlm; n}^{(\alpha\beta\gamma\eta)}(1\delta,2p) \,  {\mathcal C}_{0011;n}^{(\alpha\beta\gamma\eta)}  + \, p_i p_j p_l p_m \, {\mathcal C}_{1111}^{(\alpha\beta\gamma\eta)} \,, \nn \\
\int \frac{d^d k_1 }{(2 \pi)^d}  \frac{d^d k_2 }{(2 \pi)^d}     \frac{(k_{\alpha})_i (k_{\beta})_j (k_{\gamma})_l (k_{\eta})_m (k_{\phi})_r}{ \textrm{den}(k_1,k_2,p; \left\{\nu \right\})  }  &=& 
\sum_n  [S]_{ijlmr; n}^{(\alpha\beta\gamma\eta\phi)}(2\delta,1p) \,  {\mathcal C}_{00001;n}^{(\alpha\beta\gamma\eta\phi)} \nn \\
&&  \hspace{-4cm} + \sum_n  [S]_{ijlmr; n}^{(\alpha\beta\gamma\eta\phi)}(1\delta,3p) \,  {\mathcal C}_{00111;n}^{(\alpha\beta\gamma\eta\phi)}  
+ p_i p_j p_l p_m p_r \, {\mathcal C}_{11111}^{(\alpha\beta\gamma\eta\phi)} \,, \nn \\
\int \frac{d^d k_1 }{(2 \pi)^d}  \frac{d^d k_2 }{(2 \pi)^d}     \frac{(k_{\alpha})_i (k_{\beta})_j (k_{\gamma})_l (k_{\eta})_m (k_{\phi})_r (k_{\xi})_s}{ \textrm{den}(k_1,k_2,p; \left\{\nu \right\})  }  &=& 
\sum_n  [S]_{ijlmrs; n}^{(\alpha\beta\gamma\eta\phi\xi)}(3\delta,0p) \,  {\mathcal C}_{000000;n}^{(\alpha\beta\gamma\eta\phi\xi)} \nn \\
&& \hspace{-4cm} + \sum_n  [S]_{ijlmrs; n}^{(\alpha\beta\gamma\eta\phi\xi)}(2\delta,2p) \,  {\mathcal C}_{000011;n}^{(\alpha\beta\gamma\eta\phi\xi)} 
+ \sum_n  [S]_{ijlmrs; n}^{(\alpha\beta\gamma\eta\phi\xi)}(1\delta,4p) \,  {\mathcal C}_{001111;n}^{(\alpha\beta\gamma\eta\phi\xi)} \nn \\
&& \hspace{-4cm} + \, p_i p_j p_l p_m p_r p_s \, {\mathcal C}_{111111}^{(\alpha\beta\gamma\eta\phi\xi)} \,,
\eea
where, as usual, $[S]_{ijlm}^{(\alpha\beta\gamma\eta)}(2\delta,0p)$ denotes a tensor basis built with 2 Kronecker's deltas and no momenta, $[S]_{ijlm}^{(\alpha\beta\gamma\eta)}(1\delta,2p)$ contains only one Kronecker's delta and 2 momenta, and similarly for all the others basis.
The tensor structures are determined according to the symmetries of the original integrals, as discussed above, and are given by
\bea
\left[S \right]_{ijlm}^{(\alpha\beta\gamma\eta)}(2\delta,0p) &=& \left\{ \delta_{ij} \delta_{lm} + \delta_{il} \delta_{jm} + \delta_{im} \delta_{jl} \right\} \,, \qquad \textrm{for} \quad (\alpha\beta\gamma\eta) = (1111), (1112), (2221), (2222)  \nn  \\
\left[S \right]_{ijlm}^{(\alpha\beta\gamma\eta)}(2\delta,0p) &=& \left\{ \delta_{ij} \delta_{lm} \,, \delta_{il} \delta_{jm} + \delta_{im} \delta_{jl} \right\} \,, \qquad  \textrm{for} \quad (\alpha\beta\gamma\eta) = (1122) \nn \\
\left[S \right]_{ijlm}^{(\alpha\beta\gamma\eta)}(1\delta,2p) &=& \left\{ \delta_{ \{ ij} p_l p_{m \}} \right\} \,, \qquad  \textrm{for} \quad (\alpha\beta\gamma\eta) = (1111), (2222)  \nn \\
\left[S \right]_{ijlm}^{(\alpha\beta\gamma\eta)}(1\delta,2p) &=& \left\{ \delta_{ \{ ij} \, p_{l  \}} p_{m} \,, \delta_{ m \{ i} \, p_{j } p_{l \}} \right\} \,, \qquad  \textrm{for} \quad (\alpha\beta\gamma\eta) = (1112), (2221) \nn \\ 
\left[S \right]_{ijlm}^{(\alpha\beta\gamma\eta)}(1\delta,2p) &=& \left\{ \delta_{ij} p_l p_m \,, \delta_{lm} p_i p_j \,, p_{\{ i} \, \delta_{j \} \{ l} \, p_{m \}} \right\} \,, \qquad  \textrm{for} \quad (\alpha\beta\gamma\eta) = (1122) 
\eea
for the rank-4 integrals, 
\bea
\left[S \right]_{ijlmr}^{(\alpha\beta\gamma\eta\phi)}(2\delta,1p) &=& \left\{    \delta_{ \{ ij} \delta_{lm} p_{r \}}  \right\}  \,, \qquad  \textrm{for} \quad (\alpha\beta\gamma\eta\phi) = (11111), (22222) \nn \\
\left[S \right]_{ijlmr}^{(\alpha\beta\gamma\eta\phi)}(2\delta,1p) &=& \left\{    \delta_{ \{ ij} \delta_{lm \} } p_r \,,   \delta_{  r \{ i} \delta_{jl} p_{m \}} \right\}  \,, \qquad  \textrm{for} \quad (\alpha\beta\gamma\eta\phi) = (11112), (22221) \nn \\
\left[S \right]_{ijlmr}^{(\alpha\beta\gamma\eta\phi)}(2\delta,1p) &=& \left\{ \delta_{\{i j} p_{l\}} \delta_{mr} \,, p_{\{m} \delta_{r\} \{i} \delta_{jl \}}  \,, \delta_{l \} \{ m} \delta_{r \} \{ i} \, p_{j} \right\} \,, \qquad  \textrm{for} \quad (\alpha\beta\gamma\eta\phi) = (11122), (22211) \nn \\
\left[S \right]_{ijlmr}^{(\alpha\beta\gamma\eta\phi)}(1\delta,3p) &=& \left\{    \delta_{ \{ ij} p_{l} p_{m} p_{r \}}  \right\}  \,, \qquad  \textrm{for} \quad (\alpha\beta\gamma\eta\phi) = (11111), (22222) \nn \\
\left[S \right]_{ijlmr}^{(\alpha\beta\gamma\eta\phi)}(1\delta,3p) &=& \left\{    \delta_{ \{ ij} p_{l} p_{m \}} p_r \,,   \delta_{  r \{ i} p_{j} p_{l} p_{m \}} \right\}  \,, \qquad  \textrm{for} \quad (\alpha\beta\gamma\eta\phi) = (11112), (22221) \nn \\
\left[S \right]_{ijlmr}^{(\alpha\beta\gamma\eta\phi)}(1\delta,3p) &=& \left\{ p_i p_j p_l \delta_{mr} \,, p_{\{i} \delta_{jl\}} p_{m} p_{r} \,, p_{\{ m} \delta_{r\} \{ i} \, p_{j} p_{l \}} \right\} \,, \qquad  \textrm{for} \quad (\alpha\beta\gamma\eta\phi) = (11122), (22211) \nn \\
\eea
for the rank-5 and, finally,
\bea
\left[S \right]_{ijlmrs}^{(\alpha\beta\gamma\eta\phi\xi)}(3\delta,0p) &=& \left\{  \delta_{\{ ij} \delta_{lm} \delta_{rs \}}   \right\}  \,, \qquad  \textrm{for} \quad (\alpha\beta\gamma\eta\phi\xi) = (111111), (111112), (222221), (222222) \nn \\
\left[S \right]_{ijlmrs}^{(\alpha\beta\gamma\eta\phi\xi)}(3\delta,0p) &=& \left\{  \delta_{ m \} \{ r} \delta_{s \} \{ i} \delta_{jl }  \,, \delta_{\{ ij} \delta_{lm \}} \delta_{rs}   \right\}  \,, \qquad  \textrm{for} \quad (\alpha\beta\gamma\eta\phi\xi) = (111122), (222211)\nn \\
\left[S \right]_{ijlmrs}^{(\alpha\beta\gamma\eta\phi\xi)}(3\delta,0p) &=& \left\{ \delta^{\{i}_{\{m} \delta^{j}_{r} \delta^{l\}}_{s\}}    \,\,,   \delta_{\{ij} \delta_{l\}\{m} \delta_{rs\}}  \right\}  \,, \qquad  \textrm{for} \quad (\alpha\beta\gamma\eta\phi\xi) = (111222)\nn \\
\left[S \right]_{ijlmrs}^{(\alpha\beta\gamma\eta\phi\xi)}(2\delta,2p) &=& \left\{ \delta_{\{ ij} \delta_{lm} p_{r} p_{s\}} \right\} \,, \qquad  \textrm{for} \quad (\alpha\beta\gamma\eta\phi\xi) = (111111), (222222) \nn \\
\left[S \right]_{ijlmrs}^{(\alpha\beta\gamma\eta\phi\xi)}(2\delta,2p) &=& \left\{ \delta_{\{ ij} \delta_{lm} p_{r \}} p_{s} \,, \delta_{s \{ i} \delta_{jl} p_{m } p_{r\}} \right\} \,, \qquad  \textrm{for} \quad (\alpha\beta\gamma\eta\phi\xi) = (111112), (222221) \nn \\
\left[S \right]_{ijlmrs}^{(\alpha\beta\gamma\eta\phi\xi)}(2\delta,2p) &=& \left\{ \delta_{\{ ij} \delta_{lm \}} p_{r} p_{s} \,\,, \delta_{jl} \delta_{m\} \{r} \, p_{s\}} p_{\{i} \,\,, \delta_{m\} \{ r} \, \delta_{s\} \{i} \, p_{j} p_{l}  \,\,, 
\delta_{rs} \delta_{\{ij} p_{l} p_{m \}}  \right\} \,, \nn \\
&& \qquad\qquad\qquad \qquad\qquad\qquad\qquad \textrm{for} \quad (\alpha\beta\gamma\eta\phi\xi) = (111122), (222211) \nn \\
\left[S \right]_{ijlmrs}^{(\alpha\beta\gamma\eta\phi\xi)}(2\delta,2p) &=& \left\{  \delta_{\{ ij} \delta_{l\} \{m} \, p_{r} p_{s\}} \,\,,   \delta_{l\} \{m} \, p_{r} \, \delta_{s\} \{ i} \, p_{j}  \,\,,    \delta_{\{mr} \delta_{s\}\{i} \, p_{j} p_{l \}}  \,\,,   \delta_{jl\}} \delta_{\{ m r} p_{s\}} p_{\{i}   \right\} \,, \nn \\
&& \qquad\qquad\qquad \qquad\qquad\qquad\qquad \textrm{for} \quad (\alpha\beta\gamma\eta\phi\xi) = (111222) \nn \\
\left[S \right]_{ijlmrs}^{(\alpha\beta\gamma\eta\phi\xi)}(1\delta,4p) &=& \left\{ \delta_{\{ ij} p_{l} p_{m} p_{r} p_{s \}}  \right\} \,, \qquad  \textrm{for} \quad (\alpha\beta\gamma\eta\phi\xi) = (111111), (222222) \nn \\
\left[S \right]_{ijlmrs}^{(\alpha\beta\gamma\eta\phi\xi)}(1\delta,4p) &=& \left\{ \delta_{\{ ij} p_{l} p_{m} p_{r \}} p_{s} \,\,, \delta_{s \{ i} p_{j} p_{l} p_{m } p_{r\}} \right\} \,, \qquad  \textrm{for} \quad (\alpha\beta\gamma\eta\phi\xi) = (111112), (222221) \nn \\
\left[S \right]_{ijlmrs}^{(\alpha\beta\gamma\eta\phi\xi)}(1\delta,4p) &=& \left\{  \delta_{\{ i j} p_{l} p_{m \}} p_{r} p_{s} \,\,,    \delta_{m\} \{r} \, p_{s\}} p_{\{i} p_{j} p_{l}  \,\,,  \delta_{rs} p_{i} p_{j} p_{l} p_{m} \right\} \,, \nn \\
&& \qquad \qquad\qquad \qquad\qquad\qquad\qquad \textrm{for} \quad (\alpha\beta\gamma\eta\phi\xi) = (111122), (222211) \nn \\
\left[S \right]_{ijlmrs}^{(\alpha\beta\gamma\eta\phi\xi)}(1\delta,4p) &=& \left\{ \delta_{\{ij} p_{l\}} p_{m} p_{r} p_{s} \,\,,  \delta_{l\} \{m} \, p_{r} p_{s\}} p_{\{i} \, p_{j} \,\,, \delta_{\{mr} p_{s\}} p_{i} p_{j} p_{k}  \right\} \,, \nn \\
&& \qquad  \qquad\qquad \qquad\qquad\qquad\qquad \textrm{for} \quad (\alpha\beta\gamma\eta\phi\xi) = (111222) 
\eea
for the rank-6. Notice that the indices of the Kronecker deltas in $\left[S \right]_{ijlmrs}^{(\alpha\beta\gamma\eta\phi\xi)}(3\delta,0p)$ have been raised only to make manifest their symmetric properties. 

For the coefficients $\mathcal C$ of the tensor expansions we use the following notation
\bea
\mathcal C \equiv J \quad &\textrm{for}& \quad \textrm{den}(k_1,k_2,p; \left\{\nu \right\}) = k_1^2 (k_1 - k_2)^2 (k_2+p)^2 \nn \\
\mathcal C \equiv Y \quad &\textrm{for}& \quad \textrm{den}(k_1,k_2,p; \left\{\nu \right\}) = k_1^2 k_2^2 (k_1 - k_2)^2 (k_2+p)^2 \nn \\
\mathcal C \equiv K \quad &\textrm{for}& \quad \textrm{den}(k_1,k_2,p; \left\{\nu \right\}) = k_1^2 k_2^2 (k_1 - k_2)^2 (k_1 + p )^2 (k_2+p)^2 
\eea
where $J,Y,K$ appear, respectively, in the first three diagrams of Fig.\ref{Fig.TT2L}. The explicit expressions of such coefficients is given as a linear combination of the scalar integrals $J_0$ and $B_0^2$ and are provided in the tables below.

\begin{sidewaystable}
\begin{tabular}{|c||cc|cc|cc|cc|} \hline
& $J^{(1)}_1 $ & $ J^{(2)}_1 $ & $ J^{(11)}_{00} $ & $ J^{(11)}_{11} $ & $ J^{(22)}_{00} $ & $ J^{(22)}_{11} $ & $ J^{(12)}_{00} $ & $ J^{(12)}_{11} $ \\ \hline \hline
$J_0$ &  $ -\frac{1}{3} $ & $ -\frac{2}{3} $ & $ -\frac{p^2}{3 (3 d-4)} $ & $ \frac{d}{3 (3 d-4)} $ & $ -\frac{p^2}{3 (3 d-4)} $ & $ \frac{4 (d-1)}{3 (3 d-4)} $ & $ -\frac{p^2}{6 (3 d-4)} $ & $ \frac{2 (d-1)}{3 (3 d-4)} $ \\ \hline
$B_0^2$ &  0 & 0 & 0 & 0 & 0 & 0 & 0 & 0 \\ \hline
\end{tabular}
\\ \\
\\ \\
\\
\bigskip\bigskip
%
%
%
%
\begin{tabular}{|c||cccccccccc|} \hline
& $Y^{(1)}_1 $ & $ Y^{(2)}_1 $ & $Y^{(11)}_{00} $ & $Y^{(11)}_{11} $ & $Y^{(22)}_{00} $ & $Y^{(22)}_{11} $ & $ Y^{(12)}_{00} $ & $Y^{(12)}_{11} $ & $Y^{(111)}_{001}$ & $Y^{(111)}_{111} $\\ \hline \hline
$J_0$ & $-\frac{d-3}{(d-4) p^2}$ & $ -\frac{2 (d-3)}{(d-4) p^2}$ & $ -\frac{d-3}{3 \text{den}_3}$ & $ \frac{(d-3) d}{3 p^2 \text{den}_3} $ & $ -\frac{1}{3 (d-4)}$ & $ \frac{4 (d-3)}{3 (d-4) p^2}
  $ & $ -\frac{1}{6 (d-4)} $ & $ \frac{2 (d-3)}{3 (d-4) p^2} $ & $ \frac{d-3}{3 \text{den}_2} $ & $ -\frac{(d-3) (d+2)}{3 p^2 \text{den}_2} $ \\\hline
$B_0^2$ & $ 0 $ & $ 0 $ & $ 0 $ & $ 0 $ & $ 0 $ & $ 0 $ & $ 0 $ & $ 0 $ & $ 0 $ & $ 0 $ \\ \hline
\end{tabular}
\\
\bigskip\bigskip
\begin{tabular}{|c||cccccccccc|} \hline
& $Y^{(222)}_{001} $ & $ Y^{(222)}_{111}$ & $Y^{(112)}_{001;1} $ & $ Y^{(112)}_{001;2} $ & $Y^{(112)}_{111} $ & $ Y^{(221)}_{001;1} $ & $ Y^{(221)}_{001;2} $ & $Y^{(221)}_{111} $ & $ Y^{(1111)}_{0000}$ & $  Y^{(1111)}_{0011} $\\ \hline \hline
$J_0$ & $ \frac{2 (d-2)}{3 \text{den}_2} $ & $ -\frac{8 (d-3) (d-1)}{3 p^2 \text{den}_2} $ & $ \frac{2 d^2-9 d+8}{3 (d-1) \text{den}_2} $ & $ \frac{(d-2) d}{6 (d-1) \text{den}_2} $ & $ -\frac{2
   (d-3) d}{3 p^2 \text{den}_2}$ & $ \frac{d-2}{3 \text{den}_2} $ & $ \frac{d-2}{3 \text{den}_2} $ & $ -\frac{4 (d-3) (d-1)}{3 p^2 \text{den}_2} $ & $ \frac{(d-3) d p^2}{3 (d+1)
   \text{den}_1} $ & $ -\frac{(d-3) d (d+2)}{3 (d+1) \text{den}_1}$ \\ \hline
$B_0^2$ & $ 0 $ & $ 0 $ & $ 0 $ & $ 0 $ & $ 0 $ & $ 0 $ & $ 0 $ & $ 0 $ & $ 0 $ & $ 0 $\\ \hline
\end{tabular}
\\
\bigskip\bigskip
\begin{tabular}{|c||cccccccccc|} \hline
& $Y^{(1111)}_{1111} $ & $Y^{(2222)}_{0000}$ & $ Y^{(2222)}_{0011} $ & $ Y^{(2222)}_{1111} $ & $Y^{(1112)}_{0000} $ & $Y^{(1112)}_{0011;1} $ & $ Y^{(1112)}_{0011;2} $ & $Y^{(1112)}_{1111} $ & $Y^{(2221)}_{0000}$ & $ Y^{(2221)}_{0011;1} $\\ \hline \hline
$J_0$ & $\frac{(d-3) d (d+2) (d+4)}{3 (d+1) p^2 \text{den}_1} $ & $ \frac{d p^2}{3 \text{den}_1} $ & $ -\frac{4 (d-2) (d-1)}{3 \text{den}_1} $ & $ \frac{16 (d-3) (d-1) d}{3 p^2 \text{den}_1}
 $  &  $ \frac{(d-2) p^2}{6 \text{den}_1} $ & $ -\frac{2 d^2-7 d+2}{3 \text{den}_1} $ & $ -\frac{(d-2) (d+2)}{6 \text{den}_1}$ & $ \frac{2 (d-3) d (d+2)}{3 p^2 \text{den}_1} $ & $ \frac{d
   p^2}{6 \text{den}_1} $ & $ -\frac{2 (d-2) (d-1)}{3 \text{den}_1}$ \\ \hline
$B_0^2$ & $ 0 $ & $ 0 $ & $ 0 $ & $ 0 $ & $ 0 $ & $ 0 $ & $ 0 $ & $ 0 $ & $ 0 $ & $ 0 $ \\ \hline
\end{tabular}
\bigskip\bigskip
\begin{tabular}{|c||cccccccc|} \hline
& $Y^{(2221)}_{0011;2} $ & $ Y^{(2221)}_{1111}$ & $Y^{(1122)}_{0000;1} $ & $ Y^{(1122)}_{0000;2} $ & $Y^{(1122)}_{0011;1} $ & $ Y^{(1122)}_{0011;2} $ & $ Y^{(1122)}_{0011;3} $ & $ Y^{(1122)}_{1111} $\\ \hline \hline
$J_0$ & $ -\frac{2 (d-2) (d-1)}{3 \text{den}_1} $ & $ \frac{8 (d-3) (d-1) d}{3 p^2 \text{den}_1} $ & $ \frac{\left(2 d^2-7 d+4\right) p^2}{6 (d-1) \text{den}_1} $ & $ \frac{d^2 p^2}{12 (d-1)
   \text{den}_1} $ & $ -\frac{4 \left(d^2-4 d+2\right)}{3 \text{den}_1} $ & $ -\frac{(d-2) d}{3 \text{den}_1} $ & $ -\frac{(d-2) d}{3 \text{den}_1} $ & $ \frac{4 (d-3) d^2}{3 p^2
   \text{den}_1}$ \\ \hline
$B_0^2$ & $ 0 $ & $ 0 $ & $ 0 $ & $ 0 $ & $ 0 $ & $ 0 $ & $ 0 $ & $ 0 $ \\ \hline
\end{tabular}
\end{sidewaystable}

\begin{sidewaystable}
\begin{tabular}{|c||cccccccc|} \hline
& $K^{(1)}_1 $ & $K^{(2)}_1 $ & $K^{(11)}_{00} $ & $K^{(11)}_{11} $ & $K^{(22)}_{00} $ & $K^{(22)}_{11} $ & $K^{(12)}_{00} $ & $K^{(12)}_{11} $\\ \hline \hline
$J_0$ & $-\frac{(3 d-10) (3 d-8)}{(d-4)^2 p^4} $ & $ -\frac{(3 d-10) (3 d-8)}{(d-4)^2 p^4} $ & $ -\frac{2 (d-3) (d-2)}{p^2 \text{den}_{17}} $ & $ \frac{5 d^3-33 d^2+64 d-32}{p^4 \text{den}_{17}} $ & $ -\frac{2 (d-3) (d-2)}{p^2 \text{den}_{17}}$ & $\frac{5 d^3-33 d^2+64 d-32}{p^4 \text{den}_{17}} $ & $ -\frac{(d-3) d}{p^2 \text{den}_{17}} $ & $ \frac{2  (d-2) \left(2 d^2-9+8\right)}{p^4 \text{den}_{17}} $ \\ \hline
$B_0^2$ & $ \frac{d-3}{(d-4) p^2}$ & $ \frac{d-3}{(d-4) p^2} $ & $ \frac{d-3}{2 \text{den}_{12}} $ & $ -\frac{(d-3) d}{2 p^2 \text{den}_{12}} $ & $ \frac{d-3}{2 \text{den}_{12}} $ & $ -\frac{(d-3)
   d}{2 p^2 \text{den}_{12}} $ & $ \frac{1}{2 \text{den}_{12}} $ & $ -\frac{(d-2)^2}{2 p^2 \text{den}_{12}} $ \\ \hline
\end{tabular}
\\ \\ \\ 
%
\bigskip\bigskip
\begin{tabular}{|c||cccccccc|} \hline
& $K^{(111)}_{001} $ & $K^{(111)}_{111}$ & $K^{(222)}_{001} $ & $K^{(222)}_{111} $ & $K^{(112)}_{001;1} $ & $K^{(112)}_{001;2} $ & $K^{(112)}_{111} $ & $K^{(221)}_{001;1} $\\ \hline \hline
$J_0$ & $ \frac{(d-3) (d-2)}{p^2 \text{den}_{17}} $ & $ -\frac{3 d^3-18 d^2+29 d-8}{p^4 \text{den}_{17}} $ & $ \frac{(d-3) (d-2)}{p^2 \text{den}_{17}} $ & $ -\frac{3 d^3-18 d^2+29 d-8}{p^4
   \text{den}_{17}} $ & $ \frac{(d-3) (d-2)}{p^2 \text{den}_{17}} $ & $ \frac{(d-3) d}{2 p^2 \text{den}_{17}} $ & $ -\frac{2 d^2-9 d+8}{(d-4)^2 p^4} $ & $ \frac{(d-3) (d-2)}{p^2
   \text{den}_{17}} $ \\ \hline
$B_0^2$ & $-\frac{d-3}{4 \text{den}_{12}} $ & $ \frac{(d-3) (d+2)}{4 p^2 \text{den}_{12}} $ & $ -\frac{d-3}{4 \text{den}_{12}} $ & $ \frac{(d-3) (d+2)}{4 p^2 \text{den}_{12}} $ & $ -\frac{d-3}{4
   \text{den}_{12}} $ & $ -\frac{1}{4 \text{den}_{12}} $ & $ \frac{d-2}{4 (d-4) p^2} $ & $ -\frac{d-3}{4 \text{den}_{12}} $ \\ \hline
\end{tabular}
\\
%
\bigskip\bigskip
\begin{tabular}{|c||ccccccc|} \hline
& $K^{(221)}_{001;2} $ & $K^{(221)}_{111}$ & $K^{(1111)}_{0000} $ & $K^{(1111)}_{0011} $ & $K^{(1111)}_{1111} $ & $K^{(2222)}_{0000} $ & $ K^{(2222)}_{0011} $ \\ \hline \hline
$J_0$ & $ \frac{(d-3) d}{2 p^2 \text{den}_{17}} $ & $ -\frac{2 d^2-9 d+8}{(d-4)^2 p^4} $ & $ \frac{2 (d-3) d}{3 \text{den}_{9}} $ & $ -\frac{(d-3) \left(5 d^3-11 d^2-4 d+16\right)}{3 p^2
   \text{den}_{10}} $ & $ \frac{(d-2) \left(17 d^4-60 d^3-21 d^2+88 d+48\right)}{3 p^4 \text{den}_{10}} $ & $ \frac{2 (d-3) d}{3 \text{den}_{9}} $ & $ -\frac{(d-3) \left(5 d^3-11
   d^2-4 d+16\right)}{3 p^2 \text{den}_{10}} $ \\ \hline
$B_0^2$ &  $ -\frac{1}{4 \text{den}_{12}} $ & $ \frac{d-2}{4 (d-4) p^2} $ & $ -\frac{(d-3) p^2}{8 \text{den}_{21}} $ & $ \frac{(d-3) (d+2)}{8 \text{den}_{21}} $ & $ -\frac{(d-3) (d+2) (d+4)}{8 p^2
   \text{den}_{21}} $ & $ -\frac{(d-3) p^2}{8 \text{den}_{21}} $ & $ \frac{(d-3) (d+2)}{8 \text{den}_{21}} $ \\ \hline
\end{tabular}
\\
%
\bigskip\bigskip
\begin{tabular}{|c||ccccccc|} \hline
& $K^{(2222)}_{1111} $ & $K^{(1112)}_{0000}$ & $K^{(1112)}_{0011;1} $ & $K^{(1112)}_{0011;2} $ & $K^{(1112)}_{1111} $ & $K^{(2221)}_{0000} $ & $K^{(2221)}_{0011;1} $ \\ \hline \hline
$J_0$ &$\frac{(d-2) \left(17 d^4-60 d^3-21 d^2+88 d+48\right)}{3 p^4 \text{den}_{10}} $ & $ \frac{d^2-d-4}{3 \text{den}_{9}} $ & $ -\frac{(d-3) \left(4 d^3-7 d^2-2 d+8\right)}{3 p^2
   \text{den}_{10}} $ & $ -\frac{(d-3) d (5 d+4)}{6 p^2 \text{den}_{9}} $ & $ \frac{d \left(10 d^4-47 d^3+39 d^2+44 d-64\right)}{3 p^4 \text{den}_{10}} $ & $ \frac{d^2-d-4}{3
   \text{den}_{9}} $ & $ -\frac{(d-3) \left(4 d^3-7 d^2-2 d+8\right)}{3 p^2 \text{den}_{10}} $ \\ \hline
$B_0^2$ & $ -\frac{(d-3) (d+2) (d+4)}{8 p^2 \text{den}_{21}} $ & $ -\frac{p^2}{8 \text{den}_{21}} $ & $ \frac{d^2-2 d-2}{8 \text{den}_{21}} $ & $ \frac{d+2}{8 \text{den}_{21}} $ & $ -\frac{(d-2) d
   (d+2)}{8 p^2 \text{den}_{21}} $ & $ -\frac{p^2}{8 \text{den}_{21}} $ & $ \frac{d^2-2 d-2}{8 \text{den}_{21}} $ \\ \hline
\end{tabular}
\\
%
\bigskip\bigskip
\begin{tabular}{|c||cccccc|} \hline
& $K^{(2221)}_{0011;2} $ & $K^{2221)}_{1111}$ & $K^{(1122)}_{0000;1} $ & $K^{(1122)}_{0000;2} $ & $K^{(1122)}_{0011;1} $ & $K^{(1122)}_{0011;2} $  \\ \hline \hline
$J_0$ & $-\frac{(d-3) d (5 d+4)}{6 p^2 \text{den}_{9}} $ & $ \frac{d \left(10 d^4-47 d^3+39 d^2+44 d-64\right)}{3 p^4 \text{den}_{10}} $ & $ \frac{2 d^4-13 d^3+23 d^2+6 d-24}{3 (d-2)
   \text{den}_{10}} $ & $ \frac{d^4-d^3-6 d^2-4 d+16}{6 (d-2) \text{den}_{10}} $ & $ -\frac{(d-3) \left(5 d^3-14 d^2-4 d+16\right)}{3 (d-2) p^2 \text{den}_{9}} $ & $ -\frac{(d-3)
   \left(5 d^3-14 d^2-4 d+16\right)}{3 (d-2) p^2 \text{den}_{9}} $ \\ \hline
$B_0^2$ & $ \frac{d+2}{8 \text{den}_{21}} $ & $ -\frac{(d-2) d (d+2)}{8 p^2 \text{den}_{21}} $ & $ -\frac{\left(d^2-3 d-2\right) p^2}{8 \text{den}_{15}} $ & $ -\frac{p^2}{4 \text{den}_{15}} $ &
  $ \frac{d^2-2 d-4}{8 \text{den}_{20}} $ & $ \frac{d^2-2 d-4}{8 \text{den}_{20}} $ \\ \hline
\end{tabular}
\\
%
\bigskip\bigskip
\begin{tabular}{|c||ccccccc|} \hline
& $K^{(1122)}_{0011;3} $ & $K^{1122)}_{1111}$ & $K^{(11111)}_{00001} $ & $K^{(11111)}_{00111} $ & $K^{(11111)}_{11111} $ & $K^{(22222)}_{00001} $ & $K^{(22222)}_{00111} $  \\ \hline \hline
$J_0$ & $ -\frac{(d-3) d \left(2 d^2-d-4\right)}{3 (d-2) p^2 \text{den}_{9}} $ & $ \frac{4 d \left(2 d^3-6 d^2-3 d+4\right)}{3 p^4 \text{den}_{9}} $ & $ -\frac{(d-3) d}{3
   \text{den}_{9}} $ & $ \frac{(d-3) \left(d^3-2 d^2-d+4\right)}{p^2 \text{den}_{10}} $ & $ -\frac{11 d^5-49 d^4-3 d^3+169 d^2-8 d-240}{3 p^4 \text{den}_{10}} $ & $ -\frac{(d-3)
   d}{3 \text{den}_{9}} $ & $ \frac{(d-3) \left(d^3-2 d^2-d+4\right)}{p^2 \text{den}_{10}} $ \\ \hline
$B_0^2$ & $ \frac{d}{8 \text{den}_{20}} $ & $ -\frac{d^2}{8 (d-4) (d+1) p^2} $ & $ \frac{(d-3) p^2}{16 \text{den}_{21}} $ & $ -\frac{(d-3) (d+4)}{16 \text{den}_{21}} $ & $ \frac{(d-3) (d+4)
   (d+6)}{16 p^2 \text{den}_{21}} $ & $ \frac{(d-3) p^2}{16 \text{den}_{21}} $ & $  -\frac{(d-3) (d+4)}{16 \text{den}_{21}} $ \\ \hline
\end{tabular}
\end{sidewaystable}

\begin{sidewaystable}
\begin{tabular}{|c||ccccccc|} \hline
& $K^{(22222)}_{11111} $ & $K^{11112)}_{00001;1}$ & $K^{(11112)}_{00001;2} $ & $K^{(11112)}_{00111;1} $ & $K^{(11112)}_{00111;2} $ & $K^{(11112)}_{11111} $ & $K^{(22221)}_{00001;1} $  \\ \hline \hline
$J_0$ & $ -\frac{11 d^5-49 d^4-3 d^3+169 d^2-8 d-240}{3 p^4 \text{den}_{10}} $ & $ -\frac{(d-3) d}{3 \text{den}_{9}} $ & $ -\frac{d^2-d-4}{6 \text{den}_{9}} $ & $ \frac{(d-3) \left(2 d^3-2
   d^2-d+4\right)}{3 p^2 \text{den}_{10}} $ & $ \frac{(d-3) d^3}{2 p^2 \text{den}_{10}} $ & $ -\frac{2 d^4-9 d^3+7 d^2+8 d-16}{p^4 \text{den}_{19}} $ &  $-\frac{(d-3) d}{3
   \text{den}_{9}} $ \\ \hline
$B_0^2$ & $ \frac{(d-3) (d+4) (d+6)}{16 p^2 \text{den}_{21}} $ & $ \frac{(d-3) p^2}{16 \text{den}_{21}} $ & $ \frac{p^2}{16 \text{den}_{21}} $ & $ -\frac{d^2-d-4}{16 \text{den}_{21}} $ & $
   -\frac{d+4}{16 \text{den}_{21}} $ & $ \frac{(d-2) (d+4)}{16 p^2 \text{den}_{12}} $ & $ \frac{(d-3) p^2}{16 \text{den}_{21}} $ \\ \hline
\end{tabular}
\\ \\ \\
%
\bigskip\bigskip
\begin{tabular}{|c||ccccccc|} \hline
& $K^{(22221)}_{00001;2} $ & $K^{22221)}_{00111;1}$ & $K^{(22221)}_{00111;2} $ & $K^{(22221)}_{11111} $ & $K^{(11122)}_{00001;1} $ & $K^{(11122)}_{00001;2} $ & $K^{(11122)}_{00001;3} $  \\ \hline \hline
$J_0$ &  $-\frac{d^2-d-4}{6 \text{den}_{9}} $ & $ \frac{(d-3) \left(2 d^3-2 d^2-d+4\right)}{3 p^2 \text{den}_{10}} $ & $ \frac{(d-3) d^3}{2 p^2 \text{den}_{10}} $ & $ -\frac{2 d^4-9 d^3+7
   d^2+8 d-16}{p^4 \text{den}_{19}} $ & $ -\frac{2 d^4-13 d^3+23 d^2+6 d-24}{6 (d-2) \text{den}_{10}} $ & $ -\frac{d^2-d-4}{6 \text{den}_{9}} $ & $ -\frac{d^4-d^3-6 d^2-4 d+16}{12
   (d-2) \text{den}_{10}} $ \\ \hline
$B_0^2$ & $ \frac{p^2}{16 \text{den}_{21}} $ & $ -\frac{d^2-d-4}{16 \text{den}_{21}} $ & $ -\frac{d+4}{16 \text{den}_{21}} $ & $ \frac{(d-2) (d+4)}{16 p^2 \text{den}_{12}} $ & $ \frac{\left(d^2-3
   d-2\right) p^2}{16 \text{den}_{15}} $ & $ \frac{p^2}{16 \text{den}_{21}} $ & $ \frac{p^2}{8 \text{den}_{15}} $ \\ \hline
\end{tabular}
\\
%
\bigskip\bigskip
\begin{tabular}{|c||ccccccc|} \hline
& $K^{(11122)}_{00111;1} $ & $K^{11122)}_{00111;2}$ & $K^{(11122)}_{00111;3} $ & $K^{(11122)}_{11111} $ & $K^{(22211)}_{00001;1} $ & $K^{(22211)}_{00001;2} $ & $K^{(22211)}_{00001;3} $  \\ \hline \hline
$J_0$ & $ \frac{(d-3) d \left(d^3-3 d^2-d+4\right)}{(d-2) p^2 \text{den}_{10}} $ & $ \frac{(d-3) d \left(2 d^2-3 d-4\right)}{3 (d-2) p^2 \text{den}_{9}} $ & $ \frac{(d-3) d^2 \left(2
   d^2-d-4\right)}{6 (d-2) p^2 \text{den}_{10}} $ & $ -\frac{2 d \left(2 d^3-6 d^2-3 d+4\right)}{3 p^4 \text{den}_{19}} $ & $ -\frac{2 d^4-13 d^3+23 d^2+6 d-24}{6 (d-2)
   \text{den}_{10}} $ & $ -\frac{d^2-d-4}{6 \text{den}_{9}} $ & $ -\frac{d^4-d^3-6 d^2-4 d+16}{12 (d-2) \text{den}_{10}} $ \\ \hline
$B_0^2$ & $ -\frac{d \left(d^2-d-8\right)}{16 \text{den}_{15}} $ & $ -\frac{(d-3) d^2}{16 \text{den}_{15}} $ & $ -\frac{d^2}{16 \text{den}_{15}} $ & $ \frac{d^2}{16 p^2 \text{den}_{12}} $ & $
   \frac{\left(d^2-3 d-2\right) p^2}{16 \text{den}_{15}} $ & $ \frac{p^2}{16 \text{den}_{21}} $ & $ \frac{p^2}{8 \text{den}_{15}} $ \\ \hline
\end{tabular}
\\
%
\bigskip\bigskip
\begin{tabular}{|c||cccccc|} \hline
& $K^{(22211)}_{00111;1} $ & $K^{22211)}_{00111;2}$ & $K^{(22211)}_{00111;3} $ & $K^{(22211)}_{11111} $ & $K^{(111111)}_{000000} $ & $K^{(111111)}_{000011} $  \\ \hline \hline
$J_0$ & $ \frac{(d-3) d \left(d^3-3 d^2-d+4\right)}{(d-2) p^2 \text{den}_{10}} $ & $ \frac{(d-3) d \left(2 d^2-3 d-4\right)}{3 (d-2) p^2 \text{den}_{9}} $ & $ \frac{(d-3) d^2 \left(2
   d^2-d-4\right)}{6 (d-2) p^2 \text{den}_{10}} $ & $ -\frac{2 d \left(2 d^3-6 d^2-3 d+4\right)}{3 p^4 \text{den}_{19}} $ & $ -\frac{2 (d-3) (d+2) p^2}{9 \text{den}_{14}} $ & $
   \frac{(d-3) \left(5 d^3+11 d^2-8 d+16\right)}{9 (d+1) \text{den}_{14}} $ \\ \hline
$B_0^2$ & $ -\frac{d \left(d^2-d-8\right)}{16 \text{den}_{15}} $ & $ -\frac{(d-3) d^2}{16 \text{den}_{15}} $ & $ -\frac{d^2}{16 \text{den}_{15}} $ & $ \frac{d^2}{16 p^2 \text{den}_{12}} $ & $
   \frac{(d-3) p^4}{32 \text{den}_{11}} $ & $ -\frac{(d-3) (d+4) p^2}{32 \text{den}_{11}} $ \\ \hline
\end{tabular}
\\
%
\bigskip\bigskip
\begin{tabular}{|c||cccc|} \hline
& $K^{(111111)}_{001111} $ & $K^{(111111)}_{111111} $ & $K^{(222222)}_{000000} $ & $K^{(222222)}_{000011} $  \\ \hline \hline
$J_0$ &  $  -\frac{(d-3) \left(17 d^5+12 d^4-133 d^3+84 d^2+332 d-240\right)}{9 (d+1) p^2 \text{den}_8} $ & $ \frac{65 d^7-47 d^6-1105 d^5+667 d^4+5600 d^3-3788 d^2-11040 d+7488}{9
   (d+1) p^4 \text{den}_8} $ & $ -\frac{2 (d-3) (d+2) p^2}{9 \text{den}_{14}} $ & $ \frac{(d-3) \left(5 d^3+11 d^2-8 d+16\right)}{9 (d+1) \text{den}_{14}} $  \\ \hline
$B_0^2$ & $ \frac{(d-3) (d+4) (d+6)}{32 \text{den}_{11}} $ & $ -\frac{(d-3) (d+4) (d+6) (d+8)}{32 p^2 \text{den}_{11}} $ & $ \frac{(d-3) p^4}{32 \text{den}_{11}} $ & $ -\frac{(d-3) (d+4)
   p^2}{32 \text{den}_{11}} $  \\ \hline
\end{tabular}
\\
%
\bigskip\bigskip
\begin{tabular}{|c||ccccc|} \hline
& $K^{(222222)}_{001111} $ & $K^{(222222)}_{111111} $ & $K^{(111112)}_{000000} $ & $K^{(111112)}_{000011;1} $ & $ K^{(111112)}_{000011;2} $  \\ \hline \hline
$J_0$ & $ -\frac{(d-3) \left(17 d^5+12 d^4-133 d^3+84 d^2+332 d-240\right)}{9 (d+1) p^2 \text{den}_8} $ & $ \frac{65 d^7-47 d^6-1105 d^5+667 d^4+5600 d^3-3788 d^2-11040 d+7488}{9
   (d+1) p^4 \text{den}_8} $ & $ -\frac{\left(d^2+d-8\right) p^2}{9 \text{den}_{14}} $ & $ \frac{4 \left(d^4-8 d^2-8\right)}{9 (d+1) \text{den}_{14}} $ & $ \frac{5 d^3+d^2-38
   d+8}{18 \text{den}_{14}} $ \\ \hline
$B_0^2$ & $ \frac{(d-3) (d+4) (d+6)}{32 \text{den}_{11}} $ & $ -\frac{(d-3) (d+4) (d+6) (d+8)}{32 p^2 \text{den}_{11}} $ & $ \frac{p^4}{32 \text{den}_{11}} $ & $ -\frac{\left(d^2-8\right)
   p^2}{32 \text{den}_{11}} $ & $ -\frac{(d+4) p^2}{32 \text{den}_{11}}  $  \\ \hline
\end{tabular}
\end{sidewaystable}

\begin{sidewaystable}
\begin{tabular}{|c||cccc|} \hline
& $K^{(111112)}_{001111;1} $ & $K^{(111112)}_{001111;2} $ & $K^{(111112)}_{111111} $ & $K^{(222221)}_{000000} $  \\ \hline \hline
$J_0$ & $ -\frac{(d-3) \left(10 d^5+20 d^4-29 d^3+41 d^2+82 d-88\right)}{9 (d+1) p^2 \text{den}_8} $ & $ -\frac{(d-3) \left(17 d^4+29 d^3-42 d^2+64 d-32\right)}{18 p^2 \text{den}_8}
  $ & $ \frac{(d+2) \left(34 d^6-61 d^5-332 d^4+677 d^3-38 d^2-1696 d+1056\right)}{9 (d+1) p^4 \text{den}_8} $ & $ -\frac{\left(d^2+d-8\right) p^2}{9 \text{den}_{14}} $\\ \hline
$B_0^2$ & $ \frac{(d+4) \left(d^2-6\right)}{32 \text{den}_{11}} $ & $ \frac{(d+4) (d+6)}{32 \text{den}_{11}} $ & $ -\frac{(d-2) (d+2) (d+4) (d+6)}{32 p^2 \text{den}_{11}} $ & $ \frac{p^4}{32
   \text{den}_{11}} $  \\ \hline
\end{tabular}
\\ \\ \\
%
\bigskip\bigskip
\begin{tabular}{|c||ccccc|} \hline
& $K^{(222221)}_{000011;1} $ & $K^{(222221)}_{000011;1} $ & $K^{(222221)}_{001111;1} $ & $K^{(222221)}_{001111;2} $ & $K^{(222221)}_{111111} $   \\ \hline \hline
$J_0$ &  $ \frac{4 \left(d^4-8 d^2-8\right)}{9 (d+1) \text{den}_{14}} $  & $ \frac{5 d^3+d^2-38 d+8}{18 \text{den}_{14}} $ & $ -\frac{(d-3) \left(10 d^5+20 d^4-29 d^3+41 d^2+82
   d-88\right)}{9 (d+1) p^2 \text{den}_8} $ & $ -\frac{(d-3) \left(17 d^4+29 d^3-42 d^2+64 d-32\right)}{18 p^2 \text{den}_8} $ & $ \frac{(d+2) \left(34 d^6-61 d^5-332 d^4+677
   d^3-38 d^2-1696 d+1056\right)}{9 (d+1) p^4 \text{den}_8} $  \\ \hline
$B_0^2$ & $ -\frac{\left(d^2-8\right) p^2}{32 \text{den}_{11}} $ & $ -\frac{(d+4) p^2}{32 \text{den}_{11}} $ & $ \frac{(d+4) \left(d^2-6\right)}{32 \text{den}_{11}} $ & $ \frac{(d+4) (d+6)}{32
   \text{den}_{11}} $ & $ -\frac{(d-2) (d+2) (d+4) (d+6)}{32 p^2 \text{den}_{11}} $  \\ \hline
\end{tabular}
\\
%
\bigskip\bigskip
\begin{tabular}{|c||ccccc|} \hline
& $K^{(111122)}_{000000;1} $ & $K^{(111122)}_{000000;2} $ & $K^{(111122)}_{000011;1} $ & $K^{(111122)}_{000011;2} $ & $ K^{(111122)}_{000011;3} $  \\ \hline \hline
$J_0$ &  $  -\frac{\left(d^4+3 d^3-8 d^2-24 d+16\right) p^2}{18 \text{den}_5} $ & $ -\frac{2 \left(d^4-3 d^3-8 d^2+30 d+4\right) p^2}{9 \text{den}_5} $ & $ \frac{5 d^5-10 d^4-49 d^3+86
   d^2+32 d+32}{9 \text{den}_5} $ & $ \frac{4 d^5+d^4-44 d^3+d^2+94 d-8}{18 \text{den}_5} $ & $ \frac{5 d^6+6 d^5-45 d^4-78 d^3+24 d^2+288 d-128}{36 \text{den}_4} $ \\ \hline
$B_0^2$ & $ \frac{p^4}{16 \text{den}_7} $ & $ \frac{\left(d^2-d-10\right) p^4}{32 \text{den}_7} $ & $ -\frac{\left(d^3-d^2-10 d+8\right) p^2}{32 \text{den}_7} $ & $ -\frac{\left(d^2+d-4\right)
   p^2}{32 \text{den}_7} $ & $ -\frac{(d+4) p^2}{16 \text{den}_7} $  \\ \hline
\end{tabular}
\\
%
\bigskip\bigskip
\begin{tabular}{|c||cccc|} \hline
& $K^{(111122)}_{000011;4} $ & $K^{(111122)}_{001111;1} $ & $K^{(111122)}_{001111;2} $ & $K^{(111122)}_{001111;3} $  \\ \hline \hline
$J_0$ & $  \frac{10 d^6-39 d^5-72 d^4+423 d^3-138 d^2-480 d+224}{18 \text{den}_4} $ & $ -\frac{2 (d-3) \left(4 d^4+6 d^3-23 d^2-18 d-8\right)}{9 (d-2) p^2 \text{den}_{14}} $ & $
   -\frac{(d-3) \left(10 d^5+17 d^4-41 d^3-22 d^2-32 d+32\right)}{18 p^2 \text{den}_6} $ & $ -\frac{(d-3) \left(17 d^5-11 d^4-178 d^3+112 d^2+224 d-128\right)}{9 p^2
   \text{den}_6} $  \\ \hline
$B_0^2$ & $ -\frac{\left(d^3+d^2-12 d-16\right) p^2}{32 \text{den}_7} $ & $ \frac{d \left(d^2-8\right)}{32 \text{den}_{16}} $ & $ \frac{d (d+4)}{32 \text{den}_{16}} $ & $ \frac{(d+4)
   \left(d^2-12\right)}{32 \text{den}_{16}} $ \\ \hline
\end{tabular}
\\
%
\bigskip\bigskip
\begin{tabular}{|c||ccccc|} \hline
& $K^{(111122)}_{111111} $ & $K^{(222211)}_{000000;1} $ & $K^{(222211)}_{000000;2} $ & $K^{(222211)}_{000011;1} $ & $K^{(222211)}_{000011;2} $   \\ \hline \hline
$J_0$ &  $  \frac{2 (d+2) \left(10 d^5-8 d^4-93 d^3+47 d^2-40 d+48\right)}{9 p^4 \text{den}_8} $  &  $ -\frac{\left(d^4+3 d^3-8 d^2-24 d+16\right) p^2}{18 \text{den}_5} $ & $ -\frac{2
   \left(d^4-3 d^3-8 d^2+30 d+4\right) p^2}{9 \text{den}_5} $ & $ \frac{5 d^5-10 d^4-49 d^3+86 d^2+32 d+32}{9 \text{den}_5} $ & $ \frac{4 d^5+d^4-44 d^3+d^2+94 d-8}{18
   \text{den}_5} $ \\ \hline
$B_0^2$ & $ -\frac{d (d+2) (d+4)}{32 (d+3) p^2 \text{den}_{12}} $ & $ \frac{p^4}{16 \text{den}_7} $ & $ \frac{\left(d^2-d-10\right) p^4}{32 \text{den}_7} $ & $ -\frac{\left(d^3-d^2-10
   d+8\right) p^2}{32 \text{den}_7} $ & $ -\frac{\left(d^2+d-4\right) p^2}{32 \text{den}_7} $ \\ \hline
\end{tabular}
\\
%
\bigskip\bigskip
\begin{tabular}{|c||cccc|} \hline
& $K^{(222211)}_{000011;3} $ & $K^{(222211)}_{000011;4} $ & $K^{(222211)}_{001111;1} $ & $K^{(222211)}_{001111;2} $  \\ \hline \hline
$J_0$ &  $ \frac{5 d^6+6 d^5-45 d^4-78 d^3+24 d^2+288 d-128}{36 \text{den}_4} $ &  $ \frac{10 d^6-39 d^5-72 d^4+423 d^3-138 d^2-480 d+224}{18 \text{den}_4} $ & $ -\frac{2 (d-3) \left(4
   d^4+6 d^3-23 d^2-18 d-8\right)}{9 (d-2) p^2 \text{den}_{14}} $ & $ -\frac{(d-3) \left(10 d^5+17 d^4-41 d^3-22 d^2-32 d+32\right)}{18 p^2 \text{den}_6} $ \\ \hline
$B_0^2$ & $ -\frac{(d+4) p^2}{16 \text{den}_7} $ & $ -\frac{\left(d^3+d^2-12 d-16\right) p^2}{32 \text{den}_7} $ & $  \frac{d \left(d^2-8\right)}{32 \text{den}_{16}} $ & $ \frac{d (d+4)}{32
   \text{den}_{16}}  $ \\ \hline
\end{tabular}
\end{sidewaystable}

\begin{sidewaystable}
%
\begin{tabular}{|c||cccc|} \hline
& $K^{(222211)}_{001111;3} $ & $K^{(222211)}_{111111} $ & $K^{(111222)}_{000000;1} $ & $K^{(111222)}_{000000;2} $  \\ \hline \hline
$J_0$ &  $ -\frac{(d-3) \left(17 d^5-11 d^4-178 d^3+112 d^2+224 d-128\right)}{9 p^2 \text{den}_6} $ & $ \frac{2 (d+2) \left(10 d^5-8 d^4-93 d^3+47 d^2-40 d+48\right)}{9 p^4
   \text{den}_8} $ & $ -\frac{(d+4) \left(d^3+3 d^2-4 d-36\right) p^2}{36 \text{den}_5} $ & $ -\frac{\left(2 d^4-d^3-23 d^2+22 d+72\right) p^2}{18 \text{den}_5} $  \\ \hline
$B_0^2$ & $ \frac{(d+4) \left(d^2-12\right)}{32 \text{den}_{16}} $ & $ -\frac{d (d+2) (d+4)}{32 (d+3) p^2 \text{den}_{12}} $ & $ \frac{3 p^4}{16 (d+1) \text{den}_{13}} $ & $ \frac{(d-3) (d+2)
   p^4}{32 (d+1) \text{den}_{13}} $ \\ \hline
\end{tabular}
\\ \\ \\
%
\bigskip\bigskip
\begin{tabular}{|c||ccccc|} \hline
& $K^{(111222)}_{000011;1} $ & $K^{(111222)}_{000011;2} $ & $K^{(111222)}_{000011;3} $ & $K^{(111222)}_{000011;4} $  & $K^{(111222)}_{001111;1} $  \\ \hline \hline
$J_0$ & $\frac{d \left(5 d^4-2 d^3-57 d^2+26 d+136\right)}{18 \text{den}_5} $ & $\frac{d (d+2)^2 \left(2 d^3-3 d^2-15 d+20\right)}{18 \text{den}_4} $ & $ \frac{d \left(5 d^4-2 d^3-57
   d^2+26 d+136\right)}{18 \text{den}_5} $ & $ \frac{d \left(4 d^5-14 d^4-29 d^3+145 d^2-38 d-104\right)}{9 \text{den}_4} $ & $ -\frac{(d-3) d (d+1) \left(10 d^3-7 d^2-82
   d+88\right)}{9 p^2 \text{den}_6} $ \\ \hline
$B_0^2$ & $ -\frac{d \left(d^2+d-8\right) p^2}{32 (d+1) \text{den}_{13}} $ & $ -\frac{d (d+2) p^2}{16 (d+1) \text{den}_{13}} $ & $ -\frac{d \left(d^2+d-8\right) p^2}{32 (d+1)
   \text{den}_{13}} $ & $ -\frac{d \left(d^3-d^2-10 d+4\right) p^2}{32 (d+1) \text{den}_{13}} $ & $ \frac{d^2 \left(d^2-10\right)}{32 \text{den}_{13}}  $ \\ \hline
\end{tabular}
\\
%
\bigskip\bigskip
\begin{tabular}{|c||ccc|} \hline
& $K^{(111222)}_{001111;2} $ & $K^{(111222)}_{001111;3} $ & $K^{(111222)}_{111111} $  \\ \hline \hline
$J_0$ & $  -\frac{2 (d-3) d (d+2) \left(2 d^3+2 d^2-11 d+4\right)}{9 p^2 \text{den}_6} $ & $ -\frac{(d-3) d (d+1) \left(10 d^3-7 d^2-82 d+88\right)}{9 p^2 \text{den}_6} $ & $ \frac{4 d
   (d+2) \left(4 d^4+2 d^3-46 d^2-29 d+60\right)}{9 p^4 \text{den}_8} $ \\ \hline
$B_0^2$ & $ \frac{d^2 (d+2)}{32 \text{den}_{13}} $ & $ \frac{d^2 \left(d^2-10\right)}{32 \text{den}_{13}} $ & $ -\frac{d^2 (d+2)^2}{32 p^2 \text{den}_{18}}  $ \\ \hline
\end{tabular}
\end{sidewaystable}

\newpage

\bea
\text{den}_1 &=& (d-4) (3d-4) (3d-2) \,, \nn \\
\text{den}_2 &=& (d-4) (3d-4) \,, \nn \\
\text{den}_3 &=& (d-4) (d-1) \,, \nn \\
\text{den}_4 &=& (d-4)^2 (d-2) (d+3) (3 d-2) (d-1) (d+1) (3 d-4) \,, \nn \\
\text{den}_5 &=& (d-4)^2 (d-2) (d+1) (d+3) (3 d-4) (3 d-2) \,, \nn \\
\text{den}_6 &=& (d-4)^2 (d-2) (d-1) (d+3) (3 d-4) (3 d-2)  \,, \nn \\
\text{den}_7 &=& (d-4) (d-1) (d+1) (d-2) (d+3)  \,, \nn \\
\text{den}_8 &=& (d-4)^2 (d-1) (d+3) (3 d-4) (3 d-2)  \,, \nn \\
\text{den}_{9} &=& (d-4)^2 (d+1) (3 d-4) \,, \nn \\
\text{den}_{10} &=& (d-4)^2 (d-1) (d+1) (3 d-4) \,, \nn \\
\text{den}_{11} &=& (d-4) (d-1) (d+1) (d + 3) \,, \nn \\
\text{den}_{12} &=& (d-4) (d-1)  \,, \nn \\
\text{den}_{13} &=& (d-4) (d-2) (d-1)^2 (d+3) \,, \nn \\
\text{den}_{14} &=& (d-4)^2 (d+3) (3 d-4) (3 d-2) \,, \nn \\
\text{den}_{15} &=& (d-4) (d-1) (d-2) (d+1) \,, \nn \\
\text{den}_{16} &=& (d-4) (d-1) (d-2) (d+3) \,, \nn \\
\text{den}_{17} &=& (d-4)^2 (d-1)  \,, \nn \\
\text{den}_{18} &=& (d-4) (d-1)^2 (d+3) \,, \nn \\
\text{den}_{19} &=& (d-4)^2 (d-1) (3 d-4) \,, \nn \\
\text{den}_{20} &=& (d-4) (d-2) (d+1) \,, \nn \\
\text{den}_{21} &=& (d-4) (d-1) (d+1) \,.
\eea

\clearpage
\phantomsection
\addcontentsline{toc}{section}{References}

\providecommand{\href}[2]{#2}\begingroup\raggedright\endgroup

\end{document}